\documentclass[preprint,11pt,3p]{elsarticle}

\usepackage{amssymb}
\usepackage{amsmath}
\usepackage{amsfonts}

\usepackage{subcaption}
\usepackage{booktabs}
\usepackage{nicefrac}

\usepackage{algpseudocode}
\usepackage{algorithm}
\usepackage{algorithmicx}

\algnewcommand\algorithmicswitch{\textbf{switch}}
\algnewcommand\algorithmiccase{\textbf{case}}
\algnewcommand\algorithmicassert{\texttt{assert}}
\algnewcommand\Assert[1]{\State \algorithmicassert(#1)}%
\algdef{SE}[SWITCH]{Switch}{EndSwitch}[1]{\algorithmicswitch\ #1\ \algorithmicdo}{\algorithmicend\ \algorithmicswitch}%
\algdef{SE}[CASE]{Case}{EndCase}[1]{\algorithmiccase\ #1}{\algorithmicend\ \algorithmiccase}%
\algtext*{EndCase}%

\newcommand{\myalgnumfont}{\fontsize{8pt}{9pt}\selectfont}
\algrenewcommand{\alglinenumber}[1]{\myalgnumfont #1:}

\usepackage{float}
\usepackage{multirow}
\usepackage{xcolor,colortbl}
\usepackage{longtable}
\usepackage{url}
\usepackage[utf8]{inputenc}

\usepackage{hyperref}

\journal{Journal of \LaTeX\ Templates}

\newcommand{\ie}{i.\,e.}

\newcommand{\norm}[1]{\left\lVert#1\right\rVert}

\usepackage{mathtools}
\DeclarePairedDelimiter\ceil{\lceil}{\rceil}

\begin{document}

\begin{frontmatter}

\title{Solving large permutation flow-shop scheduling problems on GPU-accelerated supercomputers}%

\author{Jan Gmys}
\address{Inria Lille Nord Europe}

\begin{abstract}
Makespan minimization in permutation flow-shop scheduling is a well-known hard combinatorial optimization problem. 
Among the $120$ standard benchmark instances proposed by E.~Taillard in 1993, $23$ have remained unsolved for almost three decades.
In this paper, we present our attempts to solve these instances to optimality using parallel Branch-and-Bound tree search on the GPU-accelerated \textit{Jean Zay} supercomputer. We report the exact solution of $11$ previously unsolved problem instances and improved upper bounds for $8$ instances.
The solution of these problems requires both algorithmic improvements and leveraging the computing power of peta-scale high-performance computing platforms.
The challenge consists in efficiently performing parallel depth-first traversal of a highly irregular, fine-grained search tree on distributed systems composed of hundreds of massively parallel accelerator devices and multi-core processors.
We present and discuss the design and implementation of our permutation-based B\&B and experimentally evaluate its parallel performance on up to 384 V100 GPUs (2 million CUDA cores) and 3840 CPU cores.
The optimality proof for the largest solved instance requires about $64$ CPU-years of computation
---using $256$ GPUs and over $4$ million parallel search agents, the traversal of the search tree is completed in $13$ hours, exploring $339\times 10^{12}$ nodes.

\end{abstract}

\begin{keyword}
Permutation flow-shop scheduling \sep Branch-and-Bound \sep Supercomputing \sep GPU computing
\end{keyword}

\end{frontmatter}

\section{Introduction}\label{sec:intro}
Many combinatorial optimization problems (e.g. scheduling, assignment or routing problems) can be modeled by using permutations to represent candidate solutions. 
In this work, we focus on the Permutation Flowshop Scheduling Problem (PFSP) with makespan criterion.
The problem consists in scheduling $n$ jobs in identical order on $m$ machines, given the processing times $p_{jk}$ for job $J_j$ on machine $M_k$ and the constraint that a job can only start on machine $M_k$ if it is completed on all upstream machines $M_1, M_2, \ldots, M_{k-1}$.
The goal is to find a permutation (a processing order) that minimizes the completion time of the last job on the last machine, called makespan.

The problem is NP-hard for $m\geq 3$~\citep{Garey76} and exact algorithms like Branch-and-Bound (BB) can only solve small-sized instances within a reasonable amount of time.
BB performs an implicit enumeration of all possible solutions by dynamically constructing and exploring a tree. This is done using four operators: branching, bounding, selection and pruning.
For larger problem instances, the exhaustive exploration of the search space becomes practically infeasible on a sequential computer.
In this article, we present PBB$@$Cluster, a permutation-based BB (PBB) algorithm for heterogeneous clusters composed of multi-core processors and GPU accelerator devices. 
Scaling PBB$@$Cluster on hundreds of GPUs of the \textit{Jean Zay} supercomputer (\#57 in the Top500 ranking, Nov. 2020), we aim at solving hard PFSP benchmark instances to optimality that remained open for almost three decades.

Our motivation is twofold.
On the one hand, the knowledge of exact optimal solutions for benchmark instances is highly valuable, as they provide a baseline for assessing the quality of metaheuristics and other approximate methods.
In the case of the PFSP, the set of $120$ benchmark instances proposed by E.~Taillard in 1993~\citep{Taillard1993} is the most frequently used.
While instances defined by less than $20$ machines are relatively easy to solve~\citep{GmysEJOR},
most of Taillard's instances with $m=20$ machines and $n\geq50$ jobs are very hard and optimal solutions for $23$ of them remain unknown $27$ years after their introduction.

On the other hand, the efficient parallel design and implementation of backtracking/BB algorithms is challenging, mainly because this ``computational dwarf"\citep{asanovic2006landscape} is highly irregular.
Moreover, the efficient design of parallel BB is strongly influenced by the tackled problem (search space, goal and granularity) and on the targeted compute platform~\cite{bader2005parallel}.
The potential for exploiting parallelism in BB has been recognized as early as 1975 and research activity started to intensify ten years later, as parallel processing capabilities became practically available~\cite{Pruul1975}.
A survey which covers the main research lines on parallel BB from 1975 to 1994 may be found in~\cite{Gendron94}.
To place this article in a historical context, it could be useful to point out two particularly fruitful research phases.

In the early 1990s, the design of massively parallel tree-search algorithms on top of SIMD supercomputers (MasPar, CM-2, Intel Hypercube, etc.) has attracted much attention~\citep{RaoParallelDFS1987, Karypis1994UnstructuredTS}, with research interests focusing on data-parallel load balancing strategies~\citep{Fonlupt1994,reinefeld1994work}. Frequently used applications include puzzles (e.g. 15-puzzle) and games (e.g. Othello), which are characterized by regular fine-grained evaluation functions and highly irregular search trees.
Notably, backtracking and BB algorithms for modern GPUs are designed similarly, and many GPU-based parallel tree-search algorithms target fine-grained applications as well~\citep{CarneiroICA3PP,RockiMinimax,BacktrackingLessons2011}.

A decade later, with the emergence of cluster and grid computing, research focus shifted towards the design of parallel BB algorithms on top of distributed, heterogeneous and volatile platforms, targeting more coarse-grained applications of BB~\citep{Crainic2006}.
The use of well-engineered grid-enabled BB algorithms has led to breakthroughs such as the resolution, in 2002, of quadratic assignment problem (QAP) instances which had remained unsolved for over three decades (including the notorious \textit{nug30} instance)~\citep{anstreicher2002solving}.
Solving \textit{nug30} to optimality involved 7 days of computation using 650 CPUs on average and explored $12\times 10^9$ tree nodes.
The design of a BB algorithm using a new, stronger lower bound (LB) and its parallel implementation---based on Condor and Globus---on a large computational grid were vital in bringing about this achievement.
About $15$ years later,~\citet{Date2017RLT2basedPA} used an even stronger LB to re-solve \textit{nug30} on a GPU-powered cluster (Blue Waters$@$Urbana-Champaign) within $4$ days and successfully solve QAP instances with up to $42$ facilities. 
Their approach uses GPUs to accelerate the computation of strong LBs which require solving $O(n^4)$ independent linear assignment problems. 
Compared to the first exact solution in 2002, their algorithm explores a search tree several orders of magnitude smaller ($\sim 10^6$ nodes).	

The PBB$@$Cluster approach presented in this work can be viewed as a combination of concepts developed for grids and clusters of mono-core CPUs on the one hand and fine-grained massively parallel backtracking on the other.
For the PFSP, the largest attempt to exactly solve hard problem instances has been carried out by~\citet{MezmazIPDPS}, who designed and deployed a grid-enabled PBB algorithm on more than 1000 processors. 
This effort has led to the exact resolution of instance \textit{Ta056} in $25$ days ($22$ CPU-years), exploiting $328$ processors on average.
Attempts of similar scale to solve other open instances from Taillard's benchmark have been unfruitful, indicating that their solution would require algorithmic advances and/or much more processing power.
Following the work of~\citet{MezmazIPDPS}, the PFSP has been used as a test-case for several PBB algorithms targeting heterogeneous distributed systems combining multi-core CPUs and GPUs~\citep{ChakrounJPDC15,Vu2016,GmysPPAMSE}. 
However, despite reaching speed-ups between two and three orders of magnitude over sequential execution, no new solutions for the remaining Taillard benchmark instances were reported.

BB algorithms compute lower bounds (LB) on the optimal cost reachable by further exploring a partial solution to avoid unnecessary enumeration of the corresponding subspaces.
The tradeoff between the computational complexity of a LB and its potential to reduce the size of the explored tree is crucial.
For the PFSP, the strongest LB is the one proposed by \citet{Lageweg78} and the latter is used in almost all parallel BB approaches. 
However, we have shown in~\citep{GmysEJOR} that using a weaker, easy-to-compute LB from one of the first BB algorithms~\citep{Ignall1965} allows to solve large PFSP instances more efficiently, if this LB is combined with bi-directional branching.
Although weakening the LB increases the size of the explored search tree, empirical results indicate that a better overall tradeoff is achieved.
Therefore, instead of strengthening the LB---as for the QAP---we take a step in the opposite direction, i.e. compared to previous approaches, our PBB algorithm uses a weaker, more fine-grained LB. 
To give an idea of scale, with the LB used in this work, a single node evaluation can be performed in less than $10^{-6}$ seconds, and trees are composed of up to $\sim 10^{15}$ nodes.

\subsection{Contributions and related works}
The main result of this work can be summarized as follows: 
\begin{itemize}
\item $11$ out of $23$ open PFSP instances from Taillard's benchmark are solved to optimality using a scalable GPU-accelerated PBB algorithm on up to 96 quad-GPU nodes (3840 CPU cores and nearly 2 million CUDA cores) of the \textit{Jean Zay} supercomputer.
\end{itemize}
Moreover, the best-known solutions for $8$ of Taillard's instances are improved.
For the VFR benchmark~\citep{Vallada2015}, $38$ instances are solved for the first time and additionally $75$ best-known upper bounds are improved.
Scalability experiments show that PBB$@$Cluster achieves a parallel efficiency of $\sim 90\%$ on $16$, $64$ and $128$ GPUs for problem instances requiring respectively $1$, $4$ and $27$ hours of processing on a single GPU.
The largest solved instance requires over $13$ hours of processing on $256$ V100 GPUs, i.e. a total of $3400$ GPU-hours---which amounts to an estimated equivalent CPU time of $64$ years.

Although our work is the first to solve PFSP instances on a peta-scale system, we should point out that the presented results are not ``simply" a matter of brute force.
Instead, PBB$@$Cluster builds upon research efforts that stretch over several years and deal with the following challenging issues, many of which stem from the highly irregular nature of the algorithm.

\begin{itemize}
\item A key component of each search algorithm is the underlying \textbf{data structure}. PBB$@$Cluster performs up to $10^{10}$ node decompositions per second, so it is essential to define an efficient data structure for the storage and management of this ``tsunami'' of subproblems, dynamically generated at runtime. 
With a very fine-grained LB it is crucial to keep the overhead of search tree management and work distribution low, because the cost of these operations cannot be neglected (compared to node evaluations) like in coarse-grained BB algorithms.
Therefore, PBB$@$Cluster is based on an innovative data structure, called IVM~\citep{IVM-IPDPS2014}, which is dedicated to permutation problems. 
A crucial advantage of IVM, compared to conventional linked-list-based data structures, lies in its compact and constant memory-footprint, which is well-suited for GPU-based PBB implementations (PBB$@$GPU,~\citep{GmysParco}).
\item As the search tree is highly irregular, the scalability of parallel tree-exploration with millions of concurrent explorers essentially depends on the efficiency of \textbf{load balancing} mechanisms---which in turn rely on a suitable definition of work units. 
In PBB$@$Cluster, a hierarchical load balancing scheme (on the GPU and inter-node levels) with an interval-based encoding of work-units ensures that exploration-agents are kept busy.
The encoding of work units as intervals has been developed and experimented in the context of an IVM-based multi-core PBB using work-stealing with intervals of factoradics~\citep{IVM-IPDPS2014}, and successfully used in work-stealing schemes for PBB$@$GPU~\citep{GmysPPAMSE}.
The PBB$@$Grid approach of~\citet{MezmazIPDPS} uses a similar interval-encoding of work units with conventional linked-list-based data structures.
\item The algorithm is also irregular on the level of individual exploration-agents. In particular, the node evaluation function is characterized by irregular memory access patterns and diverging control flow.
This makes it difficult to take advantage \textbf{low-level parallelism} and impedes SIMD/SIMT execution efficiency. 
For the PFSP makespan and LB evaluation functions, vectorization approaches have been proposed in~\citep{BozejkoPFSP,Melab2018FGCS}. 
Mapping strategies for reducing thread divergence in PBB$@$GPU were investigated in~\citep{GmysParco}. 
In this work, we revisit the vectorization of the fine-grained LB used by PBB$@$Cluster to speed up node evaluation and exploit warp-level parallelism.
\item PBB prunes subproblems whose LB is greater than the best found solution so far. 
Therefore, it is important to \textbf{discover optimal solutions} quickly and thereby maximize the pruning rate. 
Any improvement of the incumbent solution may dramatically accelerate the exploration process and lead to superlinear speedups~\citep{deBruin1995,GmysEJOR}.
Therefore, as depth-first search (DFS) alone fails in general to find high-quality solutions, the hybridization of PBB$@$Cluster with approximate search methods is a key element for the solution of open instances.
PBB$@$Cluster uses metaheuristic searches, running on CPU processing cores, to discover high-quality solutions in parallel to and in cooperation with the GPU-based exhaustive search.
\item \textbf{Communication} patterns in PBB$@$Cluster are irregular as well: 
several unpredictable events (new best solutions, work exhaustion, checkpointing, termination detection) trigger threads and processes to exchange messages of different kinds and sizes. 
Moreover, early preliminary experiments show that, to be scalable, inter-node communications should be asynchronous on the worker-side, i.e. a primary design goal is to interrupt the GPU-based tree-traversal as little as possible.  PBB$@$Cluster uses pthreads mutual exclusion primitives for shared-memory communication and MPI at the inter-node level. While the inter-node level of PBB$@$Cluster is conceptually similar to the coordinator-worker approach in BB$@$Grid~\citep{MezmazIPDPS}, the latter is designed for mono-core processors and uses C++ socket programming. Switching to GPU-based workers and to MPI motivated us to revisit the design of the coordinator in depth, redefining work units, in particular.
\item A reliable global \textbf{checkpointing} mechanism is an indispensable component of PBB$@$Cluster. 
On the one hand, as the time required for solving a particular instance is unpredictable, node reservations may expire before the exploration is completed. On the other hand, the mean time between failure (MTBF) on large supercomputers keeps decreasing as we're entering the exascale era~\citep{cappello2009fault}.
Therefore, a minimum requirement is to be able to re-start the exploration process from the last global checkpoint without impeding correctness.
\end{itemize}

The remainder of this paper is organized as follows.
First, in Section~\ref{sec:background} we define the search space and goal associated with the PFSP, followed by a presentation of 
the sequential algorithm design, 
the IVM data structure and work units.
Section~\ref{sec:pbbgpu} presents PBB$@$GPU, the GPU-based PBB algorithm at the core of worker processes in PBB$@$Cluster.
In Section~\ref{sec:pbbcluster} we describe the design and implementation of PBB$@$Cluster's coordinator and worker processes.
Experimental results are reported in Section~\ref{sec:experiments} and 
finally, some conclusions are drawn in Section~\ref{sec:conclusion}.
Improved upper bounds and new optimal schedules for benchmark instances are provided in~\ref{appendix}.

\section{Branch-and-Bound for the PFSP}\label{sec:background}

\subsection{Problem formulation}
The flowshop scheduling problem (FSP) can be formulated as follows.
Each of $n$ jobs $J=\{J_1, J_2, \ldots, J_n\}$ has to be processed on $m$ machines $M_1, M_2, \ldots, M_m$ in that order.
The processing of job $J_j$ on machine $M_k$, takes an uninterrupted time $p_{jk}$, given by a processing time matrix.
Each machine can process at most one job at a time and jobs cannot be processed simultaneously on different machines. 

A common simplification is to consider only permutation schedules, i.e. to enforce an identical processing order on all machines, which reduces the size of the search space from $(n!)^m$ to~$n!$.
Considering minimization of the completion time of the last job on the last machine, called \textit{makespan}, the resulting problem is the permutation flow-shop problem (PFSP) with makespan criterion, denoted $F_m|prmu|C_{\max}$.
Formally, denoting 
$\pi=(\pi(1), \ldots, \pi(n))\in S_n$ a permutation of length $n$, and 
$C_{j,k}$ the completion time of job $J_j$ on machine $M_k$,
the goal is to find an optimal permutation $\pi^{\star}$ such that
\[
C_{\max}(\pi^{\star})=\underset{\pi\in S_n}{\min}C_{\max}(\pi)
\]
where $C_{\max}(\pi)=C_{\pi(n),m}$.

For $m=2$, the problem can be solved in $O(n\log n)$ steps by sorting the jobs according to Johnson’s rule~\cite{Johnson54}; for $m\geq 3$ it is shown to be NP-hard~\cite{Garey76}.
The completion times $C_{\pi(j),k}$ can be obtained recursively by
\begin{equation}\label{eq:cmax}
C_{\pi(j),k}=\max\left(
C_{\pi(j),k-1},C_{\pi(j-1),k}
\right)+
p_{\pi(j),k}
\end{equation}
where $p_{\pi(0),k}=p_{j,0}=0$ by convention. 
Thus, for a given schedule $\pi$, the makespan $C_{\max}(\pi)=C_{\pi(n),m}$ can be computed in $O(mn)$ time.

\subsection{Branch-and-Bound for permutation problems}
BB performs an implicit enumeration of the search space by dynamically constructing and exploring a tree.
The root node represents the initial problem, internal nodes represent subproblems (partial solutions) and leaves are feasible solutions (permutations).
The algorithm starts by initializing the best solution found so far (also called the incumbent) and the data-structure used for storing the tree such that it contains only the root node.
Then, the search space is explored by using four operators: selection, branching, bounding, and pruning. 

\begin{figure}[tbp]
\begin{center}
	\includegraphics[width=0.4\linewidth]{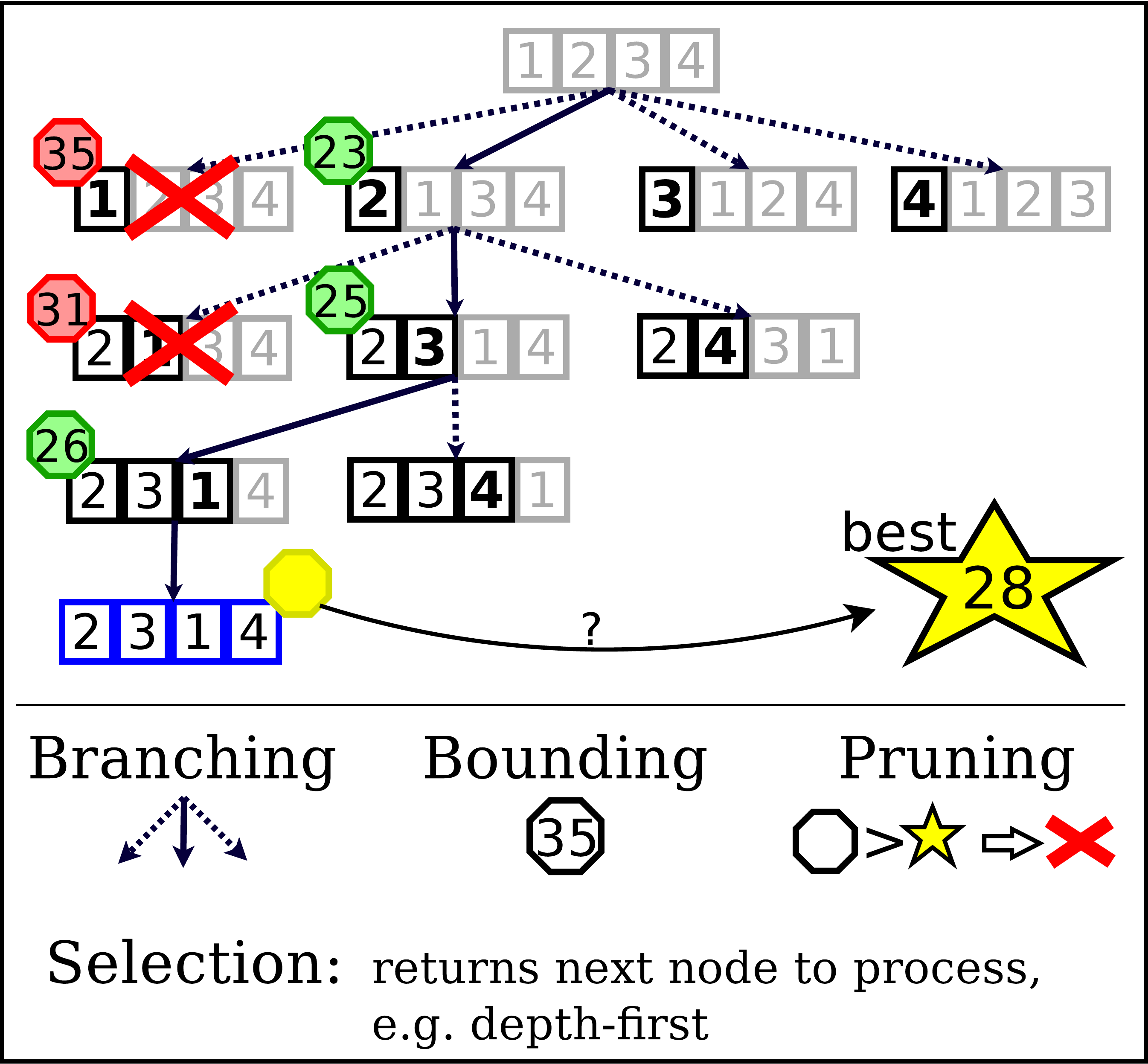}
		\caption{Illustration of a simple Branch-and-Bound algorithm for a permutation problem of size four.}
			\label{fig:simpleBB}
\end{center}
\end{figure}

Figure~\ref{fig:simpleBB} shows an illustration of the four BB-operators for a permutation problem of size four.
At each iteration, the \textbf{selection} operator returns the next subproblem to explore, starting with the root node.
The \textbf{branching} operator decomposes the subproblem into smaller disjoint subproblems. 
The \textbf{bounding} operator computes lower bounds (LB) on the optimal cost of these subproblems, in the sense that no arrangement of unscheduled jobs can yield a smaller makespan than this LB.
Based on these LBs, the \textbf{pruning} operator discards subproblems from the search that cannot lead to an improvement of the incumbent solution. 
All non-pruned subproblems are inserted into the data structure for further exploration.
In the following subsections we specify the branching and bounding operators used in the work, as well as the data structure used for storing subproblems.

\subsection{Branching rule} %
In the example of Figure~\ref{fig:simpleBB}, permutations are built from left to right, meaning that a node of depth $d$ can be represented by a prefix  partial schedule $\sigma_1$ of $d$ jobs. The \textit{forward} branching operation consists in generating $n-d$ child subproblems as follows:
\[
\text{Forward-Branch}: \sigma_1 \mapsto \{\sigma_1j,\,j\in J\setminus\sigma_1\}.
\]
A second branching type is \textit{backward} branching, which prepends unscheduled jobs to a postfix partial schedule $\sigma_2$, i.e.
\[
\text{Backward-Branch}: \sigma_2 \mapsto \{j\sigma_2,\,j\in J\setminus\sigma_2\}.
\]
Our PBB algorithm uses a \textit{dynamic} branching rule, which decides dynamically, for each decomposed node, which of the two branching types is applied.

With dynamic branching, subproblems are represented in the form $(\sigma_1,\sigma_2)$ and are decomposed as follows  
\[
\text{Dyna-Branch}: (\sigma_1,\sigma_2) \mapsto 
\begin{cases}
\{(\sigma_1j,\sigma_2),\,j\in J\setminus(\sigma_1,\sigma_2)\} & \text{if Fwd-Branch} \\
\{(\sigma_1,j\sigma_2),\,j\in J\setminus(\sigma_1,\sigma_2)\} & \text{if Bwd-Branch}
\end{cases}
\]
In order to make the branching decision, both possible children sets are generated and a heuristic chooses the one which is more likely to minimize the size of the search tree.
In this work, the branching heuristic, called \textit{MinMin}, chooses the set in which the minimal LB (among both sets) is realized less often.
Equality ties are broken by choosing the set where the sum of LBs is higher, and \textit{forward} if the sums are equal as well.

\begin{figure}[tpb]
\begin{center}
	\includegraphics[width=0.6\linewidth]{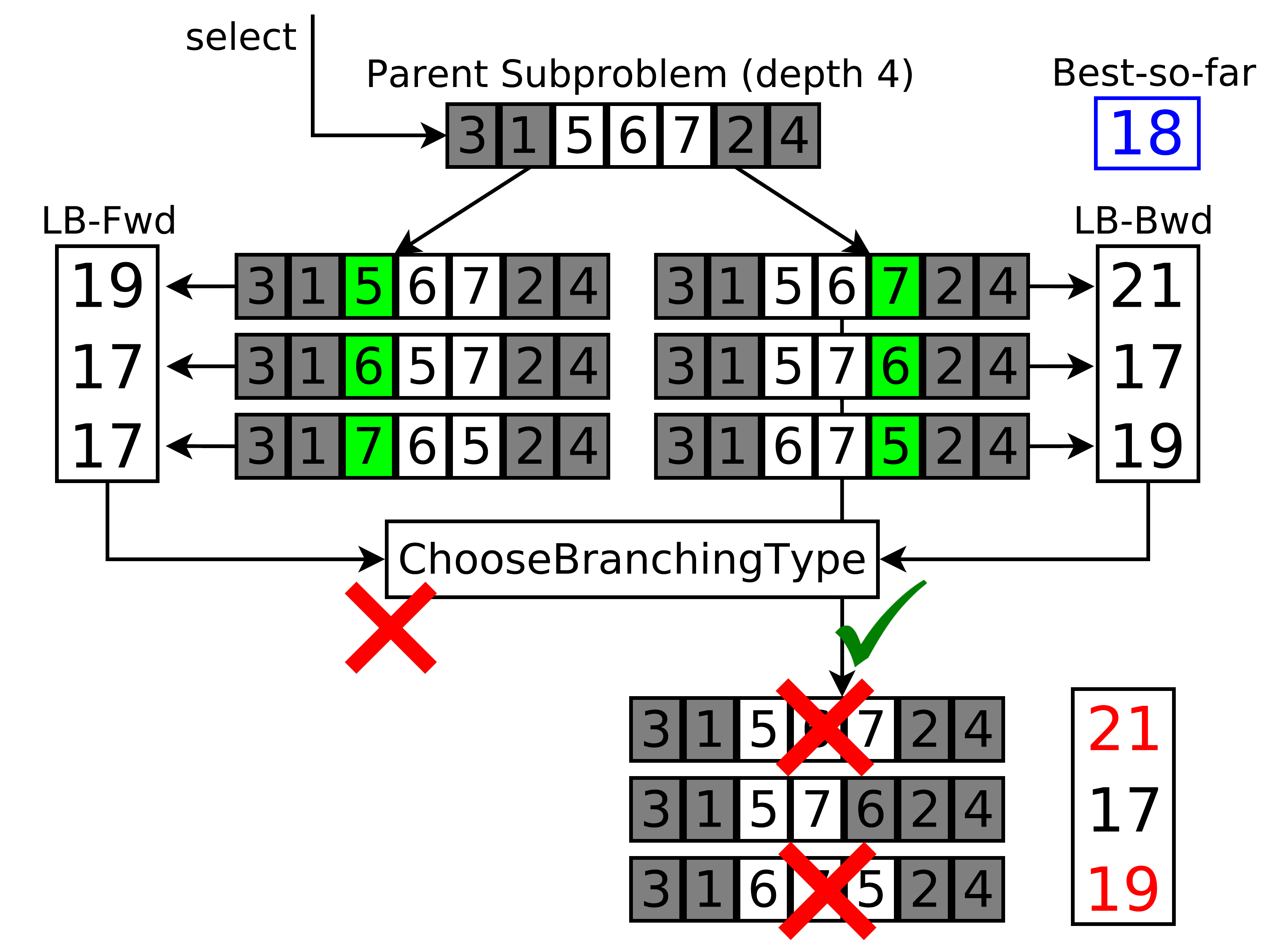}
		\caption{Illustration of a subproblem decomposition using dynamic branching}
			\label{fig:decompose}
\end{center}
\end{figure}

Figure~\ref{fig:decompose} illustrates the complete node decomposition procedure, involving branching, bounding and pruning.
In the example, the decomposed subproblem is $(\sigma_1,\sigma_2)=((3,1),(2,4))$, where jobs $\{5,6,7\}$ remain to be scheduled and the best makespan found so far is $18$.
Both children sets are evaluated, yielding $\text{LB}_{\text{Fwd}}=\{19,17,17\}$ for \textit{forward} and $\text{LB}_{\text{Bwd}}=\{21,17,19\}$ for \textit{backward}.
The smallest LB ($17$) occurs less frequently in $\text{LB}_{\text{Bwd}}$, so the \textit{MinMin} heuristic chooses \textit{backward} branching. The computed LBs are reused by the pruning operator, which, in the example, eliminates two subproblems (instead of one in the alternative branch).

\subsection{Lower Bound} %
The LB used in this work comes for the pioneering algorithms proposed independently by~\citet{Lomnicki1965} and \citet{Ignall1965}.
The so-called one-machine bound (LB1), was initially developed for BB algorithms using only forward branching, but it can be extended to the bi-directional subproblem representation, as proposed by~\citet{Potts1980}.

\begin{figure}[tpb]
\begin{center}
	\includegraphics[width=0.8\linewidth]{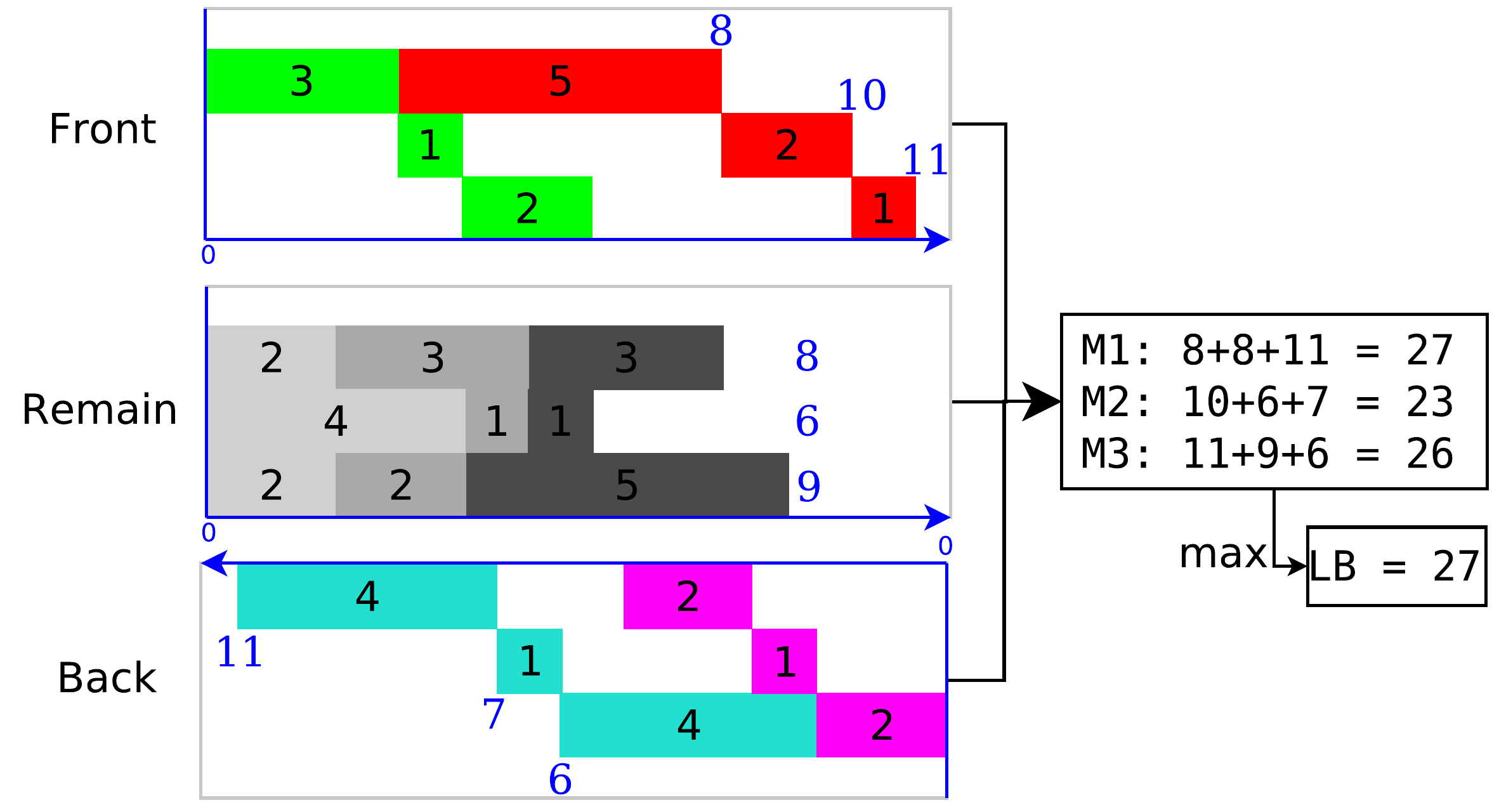}
		\caption{Lower bound computation for a subproblem $(\sigma_1=(\text{green},\text{red}),\sigma_2=(\text{cyan},\text{magenta}))$ with 3 unscheduled jobs (gray). No arrangement of the unscheduled jobs can give a better makespan than $\text{LB}=\max\{8+8+11, 10+6+7, 11+9+6\}=27$.}
			\label{fig:LB}
\end{center}
\end{figure}

The computation of LB1 for a subproblem $(\sigma_1,\sigma_2)$ can be divided into four steps, illustrated in Figure~\ref{fig:LB}.
We denote $|\sigma_1|=d_1$ and $|\sigma_2|=d_2$ the number of jobs scheduled in the prefix and suffix partial schedules respectively.
\begin{enumerate}
\item For the front, compute $C_{\sigma_1(d_1),k}$, the completion time of the last job in $\sigma_1$ on each machine, i.e. the earliest possible starting time for unscheduled jobs, given $\sigma_1$.
\item For the unscheduled jobs, compute $p(k)=\sum_{j\in J\setminus(\sigma_1,\sigma_2)}p_{j,k}$, the total remaining processing time on each machine.
\item For the back, compute $\bar{C}_{\sigma_2(d_2),k}$, the minimum time required between starting the first job in $\sigma_2$ on machine $M_k$ and the end of operations on the last machine. These values are obtained by scheduling $\sigma_2$ \textit{in reverse order} in front, using Equation~(\ref{eq:cmax}) and the reversibility property of the PFSP~\citep{Potts1980}.
\item Finally, LB1 is obtained by
\[
LB1(\sigma_1,\sigma_2)=\max_{k=1,\ldots,m}C_{\sigma_1(d_1),k}+p(k)+\bar{C}_{\sigma_2(d_2),k}.
\]
\end{enumerate}

Clearly, LB1 has the same time complexity as a makespan evaluation, i.e. $O(mn)$.
However, reusing the quantities computed for $(\sigma_1,\sigma_2)$, it is possible to deduce LB1 for a child subproblem in $O(m)$ steps.
Therefore, the computation of LB1 \textit{for all} $2\times(n-d_1-d_2)$ children of $(\sigma_1,\sigma_2)$ also requires $O(mn)$ steps.
Moreover, this incremental evaluation of the children requires only $O(1)$ additional memory per child.

LB1 is dominated by the two-machine bound LB2, proposed by~\citet{Lageweg78}, which relies on the exact resolution, for different machine-pairs, of two-machine problems using Johnson's rule.
In addition to the front/back computations, the different variants of LB2 require between $m$ and $\nicefrac{m(m-1)}{2}$ evaluations of (pre-sorted) two-machine Johnson schedules \textit{for each} child node. 
Therefore, LB2 requires between $O(mn^2)$ and $O(m^2n^2)$ time for the evaluation of all child nodes in a decomposition step. In~\citep{GmysEJOR} we found that, in combination with dynamic branching and especially for large problem instances, LB1 provides a better tradeoff between sharpness and computational effort.
To give an approximate measure of comparison, using dynamic branching and LB1, Taillard's \textit{Ta56} instance can be solved in $33$~hours on a dual-socket Intel Xeon node ($\sim 600$ CPU-hours)---with LB2 and the same branching scheme, the required CPU-time is $22$ years ($300\times$ more)~\citep{MezmazIPDPS}.

\subsection{Search strategy and data structure}\label{subsect:IVM}%
The next node to be decomposed is chosen according to a predefined selection strategy.
In this work, we consider only depth-first search (DFS), because memory requirements of best-first and breadth-first search grow exponentially with the problem size~\footnote{For example, solving \textit{Ta058} ($n=50$), the \textit{critical tree} (composed of nodes with LBs smaller than the optimal cost) contains $339\times 10^{12}$ nodes, so there exists at least one level with more than $6.7\times 10^{12}$ open subproblems. Assuming that each subproblem is stored as a sequence of $n=50$ $32$-bit integers, breadth-first exploration would require $6.7\times 10^{12}\times 50\times4$~B = $1.4$~PB of memory.}.
The data structure used for storing generated subproblems is closely related to the choice of the search strategy. 
In this work, we use the Integer-Vector-Matrix (IVM) data structure proposed in~\citep{IVM-IPDPS2014}. 
It is dedicated to permutation problems and provides a memory-efficient alternative to stacks, which are conventionally used for DFS.
The working of the IVM data structure is best introduced with an example.

\begin{figure}[tbp]
\begin{center}
	\includegraphics[width=0.9\linewidth]{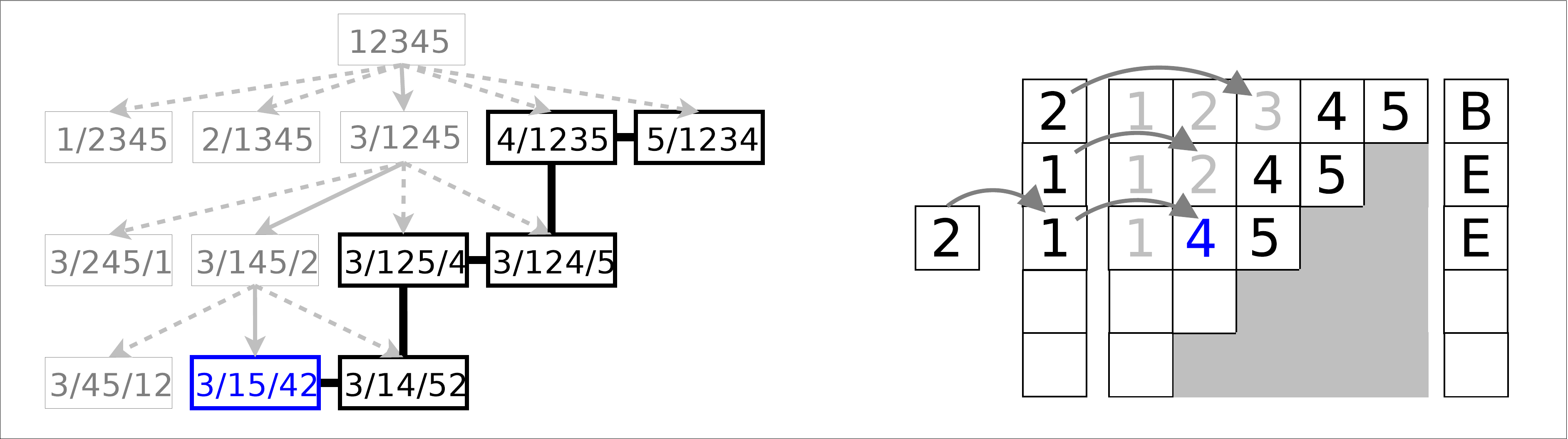}
		\caption{Tree and IVM-based representation of the search state, solving a permutation problem of size $5$.}\label{fig:tree}
\end{center}
\end{figure}

Figure~\ref{fig:tree} illustrates a pool of subproblems that could be obtained when solving a permutation problem of size $n=5$ with a DFS-PBB  algorithm using bi-directional branching.
On the left-hand side, Figure~\ref{fig:tree} shows a tree-based representation of this pool.
The parent-child relationship between subproblems is represented by dashed gray arrows.
The jobs before the first ``/'' symbol form the initial sequence $\sigma_1$, the ones behind the second ``/'' symbol form the final sequence $\sigma_2$ and jobs between the two ``/'' symbols represent the set of unscheduled jobs in arbitrary order.

On the right-hand side, IVM indicates the next subproblem to be solved.
The integer $I$ of IVM gives the level of this subproblem, using 0-based counting (at level $0$ one job is scheduled).
In this example, the level of the next subproblem is $2$.
The vector $V$ contains, for each level up to $I$, the position of the selected subproblem among its sibling nodes in the tree.
In the example, jobs $3$, $2$ and $4$ have been scheduled at levels $0$, $1$ and $2$ respectively.
The matrix $M$ contains the jobs to be scheduled at each level: all the $n$ jobs (for a problem with $n$ jobs) for the first row, the $n-1$ remaining jobs for the second row, and so on.
The data structure is completed with a binary array of length $n$ that indicates the branching type for each level.
In the example, job $3$ is scheduled at the beginning, jobs $2$  and $4$ are scheduled at the end.
Thus, the IVM structure indicates that $3/15/42$ is the next subproblem to be decomposed.

The IVM-based BB operators work as follows:
\begin{itemize}
\item To \textbf{branch} a selected subproblem, the remaining unscheduled jobs are copied to the next row of $M$ and the current level $I$ is incremented. The branching vector is set according to the branching decisions.
\item To \textbf{prune} a subproblem, the corresponding cell in $M$ should be ignored by the selection operator.  To flag a cell as ``pruned'' its value is multiplied by $-1$. With this convention the branching actually consists in copying absolute values to the next row, \ie{} the flags of remaining jobs are removed as they are copied to the next row.
\item To \textbf{select} the next subproblem, the values of $I$ and $V$ are modified such that they point to the deepest leftmost non-negative cell in $M$: the vector $V$ is incremented at position $I$ until a non-pruned cell is found or the end of the row is reached. If the end of the row is reached (\ie{} $V[I]=n-I$), then the algorithm "backtracks" to the previous level by decrementing $I$ and again incrementing $V$.
\end{itemize}

\subsection{Work units}%

Throughout the depth-first exploration, the vector $V$ behaves like a counter. 
In the example of Figure~\ref{fig:tree}, $V$ successively takes the values $00000$, $00010$, $00100$, $\ldots$, $43200$, $43210$, skipping some values due to pruning.
These $120$ values correspond to the lexicographic numbering of the $5!$ solutions in the \textit{factorial} number system~\citep{KnuthACP}.
In this mixed-radix number system, the weight of the position $k=0, 1, \ldots,$ is equal to $k!$ and the digits allowed for the $k^{th}$ position are $0, 1, \ldots, k$. 

For a problem of size $n$, each valid value of $V$ corresponds uniquely to an integer in the interval $[0,n![$.
Converting the position-vector $V$ to its decimal form allows to interpret the search as an exploration, from left to right, of the integer interval $[0,n![$.
Moreover, an initialization procedure allows to start the search at any position $a\in[0,n![$, and by comparing the position-vector $V_a$ to an end-vector $V_b$, the search can be restricted to arbitrary intervals $[a,b[\subseteq[0,n![$.

\begin{figure}[tbp]
\begin{center}
	\includegraphics[width=0.9\linewidth]{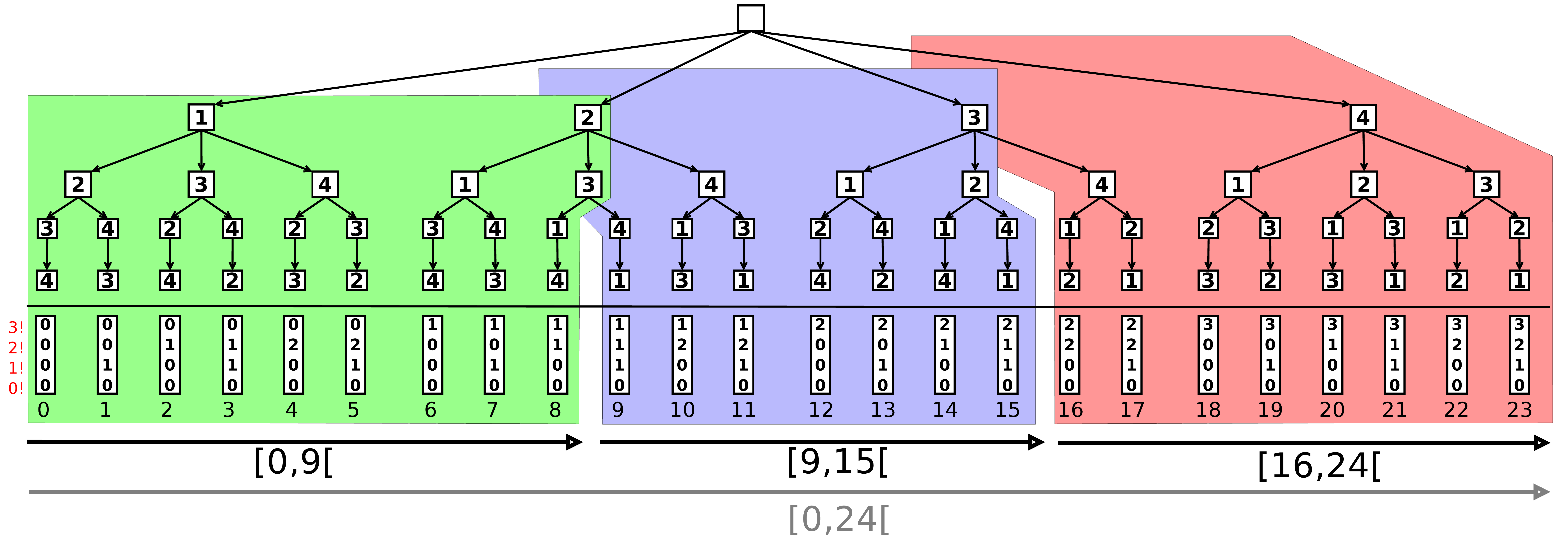}
\caption{Illustration of the search space encoded as the integer-interval $[0,4![$, for a permutation problem of size $n=4$. In this example, the search space is partitioned into three work units : $[0,9[$, $[9,15[$, $[16,24[$ (equivalently, in factoradic notation : $[0000,1100[$, $[1110,2110[$, $[2200,3210[$).}\label{fig:workunits}
\end{center}
\end{figure}

In PBB, \textbf{work units are intervals}, that can be either represented in factoradic form
$[V_a,V_b[\subseteq [(0,0,...,0),(n-1,n-2,...,2,1,0)[$ or, equivalently, in decimal form $[a,b[\subseteq[0,n![$.
Figure~\ref{fig:workunits} illustrates, for a problem of size $n=4$, the partition of the search space $[0,4![$ into three work units.

A parallel PBB algorithm is obtained as follows.
The search space $[0,n![$ is partitioned into $K$ distinct subintervals $[a_i,b_i[\subset[0,n![,\,i=1,\ldots,K$ to be explored by $K$ workers. 
As the distribution of work in $[0,n![$ is highly irregular, a work stealing approach is used~\citep{IVM-IPDPS2014}.
When a worker $i$ finishes the exploration of its interval $[a_i,b_i[$ (i.e. when $a_i=b_i$), it chooses a victim worker $j$ and ``steals'' the right half $[\frac{a_j+b_j}{2},b_j[$. 
The work stealing victim $j$ continues to explore the interval $[a_j,\frac{a_j+b_j}{2}[$.

\subsection{Parallel models for Branch-and-Bound}
The most frequently used models for the parallelization of PBB are: (1) parallel tree exploration, (2) parallel evaluation of bounds and (3) parallelization of the bounding function~\cite{Gendron94}. 

Model (1) consists in exploring disjoint parts of the search space in parallel using multiple independent BB processes.
For large trees this model yields a practically unlimited degree of parallelism (DOP). 
It requires efficient dynamic load balancing mechanisms to deal with the irregularity of the search tree, sharing of the best-found solution and a mechanism for termination detection. Model (1) can be implemented either synchronously or asynchronously.
In model (2), the children nodes generated at a given iteration are evaluated in parallel. 
The DOP is variable throughout the search as it depends on the depth of the decomposed subproblem. 
Model (3) strongly depends on the bounding function and may be nested within models (1) and (2). 
For the PFSP, model (3) refers to a low-level vectorization of the LB function.

In this work, model (1) is used hierarchically: 
on the first level the search space is distributed among asynchronous worker processes hosted on different compute nodes; 
on the second level, each worker process consists of several GPU-based, synchronous PBB sub-workers. %
On both levels, the tree is dynamically balanced among workers, best-found solutions are shared and termination conditions are handled.
On the GPU-level, each independent PBB explorer is mapped onto a CUDA warp (currently 32 threads) and exploits warp-level parallelism through a combination of models (2) and (3).

\section{GPU-based Branch-and-Bound algorithm(PBB$@$GPU)}\label{sec:pbbgpu}
The originality of our PBB$@$GPU algorithm is that all four BB operators, including work stealing, are performed on the GPU.
This differs from the other approaches that can be found in the literature, notably 
the offloading of (costly) node evaluations to GPUs~\citep{Vu2016,ChakrounJPDC2013}, and 
the generation of an initial \textit{Active Set} on the host, used as roots for concurrent GPU-based searches~\citep{RockiMinimax,CarneiroSBACPAD}.

\subsection{Outline of PBB$@$GPU}
The IVM data structure allows to bypass a major roadbloack for GPU-based tree search algorithms : the lack or poor performance of dynamic data structures (linked-lists, stacks, priority queues, etc.) in the CUDA environment.
IVM has a small and constant memory footprint---therefore, thousands of IVMs can be allocated in device memory, providing an efficient way to perform parallel DFS on the GPUs. 
Moreover, the encoding of work units as factoradic intervals allows to implement low-overhead data-parallel work-stealing mechanisms on the GPU.

\begin{figure}[tpb]
\begin{center}
	\includegraphics[width=0.6\linewidth]{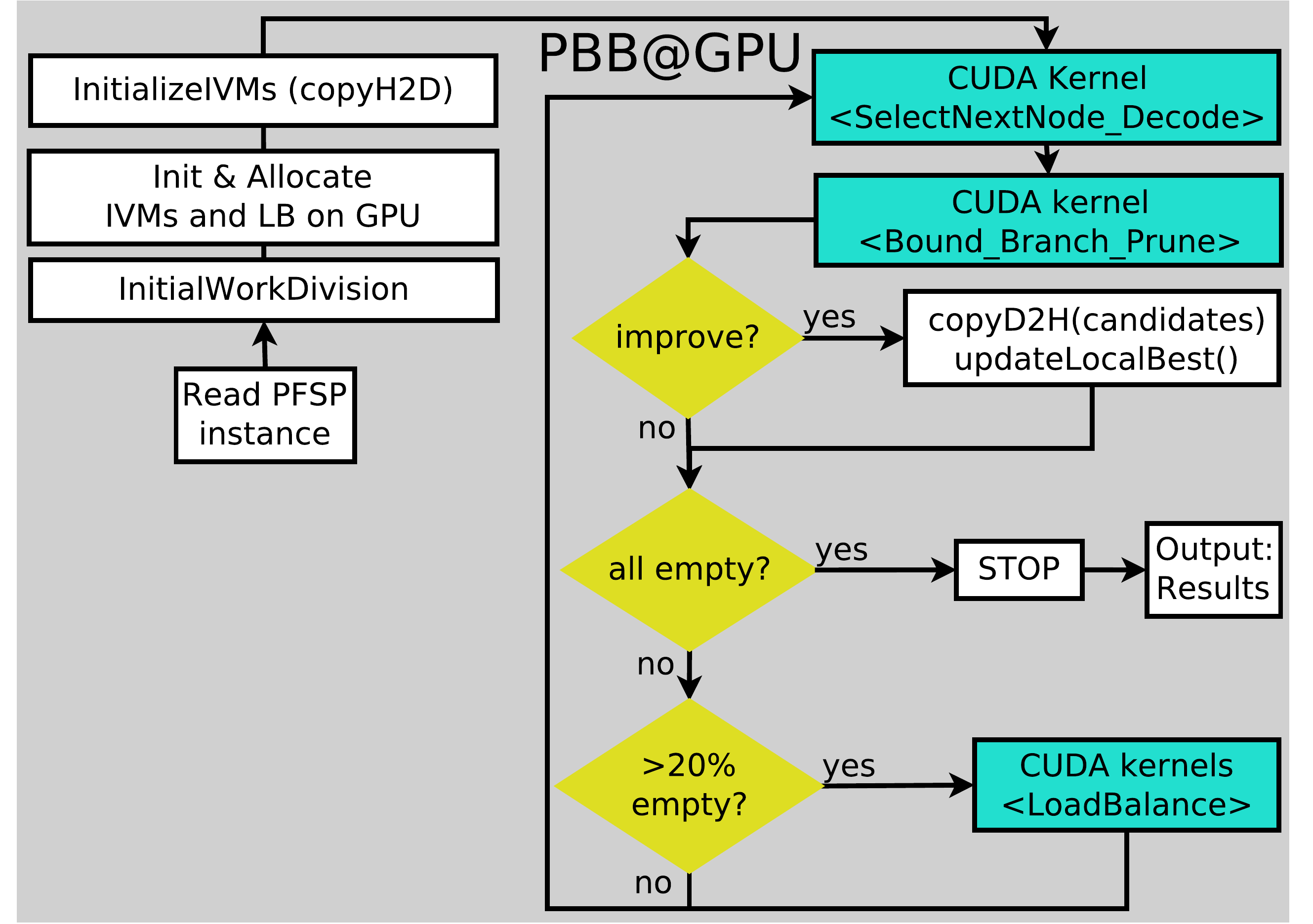}
		\caption{Outline of GPU-based PBB algorithm (PBB$@$GPU.}
			\label{fig:PBBGPU}
\end{center}
\end{figure}

Figure~\ref{fig:PBBGPU} shows a flowchart outlining the PBB$@$GPU algorithm.
After reading problem-specific input data, $K$ IVM structures are allocated in GPU memory and constant data (matrix of processing times, $n$,$m$, $\ldots$) is copied to the device. 
Then, a collection of (at most $K$) intervals is initialized on the host---for instance with a single interval $[0,n![$ or an initial partitioning of the search space $\lbrace[\frac{j\times n!}{K}, \frac{(j+1)\times n!}{K}[,\,j=0,1,\ldots,K-1\rbrace$.
If created in decimal form, the intervals are converted to factoradics in order be used as initial position- and end-vectors on the device.
After this operation, PBB$@$GPU enters the main exploration-loop, which consists of a series of CUDA kernels and a few auxiliary operations.
In Figure~\ref{fig:PBBGPU}, some details, such as kernel configurations, have been spared out.
A more detailed description of the kernels is provided in the following subsections.

The first kernel concurrently modifies the IVM structures such that they point to the next subproblem to be decomposed as described in Section~\ref{subsect:IVM}.
Then, the IVM structure is decoded, producing one subproblem of the form $[\pi,d_1,d_2]$ per IVM, where $\pi$ is a schedule with fixed jobs at positions $1,\ldots,d_1$ and $d_2,\ldots,n$.
The second kernel performs the decomposition step, as shown in Figure~\ref{fig:decompose}, for all selected parent nodes.
IVM structures are modified in parallel to apply pruning decisions.
If, during the execution of these two kernels an improving solution is found, then a device flag \texttt{newBest} is set. 
Moreover, a global device counter $0\leq$\texttt{nbActive}$\leq K$ keeps track of the number of IVMs with non-empty intervals.
Both are copied to the host at each iteration.
If the \texttt{newBest} flag is set, then the candidate solutions are copied to the host and the best solution is updated.
If the current activity level is equal to zero, then PBB$@$GPU returns the optimal solution and exploration statistics, before shutting down.
If the current activity level is below $0.8\times K$, i.e. if more than $20\%$ of IVMs are inactive, then a work stealing phase is triggered.
The goal of this trigger-mechanism is to keep the load balancing overhead low.

\subsection{Selection kernel}
As explained in Section~\ref{subsect:IVM}, the selection of the next subproblem consists in scanning through the IVM structure until the next non-eliminated node is found. This requires a variable amount of operations per IVM. 
Moreover, IVMs can be in different states (empty, active or initializing) that are treated differently in the selection kernel.
Decoding the IVM data structure also requires a variable amount of operations, depending on the depth of the current subproblem.

Mapping each IVM to exactly one GPU thread causes control flow divergence for threads within the same warp and therefore, serialized execution of divergent branches.
Experiments have shown that it is preferable to map IVMs to full warps~\cite{GmysParco}---even if that means that all threads except the warp-leader (lane $0$) are mostly inactive.
The selection kernel is thus launched with $K\times\text{warpSize}$ threads, grouped in blocks of $4$ warps.
Besides reducing thread divergence, spacing the mapping increases the amount of shared memory available per IVM, allowing to bring parts of the data structure closer to the ALUs. The 32 (warpSize) available threads per IVM are used for loading data to shared memory. 
Moreover, despite the sequential nature of the selection operator, some sub-operations (e.g. generating a new line in the IVM matrix) benefit from parallel processing.

\subsection{Bounding kernel}
In~\citet{GmysParco} we have presented a GPU-based PBB approach for the PFSP using the heavier bound LB2 (cf.~Section~\ref{sec:background}), which consumes about $99$\% of the computation time in sequential implementations.
In that situation, it is natural to focus performance optimization in the bounding kernel.
To deal with the variable number of child nodes per IVM, the LB2-based algorithm introduces an auxiliary \textit{mapping} kernel, which uses a parallel prefix sum computation to build a compact mapping of threads onto children subproblems.
Experiments show that the increased efficiency of the bounding kernel offsets the overhead incurred by the mapping kernel by far.

This design is not suitable for the more fine-grained LB1 evaluation function.
Contrary to LB2, children nodes do not need to be fully evaluated, as the LB1 values for children nodes are obtained incrementally from the partial costs of the parent.
Moreover, the computational cost of LB1 is too low to justify regularizing the workload with complex overhead operations.
As for the selection kernel,
using a compact one-thread-per-IVM mapping results in thread divergence issues and requires a very large $K$ (number of IVMs) in order to reach good device occupancy levels.

Therefore, the bounding kernel also uses a one-warp-per-IVM mapping and we extract as much low-level parallelism as possible from the LB1 evaluation function.
Like the selection kernel, the bounding kernel relies heavily on warp-synchronous programming\footnote{Before the introduction of the \texttt{\_\_syncwarp()} primitive in CUDA 9, the warp-synchronous programming style relied on the assumption that threads within a warp are \textit{implicitly} synchronized (i.e. re-converge after possible thread divergence). Recent versions of the CUDA documentation state that codes which rely on this implicit behavior are unsafe and must use explicit synchronization via the \texttt{\_\_syncwarp()} primitive introduced in CUDA 9}.
The implementation uses warp-level primitives and \textit{explicit} synchronization (\texttt{\_\_syncwarp()}) provided by the CUDA Cooperative Groups API (available since CUDA 9).

For each parent subproblem $(\sigma_1,\sigma_2)$ of depth $d_1+d_2=d$ there are $n-d$ unscheduled jobs and thus $2\times(n-d)$ subproblems to evaluate. 
As explained in Section~\ref{sec:background}, the evaluation of children nodes involves: (1) the computation of partial costs for the parent subproblem and (2) the incremental evaluation of LB1 for each child subproblem. 
Despite the data dependencies in Equation~\ref{eq:cmax}, step~(1) can be parallelized as proposed by~\citet{BozejkoPFSP} for MMX vector instructions. 
Step~(2) is embarrassingly parallel.

\begin{figure}[tpb]
\begin{center}
	\includegraphics[width=0.4\linewidth]{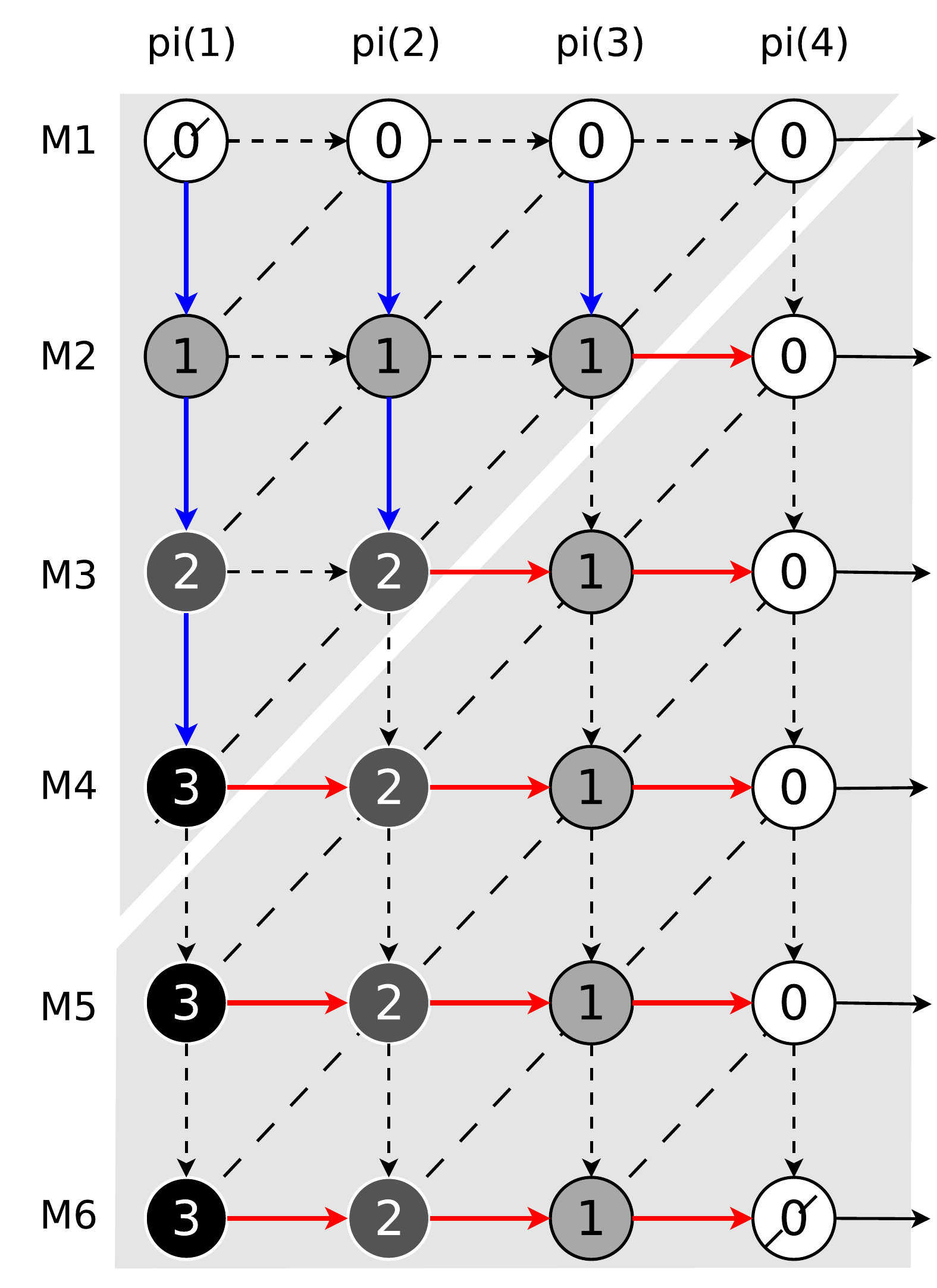}
		\caption{Illustration of a parallel makespan evaluation for a four-job (partial) schedule and $m=6$ machines.}
			\label{fig:warpLB}
\end{center}
\end{figure}

Figure~\ref{fig:warpLB} illustrates the warp-parallel evaluation of a prefix schedule of length $d_1=4$ on $m=6$ machines.
Each circle represents one max-add operation (as in Equation~\ref{eq:cmax}), and the shades of (and labels inside) the circles indicate the lane (thread index within the warp) performing the operation.
Operations connected by dashed diagonal lines are done in parallel.
The horizontal and vertical arrows represent data dependencies of two types: a thin dashed arrow indicates that the lane already holds the required value from the previous iteration, a solid arrow indicates that the required value is transferred from a neighboring lane using a warp-level \texttt{shfl\_up} (blue) or \texttt{shfl\_down} (red) operation.
These built-in warp-synchronous functions allow to bypass shared memory and perform the (partial) makespan evaluation using only per-thread registers.
Finally, the solid arrows on the right represent the storage of per-machine completion times in a (shared or global memory) array of length $m$.

In the example shown in Figure~\ref{fig:warpLB}, the $6\times 4=24$ operations are performed in $4+6-1=9$ iterations, so the theoretical speedup in this case is $\frac{24}{9}=2.7\times$. For a detailed theoretical speedup analysis we refer the reader to~\cite{BozejkoPFSP}.
The complete node decomposition, as illustrated in Figure~\ref{fig:decompose}, is performed as follows (we assume that warpSize is equal to $32$\footnote{as it has always been the case until now, but it might change in the future}).

\begin{enumerate}
\item Evaluate $C_{\sigma_1(d_1),:}$ in $\tau(m+d_1-1)$ steps using $\min(m,d_1,32)$ threads, where $\tau=\ceil*{\frac{\min(m,d_1)}{32}}$.  %
\item Evaluate $\bar{C}_{\sigma_2(d_2),:}$ in $\tau(m+d_2-1)$ steps using $\min(m,d_2,32)$ threads, where $\tau=\ceil*{\frac{\min(m,d_2)}{32}}$.%
\item \texttt{\_\_syncwarp()}
\item Compute $n-d$ \textit{Forward}-LBs in $\tau m$ steps using $\min(n-d,32)$ threads, where $\tau=\ceil*{\frac{n-d}{32}}$. %
\item Compute $n-d$ \textit{Backward}-LBs in $\tau m$ steps using $\min(n-d,32)$ threads, where $\tau=\ceil*{\frac{n-d}{32}}$.%
\item \texttt{\_\_syncwarp()}
\item Use warp-parallel min-reduce and CUDA warp-level primitives to compute the branching direction
\item Apply pruning decisions for $n-d$ subproblems in parallel.
\end{enumerate}

The computation of remaining processing time per machine is performed by subtracting from the per-machine total and integrated into steps $1$ and $2$.
The parent subproblem $[\pi,d_1,d_2]$ and $m$-element arrays representing the front, back and remaining times (see Figure~\ref{fig:LB}) are placed in shared memory. 
The processing time matrix is placed in constant memory at initialization.
The branching decision and pruning operations are carried out at full warp size modulo the depth of the subproblem.

\subsection{Work Stealing kernels}
As mentioned, a work stealing (WS) operation consists in taking the right half from a non-empty interval and assigning it to an idle worker.
To perform this operation on the GPU, device functions for elementary operations ($+$, $-$, $\div$ by scalar) on factoradic numbers are implemented.
The more challenging part is to build a mapping of empty IVMs onto exploring IVMs, such that (1) no WS victim is selected twice, (2) larger intervals are preferred and (3) the mapping is build in parallel on the device.

The $K$ IVM are seen as vertices of a hypercube, in which all empty IVMs successively poll their neighbors to acquire new work units.
For illustrative purposes, let's suppose that $K=2^{14}=16384$.
The indices of the $K$ IVMs can be written as $(\alpha_7\ldots\alpha_1)$ in base $4$.
Connecting all IVMs whose {base-4} indices differ in exactly one digit, a $4$-ary $7$-cube is obtained, where each IVM has 
$7\times(4-1)=21$ neighbors.
The victim selection is carried out in $21$ iterations during which each empty IVM tries to select 
$(\alpha_7\ldots (\alpha_i-j)\pmod 4 \ldots \alpha_1),\,i=1,\ldots,7,\,j=1,2,3$
as a work stealing victim.
A non-empty IVM can be selected if and only if (1) it is not yet selected and (2) its interval is larger than (a) the average interval-length and (b) a minimum length, fixed arbitrarily to $8!$. 
Prior to the victim selection phase, a helper kernel computes the average interval-length.
A more detailed description of the GPU-based WS mechanism can be found in~\cite{GmysPPAMSE}.

\section{Distributed GPU-based algorithm (PBB$@$cluster)}\label{sec:pbbcluster}
\subsection{PBB$@$Cluster : Coordinator process}\label{sec:coordinator}
For the inter-node level of PBB$@$Cluster, we revisit the PBB$@$Grid framework of~\citet{MezmazIPDPS} to enable the use of GPU-based (or multi-core) worker processes, instead of single-threaded workers.
Our algorithm is implemented with MPI and uses a static number of worker processes ($n_p$)---contrary to PBB$@$Grid, which uses socket programming for inter-node communication and shell-scripts to discover available resources and launch worker processes via \texttt{ssh}.

PBB$@$Cluster is based on an asynchronous coordinator-worker model with worker-initiated communications. 
At each point in time, the coordinator keeps a list of \textit{unassigned} work units and an \textit{active} list, containing copies of the work units explored by different workers.
As each worker is composed of multiple sub-workers, a cluster-level \textbf{work unit} is defined as a collection of intervals contained in $[0,n![$.
More precisely, we define a work unit $W_i$ as a finite union of $K_i$ non-overlapping intervals
\begin{equation}\label{eq:workunit}
W_i=\overset{K_i}{\underset{j=0}{\bigcup}}[A_j,B_j[\qquad\mbox{, where }\forall j: [A_j,B_j[\subseteq [0,n![ 
\mbox{ and }[A_j,B_j[\cap [A_k,B_k[=\emptyset\;,j\neq k
\end{equation}
In this definition, index $i$ is the identifier of the work unit, and $K_i<K_i^{\max}$ the number of intervals, limited by a maximum capacity $K_i^{\max}$.
The coordinator maintains (1) a list of unassigned work units $\mathcal{W}_{\text{unassigned}}$ and (2) a list of active work units $\mathcal{W}=\lbrace W^c_1,W^c_2,\ldots\rbrace$, where $W^c_i$ is the coordinator's copy of the work unit $W_i$, currently assigned to a worker.

\begin{algorithm}[!tbp]
\scriptsize
\caption{PBB@Cluster : Coordinator}\label{alg:Master}
\begin{algorithmic}[1]
\Procedure{PBB-Coordinator}{}
\State \texttt{/* not shown: allocations; initialize, build and broadcast initial solution; ...*/}
\State $\mathcal{W}_{\text{unassigned}}\gets$\textsc{GetInitialWorks()}\Comment{e.g. $[0,n![$, readFromFile(),...}\label{line:initWork}
\State $n_{\text{terminated}}\gets 0$
\While{$n_{\text{terminated}} < n_{\text{proc}}$}
\State src, tag$\gets$\textsc{MPI\_Probe}(ANY) \Comment{wait for any message from any source}\label{line:listen}
\Switch{tag}\Comment{discriminate message types}
	\Case{WORK: \texttt{/* worker checkpoint */}} 
		\State{$W_i\gets$\textsc{ReceiveWork}(src)}\Comment{wrapper for MPI\_Recv and Unpack}\label{line:recvwork}
		\State $W_{tmp}\gets$WorkerCheckpoint($W_i$)\Comment{pseudo-code provided below}\label{line:checkpointCall}
		\If{$\mathcal{W}=\emptyset \land \mathcal{W}_{\text{unassigned}}=\emptyset$}
			\State \textsc{SendEnd}(src)\Comment{no more work : send termination signal}
		\ElsIf{$W_{tmp}\neq W_i$}
			\State \textsc{SendWork}($W_{tmp}$, src)\Comment{send new or modified $W_i$ to worker}\label{line:sendwork}
		\Else
			\State \textsc{SendBest}(src)\Comment{acknowledge reception / send global best}\label{line:sendbestW}
		\EndIf	
	\EndCase  
	\Case{BEST: \texttt{/* candidate for improved global best solution */}}\label{line:from}
		\State S$\gets$\textsc{ReceiveSolution}(src)
		\State \textsc{TryImproveGlobalBest}(S) 
		\State \textsc{SendBest}(src)\label{line:sendbestB}
	\EndCase
	\Case{END: \texttt{/* worker has left computation */}}
		\State S$\gets$\textsc{ReceiveSolution}(src)
		\State \textsc{TryImproveGlobalBest}(S)
		\State $n_{\text{terminated}}$++ 
	\EndCase
\EndSwitch
\If{\textsc{DoGlobalCheckpoint}()}\Comment{global checkpoint interval elapsed?}	
	\State \textsc{SaveToDisk}($\mathcal{W}$, $\mathcal{W}_{\text{unassigned}}$)
\EndIf\label{line:to}
\EndWhile
\EndProcedure
\Procedure{WorkerCheckpoint}{$W_i$}\label{line:checkpointDef}
\State $W_i^c\gets$\textsc{FindCopy}($W_i$,$\mathcal{W}$)
\State $W_{tmp}\gets$\textsc{Intersect}$(W_i,W_i^c)$\label{line:intersect}
\If{$W_{tmp}=\emptyset$}
\State $W_{tmp}\gets$\textsc{Steal}$(i,\mathcal{W})$
\EndIf
\State $W_i^c\gets W_{tmp}$
\State \Return $W_{tmp}$
\EndProcedure
\end{algorithmic}
\end{algorithm}

A pseudo-code of the PBB$@$Cluster coordinator is shown in Algorithm~\ref{alg:Master}.
After broadcasting an initial solution (computed or read from a file), the coordinator fills $\mathcal{W}_{\text{unassigned}}$ with the initial work, e.g. the complete interval $[0,n![$, an initial decomposition, or a list of intervals read from a file (Line~\ref{line:initWork}).
Then, the coordinator starts listening for incoming messages (Line~\ref{line:listen}).
Under certain conditions, that will be detailed later, workers send checkpoint messages to the coordinator, containing 
\begin{itemize}
\item the number of nodes decomposed since the last checkpoint, 
\item $K_i^{\max}$ the maximal number of intervals the worker can handle and 
\item a work unit $W_i$ containing $K_i$ intervals and tagged with a unique identifier $i$.
\end{itemize}
After receiving a checkpoint message, the coordinator intersects the work unit $W_i$ with the copy $W_i^c$ (resulting in an empty list if $W_i^c$ doesn't exist in $\mathcal{W}$).
If the result of the intersection is empty, a new work unit of at most $K_i^{\max}$ intervals is generated, if possible, by taking intervals from $\mathcal{W}_{\text{unassigned}}$ or by splitting a work unit from $\mathcal{W}$, otherwise.
The work unit $W_{\text{tmp}}$ resulting from the intersection and/or splitting operations is placed in $\mathcal{W}$, replacing
$W_i^c$ if the latter exists. 
These operations are shown as the \texttt{workerCheckpoint} operation in Algorithm~\ref{alg:Master} (Lines~\ref{line:checkpointCall} and \ref{line:checkpointDef}).
If $W_{\text{tmp}}$ differs from the received work unit $W_i$, then $W_{\text{tmp}}$ is sent back to the worker to replace $W_i$.
If $W_{\text{tmp}}$ is identical to the received $W_i$, then there is no need to send any work back and the coordinator replies only by sending the best-found global upper bound (Line~\ref{line:sendbestW}).
The remaining tasks of the coordinator, shown in Lines~\ref{line:from} to \ref{line:to}, deal with termination detection, management of the global best solution and global checkpointing.

The sending/receiving of work units and the intersection procedure  are the most time-consuming operations of the coordinator.
We should note that it is not the \texttt{MPI\_Recv} and \texttt{MPI\_Send} calls in the work communication that are most time-consuming, but the message unpacking.
A worker checkpoint message is of type \texttt{MPI\_PACKED} and consists of a metadata header and a list of $K_i$ intervals.
As these intervals are represented by two integers of the order $\sim n!$, the GNU Multiple Precision Arithmetic Library (GMP) is used.
While the coordinator could as well work with factoradic numbers (i.e. integer arrays of length $n$), it is more convenient and faster to perform arithmetic operations (subtraction, division, addition, comparison) in decimal form.
However, due to the lack of native MPI support for GMP integers, packing intervals to the communication buffer requires converting them to raw binary format and back to \texttt{mpz\_t} at the receiving end. 
We observed that these conversions cause significant overhead.
Messages of type BEST are smaller and occur less frequently. 
Messages of type END occur only at shutdown and their only purpose is to guarantee that the coordinator doesn't exit the main loop before the termination of all workers. All received messages, except for the last one, are answered.

\paragraph{Work unit intersection}
The intersection of two intervals $[a_1,b_1[$ and $[a_2,b_2[$ is done by considering the maximum between both start points and the minimum between both end points, as shown in Equation~\ref{eq:intersect2}.
\begin{equation}\label{eq:intersect2}
[a_1,b_1[\cap[a_2,b_2[=[\max(a_1,a_2),\min(b_1,b_2)[
\end{equation}

The intersection of two work units $W_1$ and $W_2$ requires pairwise intersection of the intervals contained in both sets, as shown in Equation~\ref{eq:intersectMany}.
\begin{equation}\label{eq:intersectMany}
\begin{split}
		\left(\overset{K_1}{\underset{i=1}{\bigcup}}[a^1_i,b^1_i[\right)\cap\left(\overset{K_2}{\underset{j=1}{\bigcup}}[a^2_j,b^2_j[\right)= 
		& \overset{K_1}{\underset{i=1}{\bigcup}}\overset{K_2}{\underset{j=1}{\bigcup}}[\max(a^1_i,a^2_j),\min(b^1_i,b^2_j)[
\end{split}
\end{equation}

For arbitrary sets of intervals, $K_1\times K_2$ elementary intersections are required to compute the intersection of two interval-lists.
However, using the fact that each interval in $W_1$ intersects with at most one interval in $W_2$, and sorting intervals in increasing order, 
the operation in Equation~\ref{eq:intersectMany} can be carried out in $O(K_1+K_2)$ time.
The computational cost of work unit intersections can be further reduced by taking advantage of the following observation.
If a copy $W^c_i$ hasn't been stolen from since the last worker-checkpoint, then the intersection operation becomes trivially $W^c_i\cap W_i = W_i $. Thus, the coordinator maintains a flag for each work unit in $\mathcal{W}$ to indicate whether it has been modified since the last worker checkpoint.

\paragraph{Work unit division}
When no more unassigned works are available, the coordinator generates a new work unit by splitting the largest  work unit from $\mathcal{W}$.
For that purpose, the coordinator keeps track of the sizes of work units, defined as 
\[
\norm{W_i}=\sum_{j=1}^{K_i}(b^i_j-a^i_j).
\]

Let $W_v^c$ be the work unit selected for splitting and $K_i^{\max}$ the maximum number of intervals the requesting worker $i$ can handle.
The new work unit is generated by taking the right halves of the first $K_{new}=\min(K_i^{\max},K_v)$ intervals from $W_v^c$.
The latter is
\[
W_{new}=\overset{K_{new}}{\underset{j=1}{\bigcup}}[\frac{a_v^j+b_v^j}{2}, b_v^j[
\]
and the victim's copy of the work unit becomes
\[
W_v^c=
\left(
\overset{K_{new}}{\underset{j=1}{\bigcup}}[a_v^j, \frac{a_v^j+b_v^j}{2}[
\right)\cup
\left(
\overset{K_v}{\underset{j=K_{new}+1}{\bigcup}}[a_v^j, b_v^j[
\right)
\]
As mentioned above, after this operation, the victim's work unit is flagged as ``modified'' to make sure that the impacted worker performs a full intersection at the next checkpoint and updates its work unit.

\paragraph{Global checkpointing}
Periodically, the coordinator saves the complete lists of unassigned and active work units, $\mathcal{W}_{\text{unassigned}}$ and $\mathcal{W}$, to a file.
The total size of this file can be estimated as follows:
For a problem instance with $n=100$ jobs, the size of a work unit of $K_i=16384$ intervals is approximately
\[
16384\times 2\times \frac{log_2(100!)}{8} B = 2.1~\text{MB}
\]
so with $n_p=256$ workers the size of the checkpoint file grows to $\sim 500$~MB.
In addition, the global checkpoint must contain the best found solution. 
When restarting PBB$@$Cluster from a global checkpoint, the coordinator reads the file and places the work units in $\mathcal{W}_{\text{unassigned}}$.

\subsection{PBB$@$Cluster : Worker process}\label{sec:worker}
\begin{figure*}[!htpb]
\begin{center}
	\includegraphics[width=0.99\linewidth]{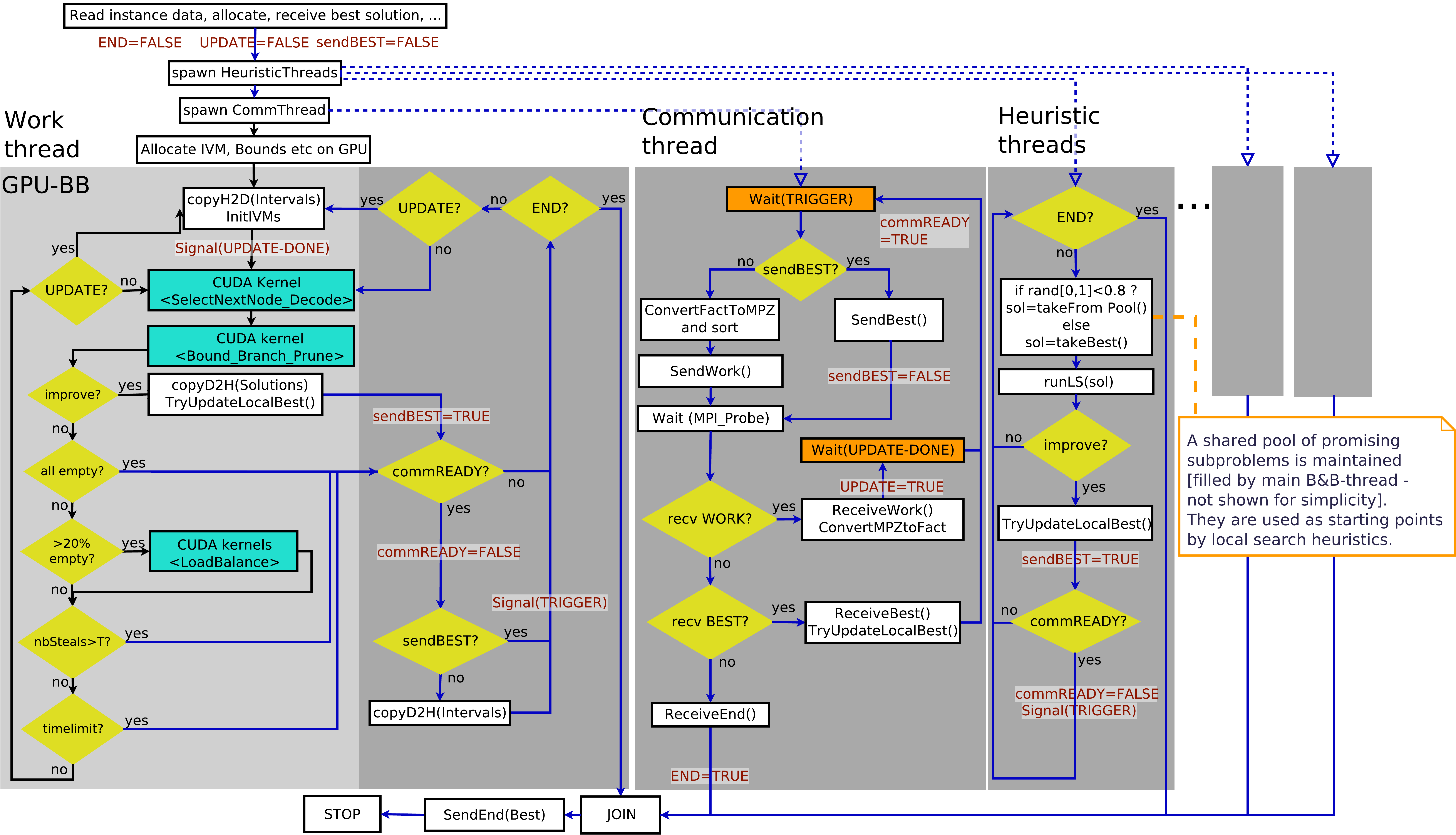}
		\caption{PBB GPU-Worker Process}
			\label{fig:worker}
\end{center}
\end{figure*}

Figure~\ref{fig:worker} shows a flowchart of a worker process, composed of a PBB$@$GPU thread (controlling the GPU), a dedicated communication thread and multiple metaheuristic threads.
The worker process is implemented using POSIX threads (\texttt{pthreads}) and the different worker components communicate through shared memory using mutexes and condition variables.
For the sake of readability, details regarding synchronization and mutual exclusion primitives are spared out in Figure~\ref{fig:worker}.
Like PBB$@$GPU, the worker process starts by allocating and initializing data structures on the CPU and GPU.
The initial best-found solution is received from the coordinator.

\paragraph{PBB$@$GPU thread}
The left part of Figure~\ref{fig:worker} corresponds to the PBB$@$GPU algorithm presented in Section~\ref{sec:pbbgpu}, including a few modifications.
Instead of stopping when all work units are empty, a checkpoint communication is initiated to acquire new work.
The following states also trigger communications with the coordinator:
\begin{itemize}
\item If the local best solution has been improved, then it should be send to the coordinator as it might improve the global best.
\item If a fixed amount of GPU-based load balancing operations were successful, the coordinator's copy of the work unit should be updated to avoid redundant exploration. This threshold is set to $\nicefrac{1}{5}$ of the sub-workers. 
\item If a fixed amount of time has elapsed since the last worker-checkpoint. The purpose of this time-limit is to ensure that the global state of the search, kept by the coordinator, is updated regularly. By default, we set this value to $30$ seconds.
\end{itemize}

As shown in Figure~\ref{fig:worker}, a dedicated thread (described below) is in charge of communicating with the coordinator.
The worker and communication threads basically interact in a producer-consumer pattern with single-item buffers.
If the communicator thread is not ready (buffers are full), then the worker checks for global termination and pending updates before resuming to exploration work.
Indeed, if no more local work is available, the worker thread quickly returns to the point where it checks the readiness of the communicator thread, effectively busy-waiting for the buffer to become free.
Otherwise, appropriate flags are set for the communicator thread, intervals are copied from the GPU to a send-buffer, and the worker returns to the PBB$@$GPU main-loop. 
The objective of this approach is that checkpoint operations or sending a new solution do not prevent the worker from making progress in the interval exploration.

\paragraph{Communication thread}
There are several reasons that motivate the use of a dedicated communication thread.
\begin{itemize}
\item As mentioned in the previous section, the communicator handles intervals in decimal form, i.e. GMP integers, and interval-lists should be sorted to simplify the intersection operation. Using a separate communication thread, this burden of pre- and post-processing work unit communications is taken off the critical path.
\item Using a dedicated communication thread is one way to actually progress the message passing asynchronously.
Similar approaches have been proposed in the literature~\citep{Vaidyanathan2015,Hoefler2008}.
\item Although quite few best solutions are discovered throughout the search, new best solutions can be found by the PBB$@$GPU thread or by heuristic search threads. 
Offloading all communication to a single dedicated thread allows to use the \texttt{MPI\_THREAD\_SERIALIZED} thread-level instead of \texttt{MPI\_THREAD\_MULTIPLE}.
\item Each message to the coordinator should be matched by an answer. In particular, if no more work is available, sending multiple subsequent work requests would cause multiple answers by the coordinator, overwriting each other.
In our opinion, from a programming point of view, assuring this constraint with conventional non-blocking routines (e.g. \texttt{MPI\_Isend} and \texttt{MPI\_IProbe}) is at least as difficult as correctly synchronizing pthreads.
\end{itemize}
On the downside, one less CPU thread is available for computations---however, on most current systems this should be negligible.
The flowchart of the communication thread is shown in the middle of Figure~\ref{fig:worker}.
In its default state, the communication thread waits on a condition variable to be triggered.
If triggered, it then either sends the local best solution or the interval-list (after converting it from factoradic to decimal and sorting it). Then, the communicator thread waits for an answer from the coordinator.

If a WORK message is received, then the interval-list is converted to the factoradic form and the availability of an update is signaled to the work-thread. 
As shown in the left part of Figure~\ref{fig:worker}, the work-thread checks at each iteration whether an update is available.
To ensure that the buffer can be safely reused, the communication thread blocks until the work-thread has copied the intervals to the device.
Upon reception of a BEST message, which is the default answer from the coordinator, the thread attempts to update the local best solution.
From coordinator to worker, BEST messages contain only the best makespan---not the corresponding schedule which is not needed by workers. 
The third possible message type is a termination message: it causes the communication thread to set the shared termination flag and join with the other worker-threads. 
Before shutting down, each worker sends a last message to the worker, containing the local best solution.

\paragraph{Heuristic threads}
The PBB$@$Cluster design presented up to this point leaves the computing power of additional CPU cores unused. 
For instance, each GPU-accelerated compute node on \textit{Jean Zay} is composed of two 20-core CPUs and 4 GPUs, meaning that 32 additional CPU cores per node can be exploited.
Preliminary experiments show that CPU-based BB-threads running on those cores only reach a fraction of the processing speed provided by the GPUs.
PBB$@$cluster therefore uses remaining CPU cores to run heuristic search algorithms.

The exact BB search and heuristic searches cooperate in the following way.
Periodically, the current subproblems of all IVMs are promoted to solutions (by fixing the unscheduled jobs according to their order in the incumbent solution), their makespans are evaluated and the best resulting schedules are added to a buffer (containing up to twice as many solutions as the number of heuristic threads).
The main purpose of this list of solutions is to provide diversified starting points for local searches that are allowed a fixed amount of time.

In principle, any kind of heuristic search can be used. However, two aspects are particularly important.
Firstly, the heuristic searches should either be stochastic or depend rather strongly on the starting solution. 
Otherwise, heuristic searches will end up finding identical solutions.
Secondly, for solving very hard instances, it is important that the searches are able to find very high-quality solutions if a long enough running time is allowed---rather than the ability to find good solutions very quickly.
Investigating the performance of different heuristic methods in the context of the hybridized PBB$@$Cluster algorithm goes beyond the scope of this paper.

In our attempts to solve hard problem instances, we mainly use an iterated local search (ILS) algorithm~\citep{Stutzle2007} using the k-insert recursive neighborhood proposed in~\citep{deroussi2010adaptation}.
We have also used a truncated best-upper-bound-first BB search (with stack-size- and time-limits), that prunes on LB1 \textit{and}
generates upper bounds from partial solutions by fixing unscheduled jobs in the order of appearance in the IVM-matrix. 
The behavior of this approach is biased by arranging jobs the first-row of the matrix according to the starting solution. Moreover, for each visited subproblem of a predefined depth, the beam-search algorithm recently proposed by~\citep{libralessoBeam} is applied on partial solutions, leaving prefix and postfix partial schedules unchanged.

Both approaches have shown good results, but no clear pattern has emerged regarding the better heuristic search to use.
Moreover, different methods for extracting solutions from the BB search should be investigated and the running time allowed for a heuristic search is fixed at $5$ minutes.
The hybridization of PBB$@$Cluster with approximate methods is still an experimental feature that requires more attention---we should note, however, that the hybridization allowed to solve instances whose resolution couldn't be achieved by the PBB algorithm alone. 

\section{Experimental evaluation}\label{sec:experiments}
First, in Section~\ref{sec:exp:hardware} some details regarding the experimental environment are provided.
In Subsection~\ref{sec:exp:singleGPU} we experimentally evaluate the performance of the single-GPU implementation on different GPUs and compare it to a multi-core approach.
In Subsection~\ref{sec:exp:scale} we study the scalability of PBB$@$cluster on up to $384$ GPUs.
In Subsection~\ref{sec:solveTa} we report on the attempted resolution of the remaining open Taillard instances and discuss the results.
New best known solutions and proofs of optimality for the VFR benchmark are given in the Appendix.

\subsection{Experimental platform}\label{sec:exp:hardware}

Most experiments are carried out on the GPU-accelerated partition of the \textit{Jean Zay} supercomputer hosted at IDRIS\footnote{
Institute du d\'{e}veloppement et des ressources en informatique scientifique (national computing centre for the French National Centre for Scientific Research (CNRS))
}.
The system has two partitions (accelerated and non-accelerated), ranked \#64 and \#108 respectively in the Top500 (November 2020). 
Accelerated nodes are equipped with \textbf{two} Intel Xeon Gold 6248 (Cascade Lake) processors and \textbf{four} Nvidia V100 SMX2 (32 GB) GPUs. Each V100 GPU has 80 streaming multiprocessors (SMs) for a total of 5120 FP32 Cuda cores clocked at 1.53 GHz (Boost Clock rate).
\textit{Jean Zay} is a HPE SGI 8600 System with Intel Omni-Path 100 GB/s interconnect.
The OS is a Red Hat 8.1 Linux distribution and the job scheduler Slurm 18.08.8.
For our experiments, we are able to reserve up to $384$ GPUs (or 96 nodes) with a maximum duration of $20$ hours for a single job.
For development, testing and medium-scale experiments we also used the GPU-equipped clusters of Grid'5000, a large-scale and flexible testbed for experiment-driven research~\footnote{\url{https://www.grid5000.fr/}}.

\subsection{Evaluation of single-GPU performance}\label{sec:exp:singleGPU}

\begin{table}[!tbp]
\scriptsize
\centering
\caption{Instances used for single-GPU performance evaluation. The search is initialized with ``initial-UB" ($\leq C_{\max}^\star$) and explores a tree of size ``tree-size", composed of nodes $\{\text{LB(node)}<\text{initial-UB}\}$
}\label{tab:exp1instances}
\begin{tabular}{cccc}
\toprule
Instance  & initial-UB & $n\times m$  & tree-size  \\
\midrule
\textit{Ta021}  & 2297  & $20\times 20$   & 495~G \\
\textit{Ta056}  & 3666  & $50\times 20$   & 1444~G   \\
\textit{Ta081}  & 6115  & $100\times 20$  & 282~G  \\
\textit{Ta101}  & 11156 & $200\times 20$  & 371~G  \\
\bottomrule
\end{tabular}
\end{table}

\begin{figure}[!btph]
\begin{center}
	\includegraphics[width=0.99\linewidth]{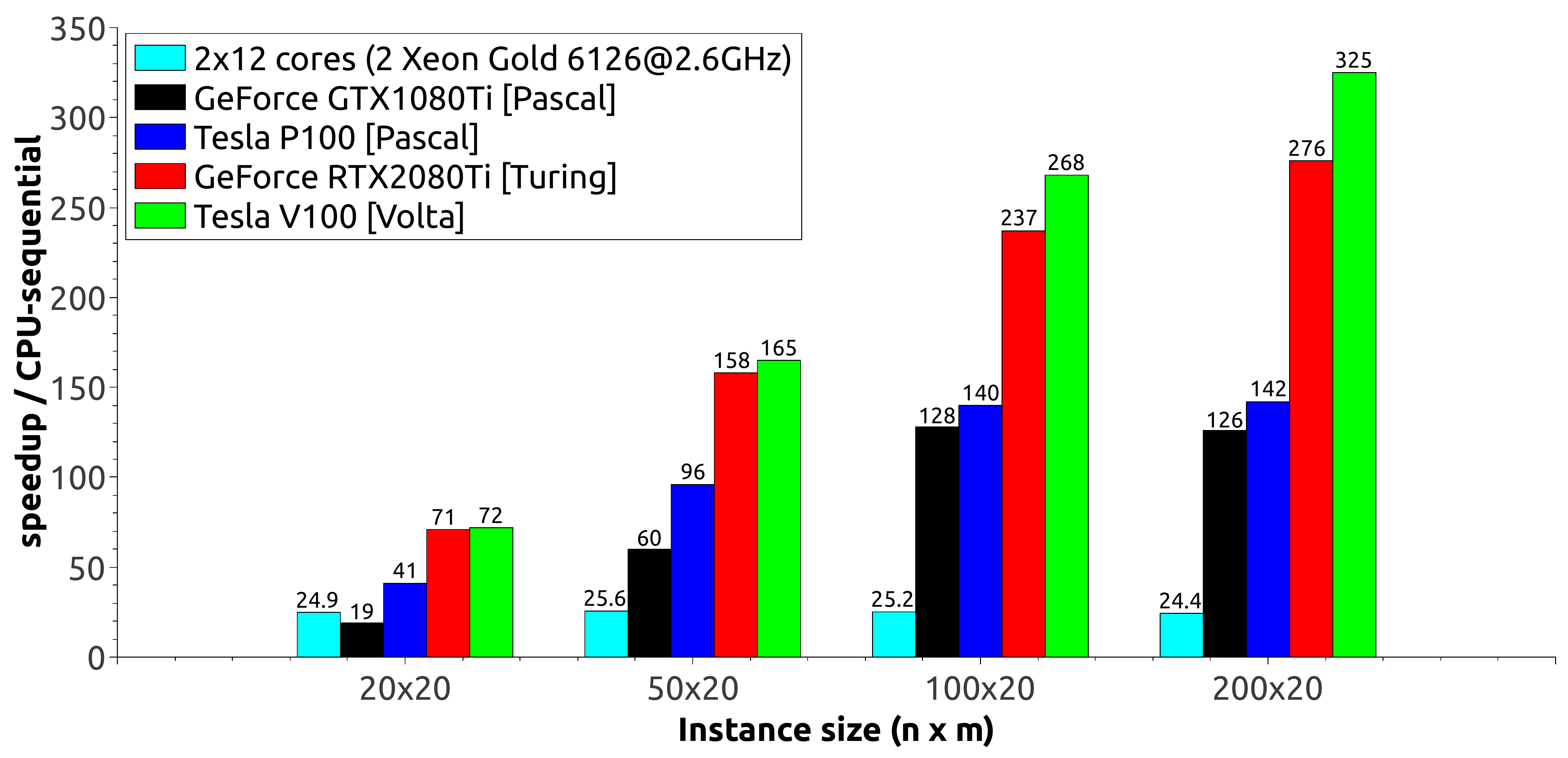}
		\caption{Performance of PBB$@$GPU compared to sequential and multi-core CPU implementations, using the benchmark instances shown in Table~\ref{tab:exp1instances}.}
			\label{fig:GPUvsMC}
\end{center}
\end{figure}

In this first experiment, the performance of PBB$@$GPU is evaluated and compared to an equivalent CPU-based PBB$@$multi-core implementation.
For the purpose of this experiment, PBB is initialized with an initial upper bound (UB) that is \textit{smaller} than the optimum---this ensures that the size of search trees is fixed, and small enough to be explored in $10$-$60$ minutes on a single CPU-core.
Instances defined by $m=20$ machines and $n=20$, $50$, $100$ and $200$ jobs are considered. The selected instances, initial UBs and the corresponding tree sizes are shown in Table~\ref{tab:exp1instances}.
The evaluation is performed with different GPUs available in the Grid'5000 testbed: two gaming devices, GTX1080Ti and RTX2080Ti, based on the Pascal and Turing microarchitectures respectively; and two data-center GPUs (previously named Tesla), the Pascal P100 and Volta V100 (PCIe versions). For all four, version 10.1 of the CUDA toolkit is used.

Figure~\ref{fig:GPUvsMC} shows the relative speed-up of multi-core and GPU-based PBB compared to a sequential execution using a single-core CPU-core. The parallel multi-core version runs on a dual-socket NUMA system composed of two Intel Gold 6126 CPUs (2x12 cores) and uses all 48 logical cores. The implementation uses pthread-based work stealing for load balancing between asynchronous exploration threads.
In preliminary experiments, we determined that $K=16384$ is a suitable value for the number of IVMs per GPU. 

One can see in Figure~\ref{fig:GPUvsMC} that the multi-core B\&B reaches speed-ups approximately equal to the number of physical CPU cores.
The dual-socket 24-core system performs better than PBB$@$GPU only for the small $20\times 20$ instance and the weakest of the four GPUs.
In all other cases, PBB$@$GPU clearly outperforms PBB$@$multi-core, reaching speed-ups between $41\times$ (for the $20\times 20$ instance using a P100 device) and $325\times$ (for $200\times 20$ instances on a V100 device) over sequential CPU execution.
In other words, for instances of size $100\times 20$ or $200\times 20$, over $1000$ CPU cores are needed to equal the processing power provided by a single quad-GPU node of Jean Zay.

One can notice a significant performance gap between the Pascal P100 and Volta V100 GPUs, and also between the Pascal- and Turing-based GeForce devices. To better understand the reasons for this improvement we profiled PBB$@$GPU executions on the four different devices using the \texttt{nvprof} command-line profiling tool.
Table~\ref{tab:PBBGPUbreakdown} details the per-iteration execution time for the two main kernels (see Figure~\ref{fig:PBBGPU}) and overhead operations (mainly load balancing kernels and data movement between host and device).
The timing of the two kernels corresponds to the average kernel execution time obtained by the \texttt{nvprof} command-line profiler.
The overhead time is obtained by dividing the total remaining walltime---excluding the two kernels---by the number of iterations.

The performance gain on the more recent GPUs mainly comes from a faster execution of the two main kernels---the V100 doubles performance compared to the P100.
The $2-3\times$ acceleration factors observed between the Pascal and Volta/Turing devices exceeds significantly what could be expected from higher core-counts and slightly increased clock speeds.
An in-depth analysis of this performance boost goes beyond the scope of this paper---however, our best guess is that PBB$@$GPU benefits substantially from the improved SIMT execution model, introduced in the Volta microarchitecture.
The latter features four warp schedulers per multiprocessor instead of two, and introduces independent thread scheduling~\cite{Volta}, which should significantly reduce execution divergence overheads and improve fine-grained synchronization. 

\newcommand{\MCSpace}{2.0em}
\begin{center}
\begin{table}
\scriptsize
\centering
\caption{Breakdown of PBB$@$GPU execution time per iteration (in microseconds per BB iteration)}\label{tab:PBBGPUbreakdown}
\resizebox{\textwidth}{!}{
\begin{tabular}{cc 
 cc
 cc@{\hspace{\MCSpace}}
 cc
 cc}
\toprule
\multirow{2}{*}{size} & \multirow{2}{*}{kernel} &  
 \multicolumn{4}{c}{time/iteration ($\mu$s)} &
 \multicolumn{4}{c}{\% of walltime} \\
 &  &  GTX1080 & P100 & V100  & RTX2080 & GTX1080 & P100 & V100  & RTX2080  \\
\midrule
\multirow{3}{*}{$20\times 20$} & LB &  
925 &  331 & 147   & 161 &
$75$ & $58$ &  $46$ &  $48$ \\
&  selectBranch & 
155 & 148 & 86 & 93 &
$13$ & $26$ & $27$ & $28$ \\
& WS/other & 
151 & 96 & 90 & 78 & 
$12$ & $17$ & $28$ & $24$ \\
\midrule 
 & \textbf{Tot} & 
\textbf{1230} & \textbf{575} & \textbf{324} & \textbf{332} & 
--- & --- & --- & --- \\
\midrule\midrule
\multirow{3}{*}{$50\times 20$} & LB &  
872 & 388 & 194 & 220 &
$70$ & $50$ & $43$ & $47$ \\
& selectBranchLB & 
311 & 290 & 170 & 181 &
25 & 38  &  38 & 38  \\
& WS/other & 
64 & 94 & 85 & 70 &
5 & 12 & 19 & 15  \\ 
\midrule 
 & \textbf{Tot} & 
\textbf{1248} & \textbf{772} & \textbf{449} & \textbf{471} & 
--- & --- & --- & --- \\
\midrule 
\midrule %
\multirow{3}{*}{$100\times 20$} & LB & 
1074 & 578 & 298 & 342 &
62 & 47 & 46 & 47 \\
& selectBranch & 
553 & 523 & 230 & 280 &
32 & 42  & 36 & 38 \\
& WS/other & 
98 & 130 & 115 & 107 &
6 & 11 &  18 & 15 \\ 
\midrule 
 & \textbf{Tot} & 
\textbf{1726} & \textbf{1232} & \textbf{642} & \textbf{729} & 
--- & --- & --- & --- \\
\midrule 
\midrule%
\multirow{3}{*}{$200\times 20$} & LB & 
1315 & 954 & 473 & 568 &
47 & 38 & 43 &  44 \\
& selectBranch & 
1320 & 1365 & 465  & 564  &
47 & 55 &  43 & 44  \\
& WS/other & 
162 & 179 &  151 & 149 &
6 & 7 &  14 & 12 \\ 
\midrule 
 & \textbf{Tot} & 
\textbf{2798} & \textbf{2498} & \textbf{1089} & \textbf{1282} & 
--- & --- & --- & --- \\
\midrule
\bottomrule
\end{tabular}
}
\end{table}
\end{center}

\subsection{Illustration of a PBB$@$Cluster run}\label{sec:exp:illustrate}
\begin{figure}[tpb]
\begin{center}
	\includegraphics[width=0.99\linewidth]{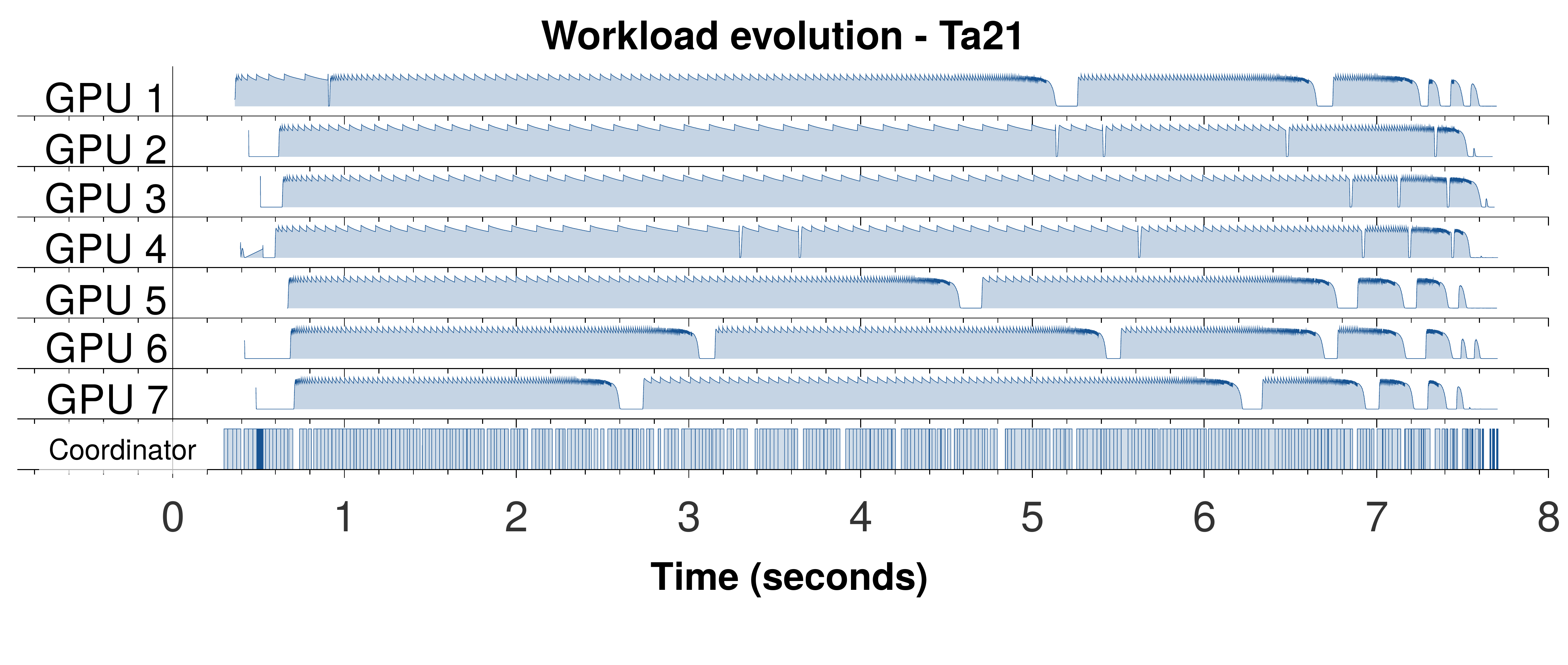}
		\caption{Evolution of workload during resolution of instance \textit{Ta021} (461M explored nodes) on 7 GPUs. The horizontal axis represents the elapsed time (in seconds).}
			\label{fig:Ta21Timeline}
\end{center}
\end{figure}

\begin{figure}[tpb]
\begin{center}
	\includegraphics[width=0.99\linewidth]{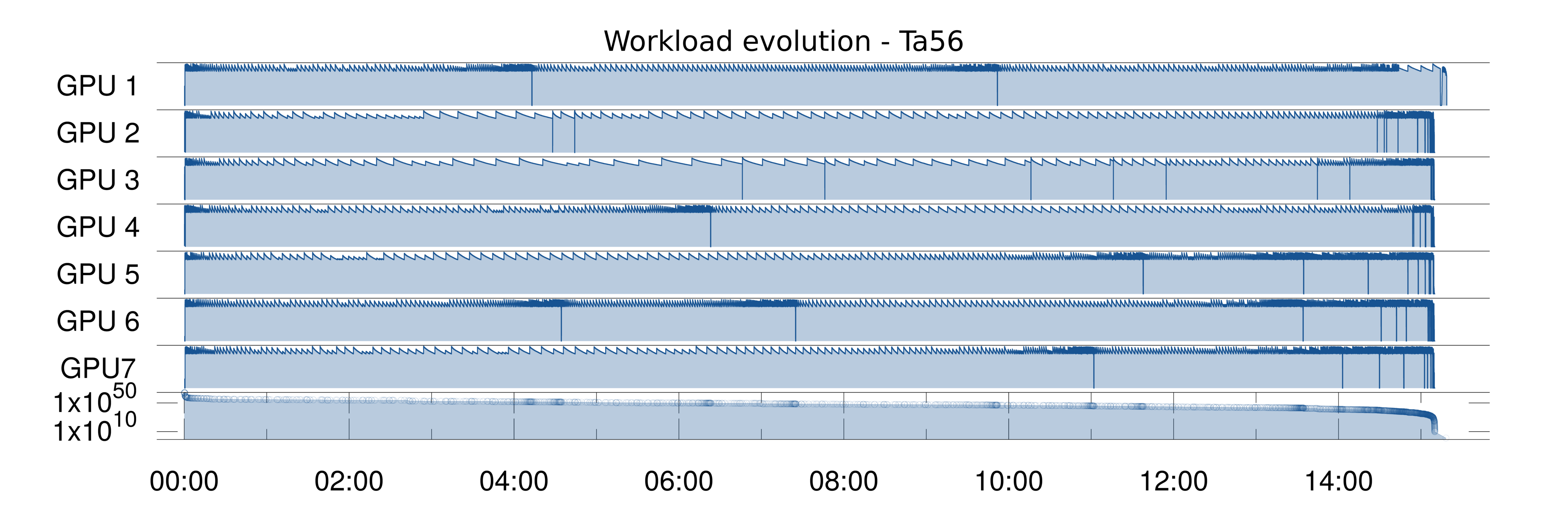}
		\caption{Evolution of workload during resolution of instance Ta56 (175G explored nodes) on 7 GPUs. The horizontal axis represents the elapsed time (in minutes).}
			\label{fig:Ta56Timeline}
\end{center}
\end{figure}

Before analyzing the scalabilty of PBB$@$Cluster on \textit{Jean Zay}, it could be useful to graphically illustrate a distributed execution of PBB$@$Cluster.
We record timestamps together with the activity level of each GPU (share of IVMs with non-empty intervals) during two runs with $7$ GPUs.
Figure~\ref{fig:Ta21Timeline} shows the activity level of GPUs throughout the resolution of instance \textit{Ta021} ($<8$~seconds) and Figure~\ref{fig:Ta56Timeline} corresponds the a resolution of \textit{Ta056} in about $15$ minutes.
The activity level of each GPU corresponds to the number of active explorers (non-empty intervals), between 0 and 16384 (100\%). 
The last row in Figure~\ref{fig:Ta21Timeline} corresponds to the activity of the coordinator, alternating between idle (0) and active (1) states. 
In Figure~\ref{fig:Ta56Timeline}, the last row shows the total remaining work of the coordinator, decreasing from $50!\approx 3\times 10^{64}$ to $0$ 

The small spikes in GPU-activity correspond to local work stealing operations, which are triggered when the ratio of active explorers decreases below $80\%$. In turn, each of these operations triggers a worker-checkpoint.
The fact that the frequency of these spikes varies throughout the exploration illustrates the irregularity of the search space.
One can notice several sharp decreases of the activity level to zero, causing workers to remain idle until they receive new work units or the global termination signal. The response time of the coordinator is critical for the parallel efficiency of PBB$@$Cluster.
One can also see that the relative worker idle time is much larger for the small \textit{Ta021} instance ($\sim$8 seconds) than for \textit{Ta056} ($\sim$15 minutes).
Notably, although the solution of \textit{Ta056} lasts about 100 times longer than \textit{Ta021}, the total number of global load balancing operations (idle workers acquiring new work) is approximately the same.
Worker idle time occurs mainly in the initial work distribution phase (because the work-intensive parts of the search space are not yet detected) and in the final phase of the exploration (because overall available work is getting scarcer).
These ramp-up and shut-down phases are limiting scalability for smaller instances, while they are negligible for large enough instances.

\subsection{Scalability experiments on Jean Zay}\label{sec:exp:scale}
\begin{table}[!tbp]
\centering
\caption{Summary of $30\times 15$ VFR instances used for scalability experiments on \textit{Jean Zay}.
For each instance the table gives the optimal makespan, size of the critical tree, exploration time on one V100 device and the corresponding single-GPU processing speed.
}\label{tab:VFRinst}
\scriptsize
\begin{tabular}{@{}cccccc@{}}
\toprule
shorthand name  & name & $C_{\max}^\star$ & NN & $T_{\text{1 GPU}}$ (hh:mm) & NN/s\\
\midrule
\textit{small}  & \textit{30\_15\_2}  & 2317 & 122~G & 0:54 & 37.6~M\\
\textit{medium} & \textit{30\_15\_5}  & 2421 & 564 G & 4:22 & 35.8~M\\
\textit{large} 	& \textit{30\_15\_9}  & 2259 & 3660 G & 27:23 & 37.1~M\\
\bottomrule
\end{tabular}
\end{table}

In this subsection we experimentally evaluate the scalability of our approach on \textit{Jean Zay} .
To perform a meaningful scalability analysis, we need to choose problem instances which are small enough to be solved within a reasonable 
 amount of time on a single device and large enough to justify the use of multiple GPUs (single-GPU execution time between $1$~hour and $1$~day). 
Moreover, the selected benchmark instances should be associated with different total workloads (tree sizes), but the granularity of the workload should be the same (i.e. the number of jobs and machines defining the instances should be identical).
Taillard's benchmark instances (except \textit{Ta056}) are either too small or too large, so we selected instances \textit{30\_15\_2}, \textit{30\_15\_5} and \textit{30\_15\_9} from the VFR benchmark, defined by $n=30$ jobs and $m=15$ machines.

\begin{figure}[!tbph]
    \centering
    \begin{subfigure}[t]{0.8\textwidth}
        \centering
		\includegraphics[width=\textwidth]{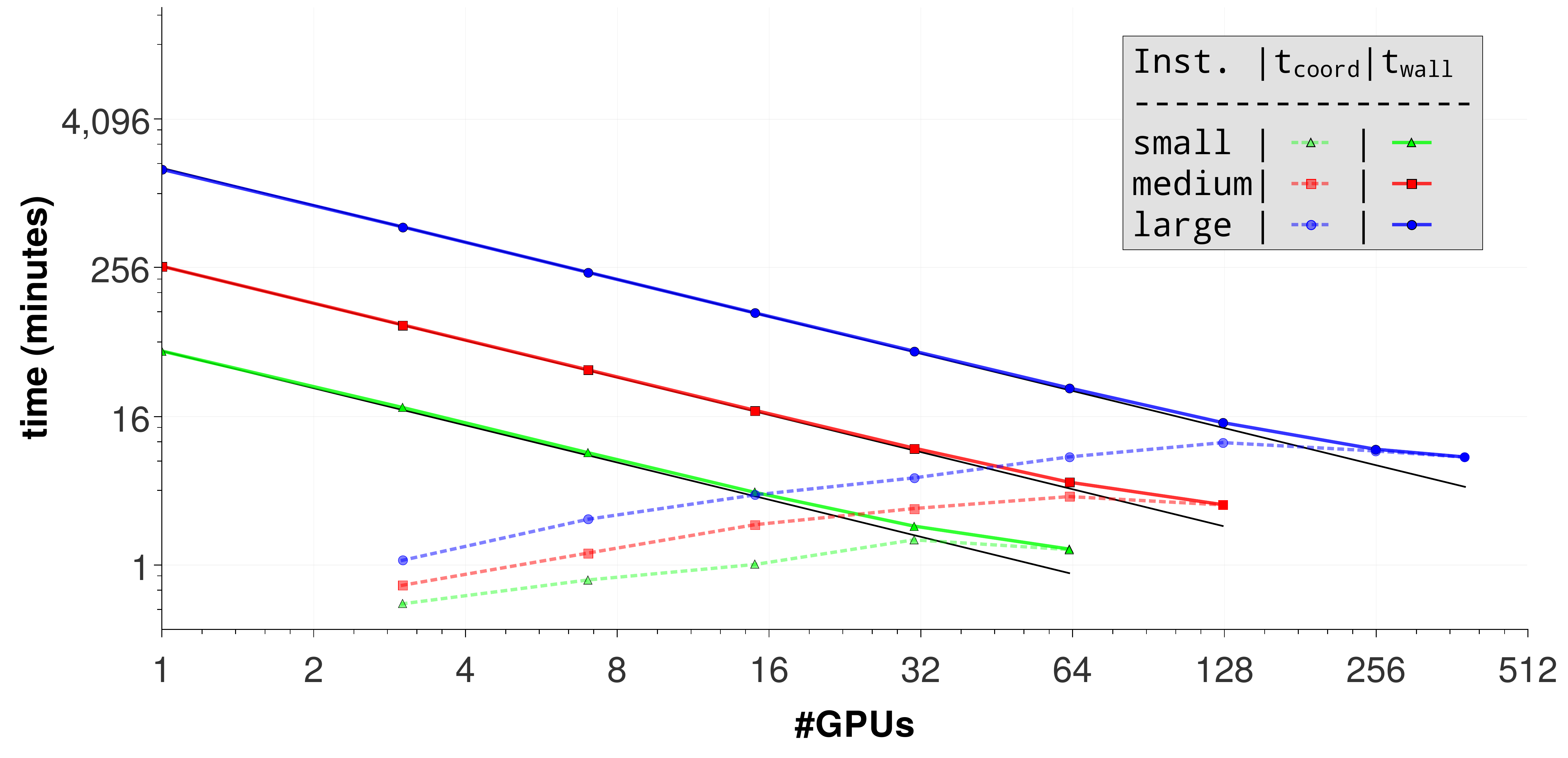}
        \caption{Elapsed walltime and coordinator time}\label{fig:VFRtime}
    \end{subfigure}%
    \hspace{1em} %
    \begin{subfigure}[t]{0.8\textwidth}
        \centering
        \includegraphics[width=\textwidth]{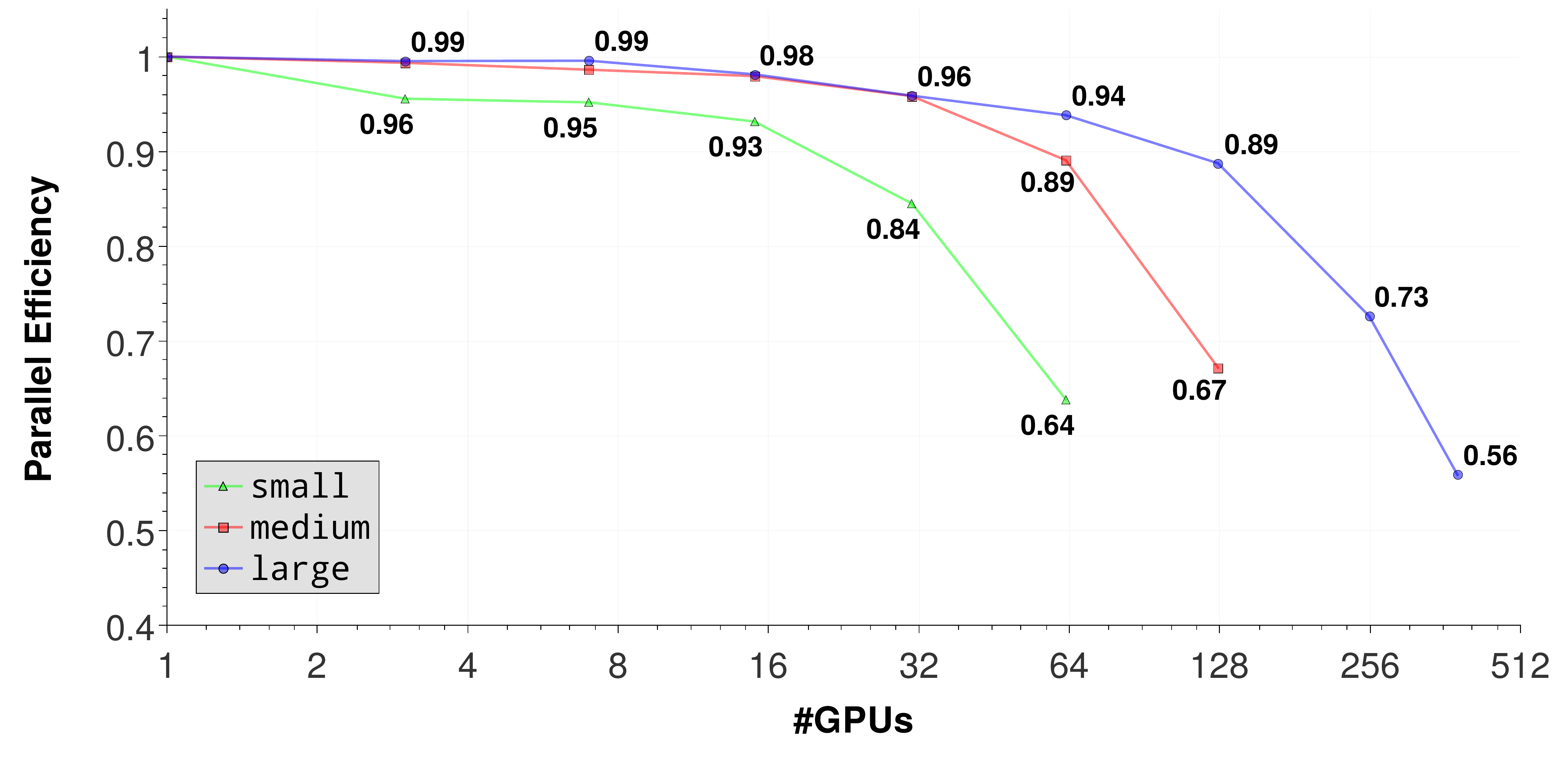}
        \caption{Parallel efficiency}\label{fig:VFReff}
    \end{subfigure}
		\caption{Evaluation of scalability on \textit{Jean Zay}}\label{fig:VFRscale}
\end{figure}

To avoid speedup anomalies, PBB$@$Cluster is initialized with the optimal solutions determined in preliminary runs.
The sizes of explored critical trees (decomposed nodes) and corresponding single-GPU walltimes are shown in Figure~\ref{tab:VFRinst}.
To simplify the presentation of results, we refer to these instances as \textit{small}, \textit{medium} and \textit{large}.
The critical tree of the \textit{small} instance is composed of $122$ billion nodes and its exploration requires $54$ minutes of processing on a single V100, which corresponds to an average processing speed of $37.6\times 10^6$ nodes per second (NN/s).
The \textit{large} instance is $30$ times larger, requiring over $27$~hours of processing at approximately the same speed.
For each of the three instances, runs are performed with $1,2,4,\ldots,2^k$ quad-GPU nodes until the observed parallel efficiency drops below $70\%$.
Four MPI processes are mapped to each node and worker processes map to GPU devices via $\texttt{MPI\_Rank}\pmod 4$ and the \texttt{cudaSetDevice} API function.
As the master process occupies one slot on node $0$, the corresponding number of GPUs is respectively $3,7,15,\ldots,2^{k+2}-1$.

For the three instances and an increasing number of GPUs, Figure~\ref{fig:VFRtime} (on top) shows the elapsed walltime ($t_{\text{wall}}$) with solid lines and the total active time of the coordinator ($t_{\text{coord}}$) with dotted lines. 
The coordinator is considered ``active'' when it is not waiting on \texttt{MPI\_Probe}, i.e. $t_{\text{coord}}$ includes the time process $0$ spends receiving messages, converting intervals and processing worker requests.
The linear scaling curve with respect to a single-GPU execution (no coordinator) is represented by black solid lines.
However, as Figure~\ref{fig:VFRtime} is drawn in $\log-\log$-scale, deviations from the ideal linear case are hard to see.
Therefore, Figure~\ref{fig:VFReff} (below) shows the corresponding parallel efficiency.

Parallel efficiency of at least $90\%$ is achieved with up to $16$, $32$, $64$ GPUs for the \textit{small}, \textit{medium} and \textit{large} instances  respectively; with $32$, $64$ and $128$, PBB$@$Cluster runs with efficiencies of $.84$, $.89$ and $.89$ respectively.
For a larger number of GPUs the parallel efficiency drops off sharply, due to saturation of the coordinator process.
Indeed, one can notice in Figure~\ref{fig:VFRtime} that for $\geq 32$, $\geq 64$ and $\geq 128$ GPUs, $t_{\text{coord}}$ is close to $ t_{\text{wall}}$, meaning that the coordinator is active nearly $100\%$ of the time and becomes a sequential bottleneck.
However, with $384$ GPUs PBB$@$Cluster still reaches a speedup of $215\times$ for the \textit{large} instance, reducing the execution time from over $27$ hours to about $7\nicefrac{1}{2}$~minutes.
The results of this experiment indicate that for larger instances ($>30$ hours on a single device), PBB$@$Cluster can efficiently exploit 
this degree of parallelism, involving over 6 million independent tree exploration agents ($K=16384$ per GPU).

\subsection{Comparison with previous solutions of instance Ta56}
As mentioned in Section~\ref{sec:intro}, the optimal makespan (3679) for the $50\times 20$ instance \textit{Ta056} is known since 2006. The solution was first obtained and proven optimal by a 25-day run of PBB$@$Grid~\citep{MezmazIPDPS}, exploiting on average $328$ CPUs.
Over the last $5$ years, we have re-solved \textit{Ta056} several times on different platforms and with different sequential designs---but with the same initial condition, that is, initialized at 3680 (optimum-plus-one).
For the purpose of comparing PBB$@$Cluster with existing large-scale parallel approaches, Table~\ref{tab:ta56previous} shows a summary of those runs, indicating the used bounding/branching operator, the approximate number of explored nodes, computing platform and walltime, as well as an estimation of the corresponding energy consumption.

\begin{table}[tbp]
\centering
\scriptsize
\caption{Summary of exact resolutions of Ta56, starting from initial solution $3680$ ($C_{\max}^\star+1$).
The energy consumption is estimated as (aggregated TDP)$\times$(walltime).
}\label{tab:ta56previous}
\begin{tabular}{ccc cp{4.2cm} cc}
\toprule
Ref. & Year & LB/branching & tree size & Platform & walltime & kWh \\
\midrule
\citep{MezmazIPDPS} & 2006 & LB2/dyn(MinSum) & $175\times 10^9$ & Avg.~$328$~CPUs (max.~1195), Grid'5000/ULille  & 25 d & 6000 \\
\citep{GmysPhD} & 2015 & LB2/dyn(MinSum) & $175\times 10^9$ & $4\times$GTX980 ($8$k Cuda cores), Univ.~Mons & 9 d & 190 \\  
\citep{GmysPhD} & 2017 & LB2/dyn(MinSum) & $175\times 10^9$ & $36\times$P100 ($130$k Cuda cores), \textit{Ouessant}, IDRIS & 9 h & 110 \\  
\citep{GmysEJOR} & 2020 & LB1/dyn(MinBranch) & $330\times 10^9$ & $2\times$E2630v3 (32 threads), Univ.~Mons & 33 h & 5.6 \\
$[\text{this}]$ & 2020 & LB1/dyn(MinMin) & $270\times 10^9$ & $128\times$V100 ($650$k Cuda cores), \textit{Jean Zay}, IDRIS  & 170s & 2 \\
\bottomrule
\end{tabular}
\end{table}

Using the same LB and branching rule as PBB$@$Grid, \textit{Ta056} was solved in $9$~days on a single quad-GPU node (Maxwell) and in $9$~hours on a GPU-accelerated cluster of $9$ IBM “Minsky” nodes (2$\times$IBM Power8+ / 4$\times$P100).
The improved sequential algorithm presented in~\cite{GmysEJOR} allows to re-solve \textit{Ta056} on a much smaller platform ($2\times 8$-core CPU) in just $33$~hours, despite exploring a larger search tree. 
The PBB$@$Cluster approach presented in this work uses the same fine-grained LB (with a slightly different branching rule) and reduces the wallclock time to less than $3$ minutes on $128$ V100 GPUs.

Although we did not perform any exact measurements with wattmeters, an estimation of energy-consumption using TDP values provided by vendors shows the progress in terms of energy-consumption.
About \nicefrac{2}{3} of the processor pool exploited by PBB$@$Grid in 2006 are AMD Opteron dual-core CPUs and the remaining \nicefrac{1}{3} are mono-core Intel Pentium 4 or Celeron CPUs, mostly at $90$nm feature size.
The high-efficiency Opterons are listed at TDPs of about $60$W and the Pentium 4 and Celerons are listed even higher.
Assuming a TDP of $30$W per core, we can estimate the total energy consumption of the PBB$@$Grid resolution of Ta56 at
$328\times 30\text{W}\times 25~\text{days}\times 24~\text{h/day}\approx 6,000~\text{kWh}$.
With the same lower bound and initial conditions, Ta56 was solved on the \textit{Ouessant} prototype cluster in $9$ hours, using $9$ nodes (2 Power8 CPUs + 36 P100 GPUs).
The energy consumption for this run can be estimated at 
$
(2\times 225\text{W} + 4\times 300\text{W})\times 9~\text{nodes}\times 9~\text{h}\approx 140~\text{kWh}
$.
The same estimation for the improved LB1-based algorithm using $32$ nodes of \textit{Jean Zay} gives
$
(2\times 150\text{W} + 4\times 300\text{W})\times 32~\text{nodes}\times \frac{170}{3600}~\text{h}\approx 2~\text{kWh}
$.
Compared to the first resolution of Ta56, the energy consumption has been reduced by three orders of magnitude.

\subsection{Resolution of open PFSP instances}\label{sec:solveTa}

In this subsection, we give feedback on our attempts to solve instances from the Taillard benchmark for which optimal solutions are unknown.
There are $9$ such instances in the $50$-job/$20$-machine class, $9$ in the $100\times 20$ group and $5$ in the $200\times 20$ group.
For the sake of clarity in the following presentation of results, let us briefly recall the possible outcomes of a PBB execution: 
\begin{itemize}
\item A solution $\pi$ with a better cost than the initial UB is found, but the algorithm does not terminate $\Rightarrow$ no proof of optimality, improved UB
\item A solution $\pi$ with a better cost than the initial UB is found and the algorithm terminates $\Rightarrow$ proof of optimality and improved UB
\item The algorithm terminates but the initial UB is not improved $\Rightarrow$ the optimal solution is larger or equal to the initial UB 
\item The algorithm does on terminate and the initial UB is not improved $\Rightarrow$ no information
\end{itemize}

\paragraph{50-job, 20-machine instances (Ta051-Ta060)}
\begin{table}
\centering
\scriptsize
\caption{Summary of solution attempts for benchmark instances \textit{Ta051}-\textit{Ta060} ($50\times 20$). Out of $9$ open instances, $4$ are solved exactly for the first time, $1$ best-known upper bound is improved, but not proven optimal.}\label{tab:firstexact50}
\begin{tabular}{c ccc cc cc}
\toprule
 \multicolumn{8}{c}{---solved---}  \\
Instance  & \#GPUs & $t_{\text{elapsed}}$ & GPUh  & NN & $\sim$CPU-time & known UB & $C_{\max}^\star$  \\
\midrule 
\textit{Ta058} &  256 & 13h17 & 3399 & 339T & 64 y & 3691 & 3691 \\
\textit{Ta053} &  128 & 7h59 & 1022 & 95T & 19 y & 3640 & 3640 \\
\textit{Ta052} &  384 & 1h54 & 729.6  & 68T &   14 y & 3704 & \textbf{3699} \\
\textit{Ta057} &  384 & 1h11 & 454.4  & 42T &   8.5 y & 3704 & 3704 \\
\midrule
 \multicolumn{8}{c}{---remain open---}  \\
 &   \multicolumn{2}{c}{Best found} & \multicolumn{5}{c}{Comment}\\
\midrule
\textit{Ta051} &  \multicolumn{2}{c}{3846} & \multicolumn{5}{l}{equal to UB from \citep{ravetti2012parallel}} \\
\textit{Ta054} &  \multicolumn{2}{c}{3719} & \multicolumn{5}{l}{equal to UB from \citep{pan2008discrete}, \cite{kizilay2019variable}} \\
\textit{Ta055} &  \multicolumn{2}{c}{3610} & \multicolumn{5}{l}{equal to UB from \citep{deroussi2006new}} \\
\textit{Ta059} &  \multicolumn{2}{c}{3741} & \multicolumn{5}{l}{equal to UB from \citep{pan2008discrete}} \\
\textit{Ta060} &  \multicolumn{2}{c}{\textbf{3755}} &  \multicolumn{5}{l}{\textbf{improved} upper bound} \\
\bottomrule
\end{tabular}
\end{table}

Table~\ref{tab:firstexact50} summarizes the execution statistics for the $9$ unsolved instances of the $50\times 20$ class---$4$ of them are solved to optimality for the first time.
The results show that, even when optimal makespans for instances in this class are available, their optimality is very hard to prove.
In all cases, the algorithm is initialized with the best-known solution from the literature.
Taking for example \textit{Ta058}, proving that no better solution than $3691$ exists required over $13$ hours of processing on $256$ GPUs, performing $339\times 10^{12}$ node decompositions. 
Based on the CPU-GPU comparison shown in Figure~\ref{fig:GPUvsMC}, this corresponds to $64$ CPU-years of sequential processing.
For instances \textit{Ta057} and \textit{Ta053}, the best-known UB is also proven optimal, exploring search trees that are 240 (resp. 540) times larger than for \textit{Ta056}.

In order to confirm the existence of a schedule with these optimal makespans, \textit{Ta057} is solved a second time, finding the same optimal solution when initialized at $C_{\max}^\star+1$.
For \textit{Ta058} and \textit{Ta053}, additional explorations with the support of heuristic searches are performed until an optimal schedule is discovered.
For \textit{Ta052}, PBB$@$Cluster finds an improved schedule and proves its optimality in less than $2$ hours, using $384$ GPUs.
Optimal permutations for these instances are given in~\ref{appendix}.

Instances \textit{Ta051}, \textit{Ta054}, \textit{Ta055}, \textit{Ta59} and \textit{Ta060} remain open, despite using $3$-$5$k GPUh per instance in solution attempts.
PBB$@$Cluster improves the initial UB provided for instance \textit{Ta060} by one unit to $3755$.
For the remaining instances, PBB$@$Cluster is restarted with a larger initial UB and stopped when it finds the best-known UB.
Enabling heuristic searches in PBB$@$Cluster, these solutions are found relatively quickly (usually within less than $1$ hour).
The goal of these additional runs is to confirm the existence of best-known solutions reported in the literature.
For the sake of completeness, corresponding permutation schedules are shown in~\ref{appendix}. 
Considering these observations, we can conjecture that the best-known UBs for the $50\times 20$ are optimal, but proofs of optimality are very hard to obtain. 
 
\paragraph{100-job, 20-machine instances (Ta081-Ta090)} 
 
\begin{table}
\centering
\scriptsize
\caption{Summary of solution attempts for benchmark instances in the \textit{Ta081}-\textit{Ta090} class ($100\times 20$). Out of $9$ open instances, $3$ are solved exactly for the first time, $6$ best-known solutions are improved.}\label{tab:firstexact100}
\begin{tabular}{cc ccc cc cc}
\toprule
  \multicolumn{8}{c}{---solved---}  \\
Instance &  \#GPUs & $t_{\text{elapsed}}$ & GPUh  & NN & $\sim$CPU-time & known UB & $C_{\max}^\star$  \\
\midrule 
\textit{Ta083} &  64 & 0h24 & 25.8 & 2.2T & 290 d & 6271 & 6252 \\
\textit{Ta084} &  32 & 0h16 & 8.5  & 427G & 95 d  & 6269 & 6254 \\
\textit{Ta090} & 128 & 78s & 3  & 168G & 31 d & 6434 & 6404 \\
\midrule
 \multicolumn{8}{c}{---remain open---}  \\
 &  old LB & new LB & $\Delta$LB & ~~~ & old UB & new UB  & $\Delta$UB\\
 \midrule
\textit{Ta081} &  
 6106 & 6135 & +0.47\% & & 6202 & 6173  & -0.47\% \\
\textit{Ta085} &  
 6262 & 6270 & +0.13\% & & 6314 & 6286  & -0.44\%\\
\textit{Ta086} &  
 6302 & 6310 & +0.13\% & & 6364 & 6331  & -0.52\%\\
\textit{Ta087} &  
 6184 & 6210 & +0.42\% & & 6268 & 6224  & -0.70\% \\
\textit{Ta088} & 
 6315 & 6327 & +0.19\% & & 6401 & 6372  & -0.45\% \\
\textit{Ta089} &  
 6204 & 6224 & +0.32\% & & 6275 & 6247  & -0.44\% \\
\midrule 
 Avg &  
  &  & \textbf{+0.28\%} & &  &   & \textbf{-0.50\%} \\
\bottomrule
\end{tabular}
\end{table} 

For the 100-job instances \textit{Ta081}-\textit{Ta090}, prior to this work the exact solution was only known for \textit{Ta082}.
We add three instances to this list : \textit{Ta083}, \textit{Ta084} and \textit{Ta090}.
Solution statistics and improved upper bounds are summarized in Table~\ref{tab:firstexact100}.
After initial, inconclusive solution attempts---without using heuristic search threads---we try to tackle instances in this group in two ways:
\begin{enumerate}
\item Using the best-known \textit{lower bound} (as reported on E.~Taillard's website~\footnote{\url{http://mistic.heig-vd.ch/taillard/problemes.dir/ordonnancement.dir/flowshop.dir/best_lb_up.txt}} as initial UB, PBB$@$Cluster proves that no better solution exists, i.e. returns without discovering a better schedule. Then, the initial UB is incremented by $+1$ and the search is restarted. Iterating over runs with an increasing initial UB, the best known lower bound is improved. This process is stopped when the algorithm finds and proves the optimality of a solution or when a fixed amount of time is elapsed. 
\item The exploration is initialized with the best-known UB and heuristic searches are used to discover better solutions.
\end{enumerate}

The first approach leads to the resolution of \textit{Ta083}, for which the previously best-known LB (6252) is optimal. 
Starting from one unit above the best-known LB (6253), an optimal schedule for \textit{Ta083} was found and proven optimal in $24$ minutes using $64$ GPUs. 
The explored search tree is composed of $2.2\times 10^{12}$ nodes, which is much smaller than the trees explored for the $50\times 20$ instances. 
Notably, once the optimal solution of \textit{Ta083} is found, the search terminates almost instantly---indeed, initialized with the optimum 6252, the exploration of the critical tree can be completed within a few seconds by a sequential PBB algorithm.

The second approach provides improved solutions for all remaining instances of this group.
However, the exact PBB search alone is not able to find these solutions and best-found solutions strongly depend on the quality of the search heuristic.
Optimality proofs are produced for two instances, \textit{Ta084} and \textit{Ta090}, and in both cases the optimal makespan is equal to the best-known LB!
The solution statistics shown in Table~\ref{tab:firstexact100} correspond only to the run that resulted in the solution of the instance.
For \textit{Ta084} and \textit{Ta090}, several PBB$@$Cluster executions were performed prior to that final run (decreasing the initial upper bound and restarted from previous checkpoints)---unfortunately, exploration statistics were lost when restarting the algorithm from a global checkpoint.

The fact that the search completes relatively quickly (for the solved instances) suggests that some of the remaining instances of the $100\times 20$ class may be solved exactly, if heuristic searches are capable of finding an optimal solution.
Indeed, contrary to the $50\times 20$ class, the hardness of the three solved $100\times 20$ instances stems from the difficulty of finding an optimal solution, while the optimality of the latter is relatively easy to prove.

For the six remaining instances, the exploration could not be completed within about $2$-$10$k GPUh of computation per instance.
Although the required remaining time is by nature unpredictable, we used the total remaining work and LB-UB gaps as indicators for the hardness of an instance, and focused efforts on more reachable instances. 
For all unsolved instances, improved upper and lower bounds on the optimal makespan are reported in Table~\ref{tab:firstexact100}.
One can see that for these instances, the previously best-known LB is not optimal (all best-known LBs are improved, on average by 0.32\%).
Taking for example \textit{Ta081}, we have $6135\leq C_{\max}^\star \leq 6173$, which narrows down the previous LB-UB interval 6106---6202. 
On average, best-known UBs for the remaining $100\times 20$ instances are improved by $-0.50\%$. 
Permutation schedules for all improved UBs are provided in~\ref{appendix}.

It should be noted that for the $100\times 20$ class, ARPD values (with respect to best-known UBs) of best-performing metaheuristics 
reported in the literature are in the order of +0.5\%~\cite{DuboisLacoste2017,kizilay2019variable}.
Our results show that, taking into account the improved UBs, actual optimality gaps of these methods are closer to +1.0\%.

\paragraph{200-job, 20-machine instances (\textit{Ta101}-\textit{Ta110})} 

\begin{table}
\centering
\scriptsize
\caption{Summary of solution attempts for benchmark instances in the \textit{Ta101}-\textit{Ta110} class ($200\times 20$). Out of $5$ open instances, $4$ are solved exactly for the first time, $1$ best-known solution is improved.}\label{tab:firstexact200}
\begin{tabular}{c ccc cc cc}
\toprule
 &   \multicolumn{7}{c}{---solved---}  \\
Instance &  \#GPUs & $t_{\text{elapsed}}$ & GPUh  & NN & $\sim$CPU-time & known UB & $C_{\max}^\star$  \\
\midrule 
\textit{Ta101}  & 32 & 0h18 & 9.6  & 225G & 130 d &  11195 & 11158 \\
\textit{Ta107}  & 32 & 0h06 & 3 &  41G & 41 d & 11360 & 11337 \\
\textit{Ta109}  & 32 & 100s & 1 &  10G &   4 d &  11192 & 11146 \\
\textit{Ta108}  & 32 & 28s & $<$1  & 2G &   80 h &  11334 & 11301 \\
\midrule
 &   \multicolumn{7}{c}{---remain open---}  \\
 &  old LB & new LB & $\Delta$LB & ~~~ & old UB & new UB  & $\Delta$ UB\\
 \midrule
\textit{Ta102} &  
 11143 & 11154 & +0.10\% & & 11203 & 11160  & -0.38\% \\
\bottomrule
\end{tabular}
\end{table} 

In the $200\times 20$ class of Taillard's benchmark, $5$ instances remain open, prior to this work.
Four of them are solved exactly and the UB of the remaining instance is improved by 0.38\%.
All four solved instances are solved by running PBB$@$Cluster successively with increasing initial UBs.
The exploration statistics shown in Table~\ref{tab:firstexact200} correspond only to the last run which results in the instance's solution.
The largest of these instances, \textit{Ta101}, is solved in $18$ minutes using 32 GPUs.
Compared to the $50$-job instances, most $200\times 20$ instances are ``rather easy" to solve---although $18$ minutes on $32$ V100 GPU still  correspond to an estimated computing time of 130 CPU-days). 
For two of the four solved instances the best-known LB is optimal (\textit{Ta107}, \textit{Ta108}).
Instance \textit{Ta102} is much harder to solve.
The range of possibly optimal makespan values was narrowed down to 11154--11160 (from 11143--11203), but 
multiple attempts consuming several thousands of GPU-hours were unsuccessful in further increasing (resp. decreasing) the lower (resp. upper) bound.

\paragraph{VFR instances}
\begin{figure}[!tbph]
\begin{center}
	\includegraphics[width=0.99\linewidth]{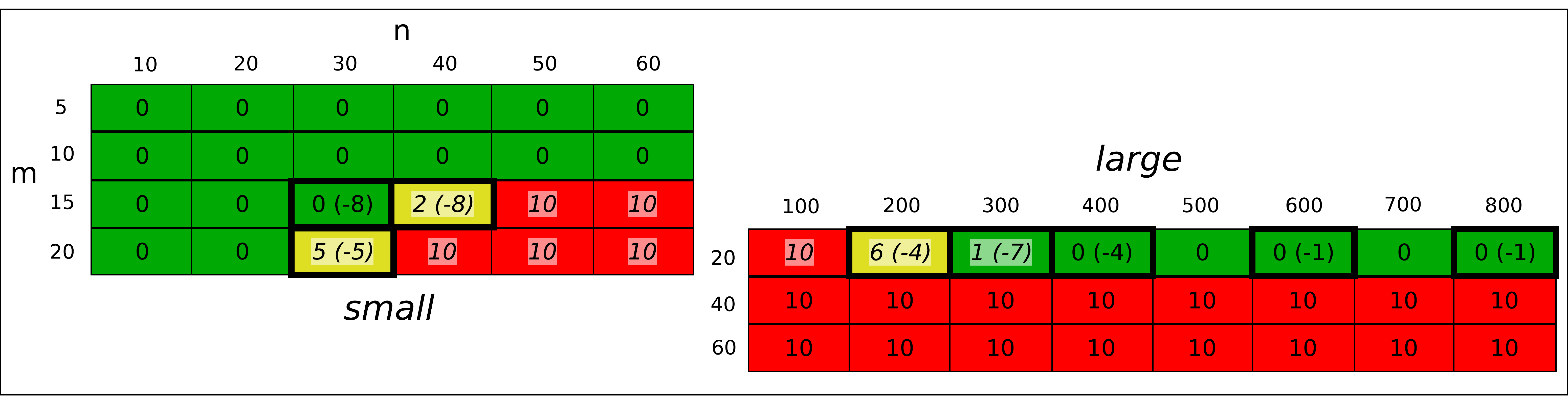}
		\caption{Summary of remaining open VFR instances. In parentheses: number of optimal solutions reported in this work. In italics: instance class for which improved upper bounds are found.}\label{fig:VFRremain}
\end{center}
\end{figure}

We also run PBB$@$Cluster on unsolved VFR instances~\citep{Vallada2015}, although less computing time per instance is spend on these attempts.
We only attempt the resolution of instances with $m=20$ machines as the quality of LBs degrades too quickly with a higher number of machines.
In~\ref{appendix}, optimality proofs and improved UBs are summarized. Overall, $38$ instances are solved exactly for the first time and $75$ improved best-known solutions are reported.
Figure~\ref{fig:VFRremain} summarizes, for each of the 48 instance classes, the number of remaining open instances and the number of optimal solutions provided in this paper.
One can see that the unsolved instances with $m=15$-$20$ machines are centered around $n=50$-$100$, which is consistent with the results obtained for Taillard's benchmark.

\section{Conclusions and future works}\label{sec:conclusion}
In this article, we have presented a hybrid Branch-and-Bound algorithm for exactly solving permutation-based combinatorial optimization problems on clusters composed of GPU-accelerated multi-core nodes (PBB$@$Cluster). 
The permutation flow-shop scheduling problem (PFSP) is used as a test-case.
Our approach solves $11$ of Taillard's benchmark instances to optimality for the first time, which is nearly half of the $23$ instances that remained open for $27$ years. Moreover, best-known upper bounds are improved for $8$ remaining instances.
PBB$@$Cluster solves instance \textit{Ta056} in less than 3 minutes on the \textit{Jean Zay} supercomputer, which is a four orders-of-magnitude improvement over the first exact solution in 2006, that required $25$ days of computation, using a grid-enabled algorithm exploiting $328$ CPUs on average.

This is not achieved through a one-shot research effort, nor by sheer brute-force computing power.
In this paper we tried to give a synthetic overview of the key building blocks and successive contributions that have led to this breakthrough, while diving into details when the latter are critical.
We have presented the design and implementation of PBB$@$Cluster, addressing challenging issues that occur at different levels, from low-level thread divergence to asynchronous inter-node communication and global checkpointing.
Starting from the best available sequential design, we have proposed solutions for mapping a highly irregular and fine-grained tree-search algorithm to modern GPU-accelerated HPC clusters, exploiting all available sources of parallelism.

We have demonstrated the efficiency of the approach through computational experiments.
Using a single V100 GPU, we observe speed-up factors up to $325\times$ compared to a single-threaded CPU-based implementation.
The scalability of PBB$@$Cluster, using millions of GPU-based concurrent tree searches on up to 384 V100 GPUs (2 million CUDA cores) has been evaluated experimentally. An instance requiring 27 hours on a single GPU, is solved in 14 minutes on $32$ quad-GPU nodes, i.e. with 90\% effiency.
The largest instance tackled in this paper (\textit{Ta058}) requires an equivalent computing power of 64 CPU-years---it is solved in $13$ hours, exploring a tree composed of $340\times 10^{12}$ nodes, which is $2000\times$ more than the largest previously solved instance \textit{Ta056}.

In the short term we plan to investigate the hybridization of exact PBB and approximate search methods, which has shown promising results.
We will also investigate the use of high-productivity PGAS-based parallel computing environments, such as Chapel, that  could greatly simplify the implementation of PBB$@$Cluster and parallel tree-search algorithms in general.

\section*{Acknowledgements}
This work was granted access to the HPC resources of IDRIS under the allocation 2020-A0070611107 made by GENCI. Some of the experiments presented in this paper were carried out using the Grid'5000 testbed, supported by a scientific interest group hosted by Inria and including CNRS, RENATER and several Universities as well as other organizations (see https://www.grid5000.fr).
Also, helpful advice by N.~Melab, M.~Mezmaz and D.~Tuyttens on the presentation of this paper is gratefully acknowledged.

\bibliographystyle{elsarticle-num-names}\biboptions{authoryear}
\bibliography{mybibfile}

\begin{thebibliography}{48}
\expandafter\ifx\csname natexlab\endcsname\relax\def\natexlab#1{#1}\fi
\providecommand{\url}[1]{\texttt{#1}}
\providecommand{\href}[2]{#2}
\providecommand{\path}[1]{#1}
\providecommand{\DOIprefix}{doi:}
\providecommand{\ArXivprefix}{arXiv:}
\providecommand{\URLprefix}{URL: }
\providecommand{\Pubmedprefix}{pmid:}
\providecommand{\doi}[1]{\href{http://dx.doi.org/#1}{\path{#1}}}
\providecommand{\Pubmed}[1]{\href{pmid:#1}{\path{#1}}}
\providecommand{\bibinfo}[2]{#2}
\ifx\xfnm\relax \def\xfnm[#1]{\unskip,\space#1}\fi
\bibitem[{Garey et~al.(1976)Garey, Johnson, and Sethi}]{Garey76}
\bibinfo{author}{M.~R. Garey}, \bibinfo{author}{D.~S. Johnson},
  \bibinfo{author}{R.~Sethi},
\newblock \bibinfo{title}{{The Complexity of Flowshop and Jobshop Scheduling}},
\newblock \bibinfo{journal}{Mathematics of Operations Research}
  \bibinfo{volume}{1} (\bibinfo{year}{1976}) \bibinfo{pages}{pp. 117--129}.
\bibitem[{Taillard(1993)}]{Taillard1993}
\bibinfo{author}{E.~Taillard},
\newblock \bibinfo{title}{Benchmarks for basic scheduling problems},
\newblock \bibinfo{journal}{European Journal of Operational Research}
  \bibinfo{volume}{64} (\bibinfo{year}{1993}) \bibinfo{pages}{278 -- 285}.
  \URLprefix
  \url{http://www.sciencedirect.com/science/article/pii/037722179390182M}.
  \DOIprefix\doi{https://doi.org/10.1016/0377-2217(93)90182-M},
  \bibinfo{note}{project Management anf Scheduling}.
\bibitem[{Gmys et~al.(2020)Gmys, Mezmaz, Melab, and Tuyttens}]{GmysEJOR}
\bibinfo{author}{J.~Gmys}, \bibinfo{author}{M.~Mezmaz},
  \bibinfo{author}{N.~Melab}, \bibinfo{author}{D.~Tuyttens},
\newblock \bibinfo{title}{A computationally efficient branch-and-bound
  algorithm for the permutation flow-shop scheduling problem},
\newblock \bibinfo{journal}{European Journal of Operational Research}
  \bibinfo{volume}{284} (\bibinfo{year}{2020}) \bibinfo{pages}{814 -- 833}.
  \URLprefix
  \url{http://www.sciencedirect.com/science/article/pii/S037722172030076X}.
  \DOIprefix\doi{https://doi.org/10.1016/j.ejor.2020.01.039}.
\bibitem[{Asanovic et~al.(2009)Asanovic, Bodik, Demmel, Keaveny, Keutzer,
  Kubiatowicz, Morgan, Patterson, Sen, Wawrzynek, Wessel, and
  Yelick}]{asanovic2006landscape}
\bibinfo{author}{K.~Asanovic}, \bibinfo{author}{R.~Bodik},
  \bibinfo{author}{J.~Demmel}, \bibinfo{author}{T.~Keaveny},
  \bibinfo{author}{K.~Keutzer}, \bibinfo{author}{J.~Kubiatowicz},
  \bibinfo{author}{N.~Morgan}, \bibinfo{author}{D.~Patterson},
  \bibinfo{author}{K.~Sen}, \bibinfo{author}{J.~Wawrzynek},
  \bibinfo{author}{D.~Wessel}, \bibinfo{author}{K.~Yelick},
\newblock \bibinfo{title}{A view of the parallel computing landscape},
\newblock \bibinfo{journal}{Commun. ACM} \bibinfo{volume}{52}
  (\bibinfo{year}{2009}) \bibinfo{pages}{56–67}. \URLprefix
  \url{https://doi.org/10.1145/1562764.1562783}.
  \DOIprefix\doi{10.1145/1562764.1562783}.
\bibitem[{Bader et~al.(2005)Bader, Hart, and Phillips}]{bader2005parallel}
\bibinfo{author}{D.~A. Bader}, \bibinfo{author}{W.~E. Hart},
  \bibinfo{author}{C.~A. Phillips},
\newblock \bibinfo{title}{Parallel algorithm design for branch and bound},
\newblock in: \bibinfo{booktitle}{Tutorials on Emerging Methodologies and
  Applications in Operations Research}, \bibinfo{publisher}{Springer},
  \bibinfo{year}{2005}, pp. \bibinfo{pages}{5--1}.
\bibitem[{Pruul et~al.(1988)Pruul, Nemhauser, and Rushmeier}]{Pruul1975}
\bibinfo{author}{E.~Pruul}, \bibinfo{author}{G.~Nemhauser},
  \bibinfo{author}{R.~Rushmeier},
\newblock \bibinfo{title}{Branch-and-bound and parallel computation: A
  historical note},
\newblock \bibinfo{journal}{Operations Research Letters} \bibinfo{volume}{7}
  (\bibinfo{year}{1988}) \bibinfo{pages}{65 -- 69}. \URLprefix
  \url{http://www.sciencedirect.com/science/article/pii/0167637788900673}.
  \DOIprefix\doi{https://doi.org/10.1016/0167-6377(88)90067-3}.
\bibitem[{Gendron and Crainic(1994)}]{Gendron94}
\bibinfo{author}{B.~Gendron}, \bibinfo{author}{T.~G. Crainic},
\newblock \bibinfo{title}{Parallel branch-and-bound algorithms: Survey and
  synthesis},
\newblock \bibinfo{journal}{Operations Research} \bibinfo{volume}{42}
  (\bibinfo{year}{1994}) \bibinfo{pages}{1042--1066}. \URLprefix
  \url{http://www.jstor.org/stable/171985}.
\bibitem[{Rao and Kumar(1987)}]{RaoParallelDFS1987}
\bibinfo{author}{V.~N. Rao}, \bibinfo{author}{V.~Kumar},
\newblock \bibinfo{title}{Parallel depth first search. part i. implementation},
\newblock \bibinfo{journal}{Int. J. Parallel Program.} \bibinfo{volume}{16}
  (\bibinfo{year}{1987}) \bibinfo{pages}{479–499}. \URLprefix
  \url{https://doi.org/10.1007/BF01389000}. \DOIprefix\doi{10.1007/BF01389000}.
\bibitem[{Karypis and Kumar(1994)}]{Karypis1994UnstructuredTS}
\bibinfo{author}{G.~Karypis}, \bibinfo{author}{V.~Kumar},
\newblock \bibinfo{title}{Unstructured tree search on simd parallel computers},
\newblock \bibinfo{journal}{IEEE Trans. Parallel Distributed Syst.}
  \bibinfo{volume}{5} (\bibinfo{year}{1994}) \bibinfo{pages}{1057--1072}.
\bibitem[{Fonlupt et~al.(1994)Fonlupt, Marquet, and Dekeyser}]{Fonlupt1994}
\bibinfo{author}{C.~Fonlupt}, \bibinfo{author}{P.~Marquet},
  \bibinfo{author}{J.-L. Dekeyser},
\newblock \bibinfo{title}{A data-parallel view of the load balancing
  experimental results on maspar mp-1},
\newblock in: \bibinfo{editor}{W.~Gentzsch}, \bibinfo{editor}{U.~Harms} (Eds.),
  \bibinfo{booktitle}{High-Performance Computing and Networking},
  \bibinfo{publisher}{Springer Berlin Heidelberg}, \bibinfo{address}{Berlin,
  Heidelberg}, \bibinfo{year}{1994}, pp. \bibinfo{pages}{338--343}.
\bibitem[{Reinefeld and Schnecke(1994)}]{reinefeld1994work}
\bibinfo{author}{A.~Reinefeld}, \bibinfo{author}{V.~Schnecke},
\newblock \bibinfo{title}{Work-load balancing in highly parallel depth-first
  search},
\newblock in: \bibinfo{booktitle}{Proceedings of IEEE Scalable High Performance
  Computing Conference}, \bibinfo{organization}{IEEE}, \bibinfo{year}{1994},
  pp. \bibinfo{pages}{773--780}.
\bibitem[{Pessoa et~al.(2016)Pessoa, Gmys, Melab, de~Carvalho~Junior, and
  Tuyttens}]{CarneiroICA3PP}
\bibinfo{author}{T.~C. Pessoa}, \bibinfo{author}{J.~Gmys},
  \bibinfo{author}{N.~Melab}, \bibinfo{author}{F.~H. de~Carvalho~Junior},
  \bibinfo{author}{D.~Tuyttens},
\newblock \bibinfo{title}{A gpu-based backtracking algorithm for permutation
  combinatorial problems},
\newblock in: \bibinfo{editor}{J.~Carretero}, \bibinfo{editor}{J.~Garcia-Blas},
  \bibinfo{editor}{R.~K. Ko}, \bibinfo{editor}{P.~Mueller},
  \bibinfo{editor}{K.~Nakano} (Eds.), \bibinfo{booktitle}{Algorithms and
  Architectures for Parallel Processing}, \bibinfo{publisher}{Springer
  International Publishing}, \bibinfo{address}{Cham}, \bibinfo{year}{2016}, pp.
  \bibinfo{pages}{310--324}.
\bibitem[{Rocki and Suda(2010)}]{RockiMinimax}
\bibinfo{author}{K.~Rocki}, \bibinfo{author}{R.~Suda},
\newblock \bibinfo{title}{Parallel minimax tree searching on gpu},
\newblock in: \bibinfo{editor}{R.~Wyrzykowski}, \bibinfo{editor}{J.~Dongarra},
  \bibinfo{editor}{K.~Karczewski}, \bibinfo{editor}{J.~Wasniewski} (Eds.),
  \bibinfo{booktitle}{Parallel Processing and Applied Mathematics},
  \bibinfo{publisher}{Springer Berlin Heidelberg}, \bibinfo{address}{Berlin,
  Heidelberg}, \bibinfo{year}{2010}, pp. \bibinfo{pages}{449--456}.
\bibitem[{Jenkins et~al.(2011)Jenkins, Arkatkar, Owens, Choudhary, and
  Samatova}]{BacktrackingLessons2011}
\bibinfo{author}{J.~Jenkins}, \bibinfo{author}{I.~Arkatkar},
  \bibinfo{author}{J.~D. Owens}, \bibinfo{author}{A.~Choudhary},
  \bibinfo{author}{N.~F. Samatova},
\newblock \bibinfo{title}{Lessons learned from exploring the backtracking
  paradigm on the gpu},
\newblock in: \bibinfo{editor}{E.~Jeannot}, \bibinfo{editor}{R.~Namyst},
  \bibinfo{editor}{J.~Roman} (Eds.), \bibinfo{booktitle}{Euro-Par 2011 Parallel
  Processing}, \bibinfo{publisher}{Springer Berlin Heidelberg},
  \bibinfo{address}{Berlin, Heidelberg}, \bibinfo{year}{2011}, pp.
  \bibinfo{pages}{425--437}.
\bibitem[{Crainic et~al.(2006)Crainic, Le~Cun, and Roucairol}]{Crainic2006}
\bibinfo{author}{T.~G. Crainic}, \bibinfo{author}{B.~Le~Cun},
  \bibinfo{author}{C.~Roucairol}, \bibinfo{title}{Parallel Branch-and-Bound
  Algorithms}, \bibinfo{publisher}{John Wiley \& Sons, Ltd},
  \bibinfo{year}{2006}, pp. \bibinfo{pages}{1--28}. \URLprefix
  \url{https://onlinelibrary.wiley.com/doi/abs/10.1002/9780470053928.ch1}.
  \DOIprefix\doi{https://doi.org/10.1002/9780470053928.ch1}.
  \href{http://arxiv.org/abs/https://onlinelibrary.wiley.com/doi/pdf/10.1002/9780470053928.ch1}{{\tt
  arXiv:https://onlinelibrary.wiley.com/doi/pdf/10.1002/9780470053928.ch1}}.
\bibitem[{Anstreicher et~al.(2002)Anstreicher, Brixius, Goux, and
  Linderoth}]{anstreicher2002solving}
\bibinfo{author}{K.~Anstreicher}, \bibinfo{author}{N.~Brixius},
  \bibinfo{author}{J.-P. Goux}, \bibinfo{author}{J.~Linderoth},
\newblock \bibinfo{title}{Solving large quadratic assignment problems on
  computational grids},
\newblock \bibinfo{journal}{Mathematical Programming} \bibinfo{volume}{91}
  (\bibinfo{year}{2002}) \bibinfo{pages}{563--588}.
\bibitem[{Date and Nagi(2019)}]{Date2017RLT2basedPA}
\bibinfo{author}{K.~Date}, \bibinfo{author}{R.~Nagi},
\newblock \bibinfo{title}{Level 2 reformulation linearization technique–based
  parallel algorithms for solving large quadratic assignment problems on
  graphics processing unit clusters},
\newblock \bibinfo{journal}{INFORMS Journal on Computing} \bibinfo{volume}{31}
  (\bibinfo{year}{2019}) \bibinfo{pages}{771--789}. \URLprefix
  \url{https://doi.org/10.1287/ijoc.2018.0866}.
  \DOIprefix\doi{10.1287/ijoc.2018.0866}.
  \href{http://arxiv.org/abs/https://doi.org/10.1287/ijoc.2018.0866}{{\tt
  arXiv:https://doi.org/10.1287/ijoc.2018.0866}}.
\bibitem[{{Mezmaz} et~al.(2007){Mezmaz}, {Melab}, and {Talbi}}]{MezmazIPDPS}
\bibinfo{author}{M.~{Mezmaz}}, \bibinfo{author}{N.~{Melab}},
  \bibinfo{author}{E.~{Talbi}},
\newblock \bibinfo{title}{A grid-enabled branch and bound algorithm for solving
  challenging combinatorial optimization problems},
\newblock in: \bibinfo{booktitle}{2007 IEEE International Parallel and
  Distributed Processing Symposium}, \bibinfo{year}{2007}, pp.
  \bibinfo{pages}{1--9}. \DOIprefix\doi{10.1109/IPDPS.2007.370217}.
\bibitem[{Chakroun and Melab(2015)}]{ChakrounJPDC15}
\bibinfo{author}{I.~Chakroun}, \bibinfo{author}{N.~Melab},
\newblock \bibinfo{title}{Towards a heterogeneous and adaptive parallel
  branch-and-bound algorithm},
\newblock \bibinfo{journal}{Journal of Computer and System Sciences}
  \bibinfo{volume}{81} (\bibinfo{year}{2015}) \bibinfo{pages}{72 -- 84}.
  \URLprefix
  \url{http://www.sciencedirect.com/science/article/pii/S0022000014000993}.
  \DOIprefix\doi{https://doi.org/10.1016/j.jcss.2014.06.012}.
\bibitem[{Vu and Derbel(2016)}]{Vu2016}
\bibinfo{author}{T.-T. Vu}, \bibinfo{author}{B.~Derbel},
\newblock \bibinfo{title}{Parallel branch-and-bound in multi-core multi-cpu
  multi-gpu heterogeneous environments},
\newblock \bibinfo{journal}{Future Generation Computer Systems}
  \bibinfo{volume}{56} (\bibinfo{year}{2016}) \bibinfo{pages}{95 -- 109}.
  \URLprefix
  \url{http://www.sciencedirect.com/science/article/pii/S0167739X15003222}.
  \DOIprefix\doi{https://doi.org/10.1016/j.future.2015.10.009}.
\bibitem[{Gmys et~al.(2017)Gmys, Mezmaz, Melab, and Tuyttens}]{GmysPPAMSE}
\bibinfo{author}{J.~Gmys}, \bibinfo{author}{M.~Mezmaz},
  \bibinfo{author}{N.~Melab}, \bibinfo{author}{D.~Tuyttens},
\newblock \bibinfo{title}{Ivm-based parallel branch-and-bound using
  hierarchical work stealing on multi-gpu systems},
\newblock \bibinfo{journal}{Concurrency and Computation: Practice and
  Experience} \bibinfo{volume}{29} (\bibinfo{year}{2017})
  \bibinfo{pages}{e4019}. \URLprefix
  \url{https://onlinelibrary.wiley.com/doi/abs/10.1002/cpe.4019}.
  \DOIprefix\doi{https://doi.org/10.1002/cpe.4019}.
  \href{http://arxiv.org/abs/https://onlinelibrary.wiley.com/doi/pdf/10.1002/cpe.4019}{{\tt
  arXiv:https://onlinelibrary.wiley.com/doi/pdf/10.1002/cpe.4019}},
  \bibinfo{note}{e4019 cpe.4019}.
\bibitem[{Lageweg et~al.(1978)Lageweg, Lenstra, and Kan}]{Lageweg78}
\bibinfo{author}{B.~J. Lageweg}, \bibinfo{author}{J.~K. Lenstra},
  \bibinfo{author}{A.~H. G.~R. Kan},
\newblock \bibinfo{title}{{A General Bounding Scheme for the Permutation
  Flow-Shop Problem}},
\newblock \bibinfo{journal}{Operations Research} \bibinfo{volume}{26}
  (\bibinfo{year}{1978}) \bibinfo{pages}{53--67}.
  \DOIprefix\doi{10.1287/opre.26.1.53}.
  \href{http://arxiv.org/abs/http://dx.doi.org/10.1287/opre.26.1.53}{{\tt
  arXiv:http://dx.doi.org/10.1287/opre.26.1.53}}.
\bibitem[{Ignall and Schrage(1965)}]{Ignall1965}
\bibinfo{author}{E.~Ignall}, \bibinfo{author}{L.~Schrage},
\newblock \bibinfo{title}{Application of the branch and bound technique to some
  flow-shop scheduling problems},
\newblock \bibinfo{journal}{Oper. Res.} \bibinfo{volume}{13}
  (\bibinfo{year}{1965}) \bibinfo{pages}{400--412}. \URLprefix
  \url{http://dx.doi.org/10.1287/opre.13.3.400}.
  \DOIprefix\doi{10.1287/opre.13.3.400}.
\bibitem[{Vallada et~al.(2015)Vallada, Ruiz, and Framinan}]{Vallada2015}
\bibinfo{author}{E.~Vallada}, \bibinfo{author}{R.~Ruiz}, \bibinfo{author}{J.~M.
  Framinan},
\newblock \bibinfo{title}{New hard benchmark for flowshop scheduling problems
  minimising makespan},
\newblock \bibinfo{journal}{European Journal of Operational Research}
  \bibinfo{volume}{240} (\bibinfo{year}{2015}) \bibinfo{pages}{666 -- 677}.
  \URLprefix
  \url{http://www.sciencedirect.com/science/article/pii/S0377221714005992}.
  \DOIprefix\doi{https://doi.org/10.1016/j.ejor.2014.07.033}.
\bibitem[{{Mezmaz} et~al.(2014){Mezmaz}, {Leroy}, {Melab}, and
  {Tuyttens}}]{IVM-IPDPS2014}
\bibinfo{author}{M.~{Mezmaz}}, \bibinfo{author}{R.~{Leroy}},
  \bibinfo{author}{N.~{Melab}}, \bibinfo{author}{D.~{Tuyttens}},
\newblock \bibinfo{title}{A multi-core parallel branch-and-bound algorithm
  using factorial number system},
\newblock in: \bibinfo{booktitle}{2014 IEEE 28th International Parallel and
  Distributed Processing Symposium}, \bibinfo{year}{2014}, pp.
  \bibinfo{pages}{1203--1212}. \DOIprefix\doi{10.1109/IPDPS.2014.124}.
\bibitem[{Gmys et~al.(2016)Gmys, Mezmaz, Melab, and Tuyttens}]{GmysParco}
\bibinfo{author}{J.~Gmys}, \bibinfo{author}{M.~Mezmaz},
  \bibinfo{author}{N.~Melab}, \bibinfo{author}{D.~Tuyttens},
\newblock \bibinfo{title}{A gpu-based branch-and-bound algorithm using
  integer-vector-matrix data structure},
\newblock \bibinfo{journal}{Parallel Comput.} \bibinfo{volume}{59}
  (\bibinfo{year}{2016}) \bibinfo{pages}{119--139}. \URLprefix
  \url{https://doi.org/10.1016/j.parco.2016.01.008}.
  \DOIprefix\doi{10.1016/j.parco.2016.01.008}.
\bibitem[{Bo\.{z}ejko(2009)}]{BozejkoPFSP}
\bibinfo{author}{W.~Bo\.{z}ejko},
\newblock \bibinfo{title}{Solving the flow shop problem by parallel
  programming},
\newblock \bibinfo{journal}{Journal of Parallel and Distributed Computing}
  \bibinfo{volume}{69} (\bibinfo{year}{2009}) \bibinfo{pages}{470 -- 481}.
  \URLprefix
  \url{http://www.sciencedirect.com/science/article/pii/S0743731509000215}.
  \DOIprefix\doi{https://doi.org/10.1016/j.jpdc.2009.01.009}.
\bibitem[{Melab et~al.(2018)Melab, Gmys, Mezmaz, and Tuyttens}]{Melab2018FGCS}
\bibinfo{author}{N.~Melab}, \bibinfo{author}{J.~Gmys},
  \bibinfo{author}{M.~Mezmaz}, \bibinfo{author}{D.~Tuyttens},
\newblock \bibinfo{title}{Multi-core versus many-core computing for many-task
  branch-and-bound applied to big optimization problems},
\newblock \bibinfo{journal}{Future Generation Computer Systems}
  \bibinfo{volume}{82} (\bibinfo{year}{2018}) \bibinfo{pages}{472 -- 481}.
  \URLprefix
  \url{http://www.sciencedirect.com/science/article/pii/S0167739X16308706}.
  \DOIprefix\doi{https://doi.org/10.1016/j.future.2016.12.039}.
\bibitem[{de~Bruin et~al.(1995)de~Bruin, Kindervater, and
  Trienekens}]{deBruin1995}
\bibinfo{author}{A.~de~Bruin}, \bibinfo{author}{G.~A.~P. Kindervater},
  \bibinfo{author}{H.~W. J.~M. Trienekens},
\newblock \bibinfo{title}{Asynchronous parallel branch and bound and
  anomalies},
\newblock in: \bibinfo{editor}{A.~Ferreira}, \bibinfo{editor}{J.~Rolim} (Eds.),
  \bibinfo{booktitle}{Parallel Algorithms for Irregularly Structured Problems},
  \bibinfo{publisher}{Springer Berlin Heidelberg}, \bibinfo{address}{Berlin,
  Heidelberg}, \bibinfo{year}{1995}, pp. \bibinfo{pages}{363--377}.
\bibitem[{Cappello(2009)}]{cappello2009fault}
\bibinfo{author}{F.~Cappello},
\newblock \bibinfo{title}{Fault tolerance in petascale/exascale systems:
  Current knowledge, challenges and research opportunities},
\newblock \bibinfo{journal}{The International Journal of High Performance
  Computing Applications} \bibinfo{volume}{23} (\bibinfo{year}{2009})
  \bibinfo{pages}{212--226}.
\bibitem[{Johnson(1954)}]{Johnson54}
\bibinfo{author}{S.~M. Johnson},
\newblock \bibinfo{title}{{Optimal two- and three-stage production schedules
  with setup times included}},
\newblock \bibinfo{journal}{Naval Research Logistics Quarterly}
  \bibinfo{volume}{1} (\bibinfo{year}{1954}) \bibinfo{pages}{61--68}.
  \DOIprefix\doi{10.1002/nav.3800010110}.
\bibitem[{Lomnicki(1965)}]{Lomnicki1965}
\bibinfo{author}{Z.~A. Lomnicki},
\newblock \bibinfo{title}{A ``branch-and-bound'' algorithm for the exact
  solution of the three-machine scheduling problem},
\newblock \bibinfo{journal}{Journal of the Operational Research Society}
  \bibinfo{volume}{16} (\bibinfo{year}{1965}) \bibinfo{pages}{89--100}.
  \URLprefix \url{https://doi.org/10.1057/jors.1965.7}.
  \DOIprefix\doi{10.1057/jors.1965.7}.
\bibitem[{Potts(1980)}]{Potts1980}
\bibinfo{author}{C.~Potts},
\newblock \bibinfo{title}{An adaptive branching rule for the permutation
  flow-shop problem},
\newblock \bibinfo{journal}{European Journal of Operational Research}
  \bibinfo{volume}{5} (\bibinfo{year}{1980}) \bibinfo{pages}{19 -- 25}.
  \URLprefix
  \url{http://www.sciencedirect.com/science/article/pii/0377221780900697}.
  \DOIprefix\doi{https://doi.org/10.1016/0377-2217(80)90069-7}.
\bibitem[{Knuth(1997)}]{KnuthACP}
\bibinfo{author}{D.~Knuth},
\newblock \bibinfo{title}{{The Art of Computer Programming, Volume 2:
  Seminumerical Algorithms}},
\newblock \bibinfo{journal}{Reading, Ma}  (\bibinfo{year}{1997})
  \bibinfo{pages}{192}. \bibinfo{note}{ISBN=9780201896848}.
\bibitem[{Chakroun et~al.(2013)Chakroun, Melab, Mezmaz, and
  Tuyttens}]{ChakrounJPDC2013}
\bibinfo{author}{I.~Chakroun}, \bibinfo{author}{N.~Melab},
  \bibinfo{author}{M.~Mezmaz}, \bibinfo{author}{D.~Tuyttens},
\newblock \bibinfo{title}{Combining multi-core and gpu computing for solving
  combinatorial optimization problems},
\newblock \bibinfo{journal}{Journal of Parallel and Distributed Computing}
  \bibinfo{volume}{73} (\bibinfo{year}{2013}) \bibinfo{pages}{1563 -- 1577}.
  \URLprefix
  \url{http://www.sciencedirect.com/science/article/pii/S0743731513001615}.
  \DOIprefix\doi{https://doi.org/10.1016/j.jpdc.2013.07.023},
  \bibinfo{note}{heterogeneity in Parallel and Distributed Computing}.
\bibitem[{Carneiro et~al.(2011)Carneiro, Muritiba, Negreiros, and
  de~Campos}]{CarneiroSBACPAD}
\bibinfo{author}{T.~Carneiro}, \bibinfo{author}{A.~Muritiba},
  \bibinfo{author}{M.~Negreiros}, \bibinfo{author}{G.~L. de~Campos},
\newblock \bibinfo{title}{A new parallel schema for branch-and-bound algorithms
  using gpgpu},
\newblock in: \bibinfo{booktitle}{Computer Architecture and High Performance
  Computing, Symposium on}, \bibinfo{publisher}{IEEE Computer Society},
  \bibinfo{address}{Los Alamitos, CA, USA}, \bibinfo{year}{2011}, pp.
  \bibinfo{pages}{41--47}. \URLprefix
  \url{https://doi.ieeecomputersociety.org/10.1109/SBAC-PAD.2011.20}.
  \DOIprefix\doi{10.1109/SBAC-PAD.2011.20}.
\bibitem[{{Vaidyanathan} et~al.(2015){Vaidyanathan}, {Kalamkar}, {Pamnany},
  {Hammond}, {Balaji}, {Das}, {Park}, and {Joó}}]{Vaidyanathan2015}
\bibinfo{author}{K.~{Vaidyanathan}}, \bibinfo{author}{D.~D. {Kalamkar}},
  \bibinfo{author}{K.~{Pamnany}}, \bibinfo{author}{J.~R. {Hammond}},
  \bibinfo{author}{P.~{Balaji}}, \bibinfo{author}{D.~{Das}},
  \bibinfo{author}{J.~{Park}}, \bibinfo{author}{B.~{Joó}},
\newblock \bibinfo{title}{Improving concurrency and asynchrony in multithreaded
  mpi applications using software offloading},
\newblock in: \bibinfo{booktitle}{SC '15: Proceedings of the International
  Conference for High Performance Computing, Networking, Storage and Analysis},
  \bibinfo{year}{2015}, pp. \bibinfo{pages}{1--12}.
  \DOIprefix\doi{10.1145/2807591.2807602}.
\bibitem[{{Hoefler} and {Lumsdaine}(2008)}]{Hoefler2008}
\bibinfo{author}{T.~{Hoefler}}, \bibinfo{author}{A.~{Lumsdaine}},
\newblock \bibinfo{title}{Message progression in parallel computing - to thread
  or not to thread?},
\newblock in: \bibinfo{booktitle}{2008 IEEE International Conference on Cluster
  Computing}, \bibinfo{year}{2008}, pp. \bibinfo{pages}{213--222}.
  \DOIprefix\doi{10.1109/CLUSTR.2008.4663774}.
\bibitem[{Ruiz and St{\"{u}}tzle(2007)}]{Stutzle2007}
\bibinfo{author}{R.~Ruiz}, \bibinfo{author}{T.~St{\"{u}}tzle},
\newblock \bibinfo{title}{A simple and effective iterated greedy algorithm for
  the permutation flowshop scheduling problem},
\newblock \bibinfo{journal}{European Journal of Operational Research}
  \bibinfo{volume}{177} (\bibinfo{year}{2007}) \bibinfo{pages}{2033 -- 2049}.
  \URLprefix
  \url{http://www.sciencedirect.com/science/article/pii/S0377221705008507}.
  \DOIprefix\doi{https://doi.org/10.1016/j.ejor.2005.12.009}.
\bibitem[{Deroussi et~al.(2010)Deroussi, Gourgand, and
  Norre}]{deroussi2010adaptation}
\bibinfo{author}{L.~Deroussi}, \bibinfo{author}{M.~Gourgand},
  \bibinfo{author}{S.~Norre},
\newblock \bibinfo{title}{Une adaptation efficace des mouvements de lin et
  kernighan pour le flow-shop de permutation},
\newblock in: \bibinfo{booktitle}{8eme Conférence Internationale de
  MOdélisation et SIMulation: MOSIM, Tunisie}, \bibinfo{year}{2010}.
  \URLprefix
  \url{http://citeseerx.ist.psu.edu/viewdoc/download?doi=10.1.1.167.2662&rep=rep1&type=pdf}.
\bibitem[{Libralesso et~al.(2020)Libralesso, Focke, Secardin, and
  Jost}]{libralessoBeam}
\bibinfo{author}{L.~Libralesso}, \bibinfo{author}{P.~A. Focke},
  \bibinfo{author}{A.~Secardin}, \bibinfo{author}{V.~Jost},
  \bibinfo{title}{{Iterative beam search algorithms for the permutation
  flowshop}}, \bibinfo{year}{2020}. \URLprefix
  \url{https://hal.archives-ouvertes.fr/hal-02937115}, \bibinfo{note}{working
  paper or preprint}.
\bibitem[{{Choquette} et~al.(2018){Choquette}, {Giroux}, and {Foley}}]{Volta}
\bibinfo{author}{J.~{Choquette}}, \bibinfo{author}{O.~{Giroux}},
  \bibinfo{author}{D.~{Foley}},
\newblock \bibinfo{title}{Volta: Performance and programmability},
\newblock \bibinfo{journal}{IEEE Micro} \bibinfo{volume}{38}
  (\bibinfo{year}{2018}) \bibinfo{pages}{42--52}.
  \DOIprefix\doi{10.1109/MM.2018.022071134}.
\bibitem[{{Gmys}(2017)}]{GmysPhD}
\bibinfo{author}{J.~{Gmys}}, \bibinfo{title}{Heterogeneous cluster computing
  for many-task exact optimization - Application to permutation problems},
  Ph.D. thesis, Université de Mons / Université de Lille,
  \bibinfo{year}{2017}. \URLprefix \url{https://hal.inria.fr/tel-01652000}.
\bibitem[{Ravetti et~al.(2012)Ravetti, Riveros, Mendes, Resende, and
  Pardalos}]{ravetti2012parallel}
\bibinfo{author}{M.~G. Ravetti}, \bibinfo{author}{C.~Riveros},
  \bibinfo{author}{A.~Mendes}, \bibinfo{author}{M.~G. Resende},
  \bibinfo{author}{P.~M. Pardalos},
\newblock \bibinfo{title}{Parallel hybrid heuristics for the permutation flow
  shop problem},
\newblock \bibinfo{journal}{Annals of Operations Research}
  \bibinfo{volume}{199} (\bibinfo{year}{2012}) \bibinfo{pages}{269--284}.
\bibitem[{Pan et~al.(2008)Pan, Tasgetiren, and Liang}]{pan2008discrete}
\bibinfo{author}{Q.-K. Pan}, \bibinfo{author}{M.~F. Tasgetiren},
  \bibinfo{author}{Y.-C. Liang},
\newblock \bibinfo{title}{A discrete differential evolution algorithm for the
  permutation flowshop scheduling problem},
\newblock \bibinfo{journal}{Computers \& Industrial Engineering}
  \bibinfo{volume}{55} (\bibinfo{year}{2008}) \bibinfo{pages}{795--816}.
\bibitem[{Kizilay et~al.(2019)Kizilay, Tasgetiren, Pan, and
  Gao}]{kizilay2019variable}
\bibinfo{author}{D.~Kizilay}, \bibinfo{author}{M.~F. Tasgetiren},
  \bibinfo{author}{Q.-K. Pan}, \bibinfo{author}{L.~Gao},
\newblock \bibinfo{title}{A variable block insertion heuristic for solving
  permutation flow shop scheduling problem with makespan criterion},
\newblock \bibinfo{journal}{Algorithms} \bibinfo{volume}{12}
  (\bibinfo{year}{2019}) \bibinfo{pages}{100}.
\bibitem[{Deroussi et~al.(2006)Deroussi, Gourgand, and Norre}]{deroussi2006new}
\bibinfo{author}{L.~Deroussi}, \bibinfo{author}{M.~Gourgand},
  \bibinfo{author}{S.~Norre}, \bibinfo{title}{New effective neighborhoods for
  the permutation flow shop problem}, \bibinfo{type}{Technical Report}
  \bibinfo{number}{LIMOS/RR-06-09}, LIMOS, Université Clermont Auvergne,
  \bibinfo{year}{2006}. \URLprefix
  \url{https://hal.archives-ouvertes.fr/hal-00678053/}.
\bibitem[{Dubois-Lacoste et~al.(2017)Dubois-Lacoste, Pagnozzi, and
  Stützle}]{DuboisLacoste2017}
\bibinfo{author}{J.~Dubois-Lacoste}, \bibinfo{author}{F.~Pagnozzi},
  \bibinfo{author}{T.~Stützle},
\newblock \bibinfo{title}{An iterated greedy algorithm with optimization of
  partial solutions for the makespan permutation flowshop problem},
\newblock \bibinfo{journal}{Computers \& Operations Research}
  \bibinfo{volume}{81} (\bibinfo{year}{2017}) \bibinfo{pages}{160 -- 166}.
  \URLprefix
  \url{http://www.sciencedirect.com/science/article/pii/S030505481630329X}.
  \DOIprefix\doi{https://doi.org/10.1016/j.cor.2016.12.021}.

\end{thebibliography}

\appendix
\section{Solutions for benchmark instances}\label{appendix}

\centering
\tiny
\begin{longtable}{cccp{12cm}}
\caption{Best-known solutions for instances \textit{Ta051}-\textit{Ta060} (50 jobs / 20 machines). Makespans shown in a box (\fbox{$C_{\max}$}) are optimal. Bold-faced instance-names indicate that the upper bound is improved and/or that the instance is solved for the first time in this paper.}
\\ \hline
Inst & old LB-UB~\footnote{\url{http://mistic.heig-vd.ch/taillard/problemes.dir/ordonnancement.dir/flowshop.dir/best_lb_up.txt} and other sources.}
 & best $C_{\max}$ & Permutation schedule  \\
\midrule
\textit{Ta051} & 3771-3846 & 3846 & 20  31  39  27  43  15  44  11   8  45  35  37   6  17  34  28   7  14  42  33  40  24   5  29  10   2  18  47  48  21  46   1  16  49  12  23  22  36  32  38  19   9  26  25  13  41  30   4  50   3 \\ \midrule
\textbf{Ta052} & 3668-3704 & \fbox{\textbf{3699}} &
 33  20  41  43  32  38  36  18  39  29  42  17  11  16  13  31   1  50  46  47  37  40  28  14  49  12  45   5   2  23   4  25  15  35  44  19  48  26  24  10  21  30   6   3   8  22  34   7  27   9 \\ \midrule
\textbf{Ta053} & 3591-3640 & \fbox{\textbf{3640}} & 
24 4 10 28 21 8 37 46 16 22 31 5 39 2 32 11 25 49 47 20 15 48 26 3 35 17 14 43 27 45 9 1 19 50 30 6 36 34 29 42 23 33 41 12 7 18 40 44 13 38\\ \midrule
\textit{Ta054} & 3635-3719 & 3719 &
  5  21  11  14  36  30  13  24  12   7  45  19  35  20  31  25  37   3  44  33  32  50  48  43  49  29  46  23  10  40  15  38   9  17  42  22   6  39  26  47   4  27  18   8   2  41  34   1  16  28 \\ \midrule
\textit{Ta055} & 3553-3610 & 3610 &
 40  48   4   2  19  31  50  28  20  49  34   5  23  21  32  25  43  45  44  18  26  36  33  42  27  16  41  14   8  47  39  38  10   6  22  17  30  12  13   3  37   9   7   1  46  24  15  29  35  11\\ \midrule
\textit{Ta056} & \fbox{3679} & \fbox{3679} &
 14  37   3  18   8  50   5  42  33  40   4  45  17  27  20  21  13  49  43  11  10  41  24  15  16  19  44  32  26  28  46   1  36  39  47  25  30   7   2  31  23   6  48  22  29  34   9  35  38  12 \\ \midrule
\textbf{Ta057} & 3672-3704 & \fbox{\textbf{3704}} & 
 4  23  15   1  12  10  13  20  17  38   2  49  19   8  33  45  11  31  41  22  50  47  21  14  34  30  48  27  39  32   5  29  46  35  40  28  37  25  24   3   7   9  18  42  36  44   6  26  43  16 \\ \midrule
\textbf{Ta058} & 3627-3691 & \fbox{\textbf{3691}} & 
39 32 18 7 4 20 29 31 6 8 48 19 33 12 27 30 38 26 15 36 47 21 35 10 2 17 41 5 9 28 3 25 16 1 24 37 49 42 45 22 11 50 40 46 13 43 14 44 34 23\\ \midrule
\textit{Ta059} & 3645-3741 & 3741 &
  3  14   8  37  22  32  12  46  16   9  41  30  38  24  10   1  18  17  34  50  28  36  40  29  26  47   6   7  13  27  33  39  23  11  49  45   4   5  43  48  21  31  42  19  25   2  20  15  44  35\\ \midrule
\textbf{Ta060} & 3696-3756 & \textbf{3755} &
 33  12  19   8   3  22  15  23   2   9  40   1  11  21  36  32  25  47  31  16  37  10  42  18  50  27  29  13  44  14  38  34  17  28  39   6  26  49  46   5  24  41  20  30  35   7  48  45  43   4 \\
\bottomrule
\end{longtable}

\centering
\tiny
\begin{longtable}{cccp{12cm}}
\caption{Best-known solutions for instances \textit{Ta081}-\textit{Ta090} (100 jobs / 20 machines). Makespans shown in a box (\fbox{$C_{\max}$}) are optimal. Bold-faced instance-names indicate that the upper bound is improved and/or that the instance is solved for the first time.} 
\\ \hline %
Inst & old LB-UB~\footnote{\url{http://mistic.heig-vd.ch/taillard/problemes.dir/ordonnancement.dir/flowshop.dir/best_lb_up.txt} and other sources.} & LB-UB & Permutation schedule  \\
\midrule
\textbf{Ta081} & 6106-6202 & 6134-\textbf{6173} & 
 54  78  59  74  79   1  83  89  31   3  51  50  80  99  61  25   9  35  76  40  82  70  12  85   5  16  91  81   6  60  56  33  22  10  44 100  63  65  21  96  28  55  62  86  94  18  24  30  53  20   2  45  47  48  23  27   4  11  41  92  29  88  46  73  64  71  36  57  93  32  67  98  90  77  37  87  42  26  66  39  95  52  68  58  72  84  34  38  43  69  19  15  14  75  17  97   8  49  13   7 \\ \midrule
\textit{Ta082} & \fbox{6183} & \fbox{6183} &
 50  49  95  65  32  27  87  66  80  52  69  90  35  82  72  89  19  31  10  40  14  96  62  79  78   2  33  59  75  93  48  77  13  71   9  70  54  22   1  36   5   7  34  84  91  46  68 100  61  98  53  20  47  76  92  58  43  15  45  99  26  23  55  42  73  38  11   4  85  37  86  97  74   8  41  51   3  63  64  60  83  30  24  25  56  16  88  67  28  17   6  44  18  21  12  94  29  81  39  57 \\ \midrule
\textbf{Ta083} & 6252-6271 & \fbox{\textbf{6252}} &
 10  41  54  87  67  56  11  86  29  51  76  24  64  42   8  57  37  58  31  23  48   1   4  50  97  63  94  61  88  80   5  46  33  98  28  32  43  36  47  78   9  40  44  77   2  70  22  72  84  20 81  49  90  91  35  26  69   6  52  21  66  25   7  39  17  85  73  53  12  89  34   3  55  96  59  27  82  79  75  68  18  15  19  99  71  30 100  13  60  62  16  74  95  38  93  65  45  83  14  92 \\ \midrule
\textbf{Ta084} & 6254-6269 & \fbox{\textbf{6254}} & 
 74  98  45  44  25  79  38  57  58  23  67  89  43  29  90  51  88  60  84  50  69  65   1  15   5  71  92  46  95  26  56  96  52   8   7  91  16  55  86  77  22  78  54  87  64  63  76  32  41  68  11  99  13  59  72   3  47  28  37  70  30  93  31  62  85   6   4   9  73   2  61  81  39  17  20  35  34  19  94 100  24  42  83  21  10  75  82  66  27  18  12  49  97  48  14  40  53  33  80  36 \\ \midrule
\textbf{Ta085} & 6262-6314 & 6270-\textbf{6285} &
 64  10  98  51  50  97  12  32  16  56  14  11  72  30  38  61  70  74  33  85  76  58  62   1  53  69  41  28  37   3  57  52  95  15  17  39  90  88  94  65  18   2  20   9  46  87  60  71   5   8 45  89   6   4  23  31  21  92  40  86  22  93  82  36  26  63  25  99  55  80  44  66  29  34  35  49  68  59  42  54  81  13  27  96   7  77  48 100  75  78  84  24  91  83  79  73  43  19  47  67 \\ \midrule
\textbf{Ta086} & 6302-6364 & 6307-\textbf{6331} & 
 83  32   2  12  21  33  39  92  37  69   1  86  31  82  88  89  56  72  38  25  80  76  71   9  65  55   7  34  10  66  94  23  75  52  58  26   8  99  63  40  60 100  98  81  95  68  48  97  16  45  77  17  20  91  84   6   5  14  62   3  87  35  15  28  57  30  79  46   4  67  13  64  24  36  85  78  27  50  54  74  93  41  70  44  18  53  22  49  11  51  73  19  43  61  29  59  47  96  42  90 \\ \midrule
\textbf{Ta087} & 6184-6268 & 6216-\textbf{6223} &
93 85 62 96 95 67 94 32 49 73 27 88 60 19 20 53 16 23 64 84 14 51 98 39 45 48 89 57 33 28 59 12 78 13 42 29 43 69 22 90 31 34 91 8 55 36 61 21 77 17 41 52 79 80 65 3 82 71 24 47 11 38 50 72 40 58 100 97 99 9 4 7 66 46 6 54 87 18 10 56 30 70 81 15 63 37 25 5 2 1 92 26 74 35 44 83 68 86 76 75 \\ \midrule
\textbf{Ta088} & 6315-6401 & 6331-\textbf{6385} & 
39  62  29  90  31  87  36  22  71   5  78  72  45  81  12  24  69  55   1  91  70  58  14  44  56  67  93  10  25   7   9  52  83  37  57  41  77  73  96  59  23  28 100  35  88  27  17  94  49  51  18  75  66  86  64   4  50  19  74   6   3  98  60  68  40  80  95  53  48  89  30  47  65  85  46  76  54  33  42  34  82  11  16  63  79  84   8  43  32  38  21  13   2  99  61  92  26  15  97  20 \\ \midrule
\textbf{Ta089} & 6204-6275 & 6232-\textbf{6247} & 
88  15   2  74  60  43  77  17  42  89  95  68  47  90  21  24  62  50  96  81  94   4  41  80  19  16  54  20  39  82  12  97  38  30  46  63  79  10  23  78  61  65  40  55  58  26  84  37  59  70  73  11  25  86  92  34  32  18   7  53  76  71  27  87  48  33  14  56  69  83 100  31   6   8  13  85  93  51  72  57   3  99  45  52   1   9  36  67  44  66  75  28  64  35  29  49  22   5  91  98 \\ \midrule
\textbf{Ta090} & 6404-6434 & \fbox{\textbf{6404}} & 
83 11 54 1 67 6 24 48 52 77 51 62 100 26 90 3 87 12 38 35 96 20 92 40 60 34 70 43 21 27 78 36 84 10 65 47 14 81 94 32 74 31 25 98 69 86 95 56 46 15 37 89 99 4 68 72 5 82 75 80 88 29 50 97 13 71 7 19 2 59 41 91 61 9 23 45 42 33 22 85 49 18 58 39 16 30 17 79 64 57 76 8 55 63 44 66 73 28 53 93\\
\bottomrule
\end{longtable}

\centering
\tiny
\begin{longtable}{cccp{12cm}}
\caption{Best-known solutions for instances \textit{Ta101}-\textit{Ta110} (200 jobs / 20 machines). Makespans shown in a box (\fbox{$C_{\max}$}) are optimal. Bold-faced instance-names indicate that the upper bound is improved and/or that the instance is solved for the first time.} 
\\ \hline
Inst & old LB-UB & LB-UB & Permutation schedule  \\
\midrule
\textbf{Ta101} & 11152-11195 & \fbox{11158} & 
 83 151 170  94 138  89  78 137 163 152 166 140 124  61  95  42 111  19 121  62  76  24 198  33 188  26 131  96 109 160 126 120  59  69 113  56 136 123  93  66  22  81  13 144  67 146 178 150  28  57 103 130  48   9 125  14 187 105 133  85 184   5 147 194  27 135  49 186  46   7 107 142 148 162  36  79 129  88 158   2 143  91 122  80 155  47  34  12  54  40  35 182  25  64 106  72 100 101 156  51  43  39 102  52 180 149 189 153 168 173 157 139  70   3 119 169  77  32 199 175  16 134   4  11   8   1  84 176  29 110  41  71  17 171 116  18 164 127  38 161  73  10  20  98 177  74  50   6  31  21  58 190  15  87  75 195 104  99 181 172  37 128 132 165 197  30 179  23 112  55 167 141  97 196 108 200  92  63  53  68 118  45  90 191 159 185 117 114 115 154 193  65 174  60 145  86 183  44 192  82 \\ \midrule
 \textbf{Ta102} & 11143-11203 & \textbf{11152-11160} & 
 56 184 171 11 25 163 54 169 50 118 149 132 60 91 19 94 179 151 119 155 92 45 176 13 49 26 57 74 32 168 48 128 16 28 113 159 178 173 138 31 105 197 52 156 146 111 58 96 15 122 47 144 134 30 182 103 152 69 38 185 150 5 77 116 183 187 100 174 88 181 8 43 84 162 41 170 108 59 193 4 23 115 107 20 17 148 80 78 104 35 157 186 166 70 29 120 175 136 112 46 97 65 114 93 24 9 66 51 192 195 42 142 95 129 39 135 154 188 14 190 167 68 137 139 198 33 130 189 34 71 76 158 99 10 161 147 62 143 75 124 1 172 177 36 121 98 64 72 191 6 125 2 110 160 21 55 63 102 79 85 131 200 27 164 123 101 22 199 81 40 141 126 90 165 61 86 83 18 127 117 196 109 145 73 7 37 153 12 140 82 106 194 133 87 89 53 67 3 44 180\\ \midrule
 \textbf{Ta107} & 11337-11360 & \fbox{\textbf{11337}} &
168 200 190  66 175  10  44 109  65 161  23 141 102 193 182  27 125 166  68 140  28   9  59   3  99  83 165  89  57 159  90 163 149 171 111 117  18 154  25 194  98 131  87  64 136   4 196 138 169 164  75  19  91  80 129 181  62  45 124 137 110  74 100  17  47  50 156 184 143  70  84  56  37  78  14  32 142  35  72  86  77 105  42 112 157 151 167 123  67  20  33  95 144  51  76  63 183   8 114  85  24 128   6  82  46 153  39  31  88  93  61  81 145  54 162 197  52 107 172 139  58 133  38 118 158 189  94 134 185 179   7 101 150  21 191 180 177  11 135 127 178  60 148  96 195 115  69 119  30 147  73  15  49   1  48  55 130 132  13 176  40  53 113 121  41  92 188 198   5  34  71 170  79 104 186  36 106 174 155 152  16  97 160  26 146 126 187  12  43  22 122 173  29 103   2 199 116 108 192 120 \\ \midrule
\textbf{Ta108} & 11301-11334 & \fbox{\textbf{11301}} & 
 52 192 105  39  30 196  76   6 121 136 112 157  21  11 200  10 191  91  73 102 155  61  24 174 142 167  28 119 129  96 126  62 182  87 149  44  74  86 194  32 133  26 115 139  54 188 114  88 137 154 148  98  27  95 124  64 198  17  72 123 199 146   7  93 122  55  56 173 140 164  42   2  49 165  18  92 159  63  29 153 113 107 111 169 131  67   8  47 179 187 117  82  75  15  71 162 104 145 161   1  41 181  48  90 100  13  79 180 183  20  77  59  40 189 166 135  84 118 178  35 147  34  23 184  12 103 134 163 195 132   9 125   5 128 172 143   3 151  46 120  83 171  31 158 170 101  51  66 144 193  65  89  43  16  85 130  37 175  22  19 177 138  25 141  78  50  38  36  68  45  53 116  69  57  94 168 160  60  58 185   4 109 197  33 176 186 110 106  14 190  97 108 156 127  81 150  80 152  99  70 \\ \midrule
 \textbf{Ta109}  & 11145-11192 & \fbox{\textbf{11146}} &
 55  10 166 190  32 199  19  23  25 101 108  77 106  72 111  37 170 176  57   4  91   6  21 100  70  29 123  16  17  79 121  41 198  27 103  47 194 120  74  69 186 113  38  61 196 175 116  68 181  76 177 126 185  86 136  28  83 197 132   3 112 167  75 154  54 139 169 163  12  60 153  80 157   9 109  89 133  39 155 178 141  88 191  43  44 125  59  53 137  31  81 118 149  48 143   7 127 182  97 193  33  62  35  49  90  52 200 195 184 104 102 188  95  46 187 159  66  15  42 140 147  65 128   5 183  13  85  63  26  11 179   2 129 156 115 142  34  24 144 161 192 165  22 172  45  73 189 162 131 150 107 138 105   8 180 171  51  18 119  87  96 146  78  99  82  36 114  67  56 164  98 122  14  93 134  64  94   1 152 110 160 151 168 158 174  84 173  50  71 145  30 130  40 124 148  20 117  58  92 135 \\
\bottomrule
\end{longtable}

\newcommand{\wbox}{\fcolorbox{white}{white}}

\newcommand{\ColSpace}{4.0em}
\centering
\tiny
\begin{longtable}{
 cc@{\hspace{\ColSpace}}
 cc@{\hspace{\ColSpace}}
 cc@{\hspace{\ColSpace}} cc }
\caption{Improved upper bounds and optimal makespans for \textit{VFR-small} instances. 
Makespans shown in a box (\fbox{$C_{\max}$}) are optimal. 
Bold-faced instance-names indicate that the upper bound is improved and/or that the instance is solved for the first time (reference UB~\citep{Vallada2015})
} 
\\ \hline
Instance & $C_{\max}$ & Instance & $C_{\max}$ & Instance & $C_{\max}$ & Instance & $C_{\max}$   \\
\midrule 
\textbf{30\_15\_1} & \fbox{2378} & 
\textbf{40\_15\_1} & \fbox{3004} & 
\textbf{50\_15\_1} & 3305          &  
\textbf{60\_15\_1} & 3926  \\
\textbf{30\_15\_2} & \fbox{2317} & 
\textbf{40\_15\_2} & \fbox{2816} & 
\textbf{50\_15\_2} & 3342   & 
\textbf{60\_15\_2} & 3865   \\
\textbf{30\_15\_3} & \fbox{2304} & 
\textbf{40\_15\_3} & \fbox{2904}  &
\textbf{50\_15\_3} & 3292  & 
\textbf{60\_15\_3}  & 3859  \\
\textbf{30\_15\_4} & \fbox{2444} &  
\textbf{40\_15\_4} & \fbox{2915}  & 
\textbf{50\_15\_4} & 3510  &  
\textbf{60\_15\_4}  & 3692  \\
\textbf{30\_15\_5} & \fbox{2421} &  
\textbf{40\_15\_5} & \fbox{2941}  & 
\textbf{50\_15\_5} & 3332  &  
\textbf{60\_15\_5}  & 3858  \\
\textbf{30\_15\_6} & \fbox{2306} &
\textbf{40\_15\_6} & \fbox{2804}  & 
\textbf{50\_15\_6} & 3341  &  
\textbf{60\_15\_6}  & 3868  \\
\textbf{30\_15\_7} & \fbox{2316} &  
\textbf{40\_15\_7} & \fbox{2863}  & 
\textbf{50\_15\_7} & 3475  &  
\textbf{60\_15\_7}  & 3791  \\
30\_15\_8 & \fbox{2366} &  
\textbf{40\_15\_8} & \fbox{2896} & 
\textbf{50\_15\_8} & 3420  &  
\textbf{60\_15\_8}  & 3727  \\
\textbf{30\_15\_9} & \fbox{2259} &  
\textbf{40\_15\_9} & 2705  & 
\textbf{50\_15\_9} & 3194  &  
\textbf{60\_15\_9}  & 3784  \\
{30\_15\_10} & \fbox{2385} &  
40\_15\_10 & 2945  & 
\textbf{50\_15\_10} & 3394 &  
\textbf{60\_15\_10} & 3882  \\
\midrule
\textbf{30\_20\_1} & \fbox{2643} &  
\textbf{40\_20\_1} & 3317 &
\textbf{50\_20\_1} & 3683         &
\textbf{60\_20\_1}  & 4144  \\
30\_20\_2 & \wbox{2835} &  
\textbf{40\_20\_2} & 3224 & 
\textbf{50\_20\_2} & 3704 &  
\textbf{60\_20\_2}  & 4274   \\
\textbf{30\_20\_3} & \fbox{2783} &
\textbf{40\_20\_3} & 3224  &
\textbf{50\_20\_3} & 3773  &  
\textbf{60\_20\_3}  & 4341  \\
\textbf{30\_20\_4} & \fbox{2680} &  
\textbf{40\_20\_4} & 3227  & 
\textbf{50\_20\_4} & 3702  &  
\textbf{60\_20\_4}  & 4175  \\
{30\_20\_5} & \wbox{2672} &  
\textbf{40\_20\_5} & 3050  & 
\textbf{50\_20\_5} & 3622  & 
\textbf{60\_20\_5}  & 4180  \\
\textbf{30\_20\_6} & \fbox{2715} &  
\textbf{40\_20\_6} & 3184  & 
\textbf{50\_20\_6} & 3779  &  
\textbf{60\_20\_6}  & 4184  \\
{30\_20\_7} & \wbox{2712} &  
40\_20\_7 &   & 
\textbf{50\_20\_7} & 3689  & 
\textbf{60\_20\_7}  & 4251  \\
{30\_20\_8} & \wbox{2812} &  
\textbf{40\_20\_8} & 3261  & 
\textbf{50\_20\_8} & 3775  &  
\textbf{60\_20\_8}  & 4171  \\
\textbf{30\_20\_9} & \fbox{2795} &  
\textbf{40\_20\_9} & 3332  & 
\textbf{50\_20\_9} & 3799  & 
\textbf{60\_20\_9}  & 4198  \\
{30\_20\_10} & \wbox{2805} &  
\textbf{40\_20\_10} & 3115  & 
\textbf{50\_20\_10} & 3756 &  
\textbf{60\_20\_10} & 4186  \\
\bottomrule
\end{longtable}

\centering
\tiny
\begin{longtable}{
cc@{\hspace{\ColSpace}}
cc@{\hspace{\ColSpace}}
cc@{\hspace{\ColSpace}}
cc }
\caption{Improved upper bounds and optimal makespans for \textit{VFR-large} instances. 
Makespans shown in a box (\fbox{$C_{\max}$}) are optimal. 
Bold-faced instance-names indicate that the upper bound is improved and/or that the instance is solved for the first time (reference UB from~\citep{Vallada2015})
} 
\\ \hline
Instance & $C_{\max}$ & Instance & $C_{\max}$ & Instance & $C_{\max}$ & Instance & $C_{\max}$   \\
\midrule 
\textbf{100\_20\_1} & 6121 &  
\textbf{200\_20\_1} & 11181 & 
\textbf{300\_20\_1} & \fbox{15996} &  
\textbf{400\_20\_1} & \fbox{20952}  \\
\textbf{100\_20\_2} & 6224 &  
\textbf{200\_20\_2} & 11254 & 
\textbf{300\_20\_2} & 16409 &  
400\_20\_2 & \fbox{21346}  \\
\textbf{100\_20\_3} & 6157 &  
\textbf{200\_20\_3} & 11233 & 
\textbf{300\_20\_3} & \fbox{16010} &  
\textbf{400\_20\_3} & \fbox{21379}  \\
\textbf{100\_20\_4} & 6173 &  
\textbf{200\_20\_4} & 11090 & 
\textbf{300\_20\_4} & \fbox{16052} &  
\textbf{400\_20\_4} & \fbox{21125}  \\
\textbf{100\_20\_5} & 6221 &  
\textbf{200\_20\_5} & \fbox{11076} & 
\textbf{300\_20\_5} & \fbox{21399} &  
400\_20\_5 & \fbox{16245}  \\
\textbf{100\_20\_5} & 6247 &  
\textbf{200\_20\_6} & 11208 & 
300\_20\_6 & \fbox{16021}&  
\textbf{400\_20\_6} & \fbox{21075}  \\
\textbf{100\_20\_7} & 6358 &  
\textbf{200\_20\_7} & \fbox{11266} & 
\textbf{300\_20\_7} & \fbox{16188} &  
400\_20\_7 & \fbox{21507}  \\
\textbf{100\_20\_8} & 6023 &  
\textbf{200\_20\_8} & \fbox{11041} & 
\textbf{300\_20\_8} & \fbox{16287} &  
400\_20\_8 & \fbox{21198}  \\
\textbf{100\_20\_9} & 6286 &  
\textbf{200\_20\_9} & 11008 & 
\textbf{300\_20\_9} & \fbox{16203} & 
400\_20\_9 & \fbox{21236}  \\
\textbf{100\_20\_10} & 6048 &  
\textbf{200\_20\_10} & \fbox{11193} & 
300\_20\_10 & \fbox{16780} &  
400\_20\_10 & \fbox{21456}  \\
\midrule
\textbf{600\_20\_5} & \fbox{31323} &
\textbf{800\_20\_7} & \fbox{41342} &
& &
&  \\
\bottomrule
\end{longtable}

\end{document}